\documentclass[review, preprint, 12pt]{elsarticle}
\usepackage{paralist}
\usepackage{setspace}
\usepackage[margin = 2.7cm]{geometry}

\usepackage[]{caption}
\usepackage{amsthm}
\usepackage[psamsfonts]{amsfonts}
\usepackage{amsmath}
\usepackage{amssymb}

\usepackage{enumerate}

\usepackage{epsfig}
\usepackage{bbm}
\usepackage{verbatim}
\usepackage{graphics}
\usepackage{graphicx}
\usepackage{curves}

\usepackage{algorithmic}
\usepackage{algorithm}
\usepackage{amsfonts}
\usepackage{url}
\usepackage[mathscr]{euscript}

\renewcommand{\l}{\left}
\renewcommand{\r}{\right}

\newcommand{\argmin}{\operatornamewithlimits{argmin}}

\newcommand{\conv}{\operatorname{conv}}

\newcommand{\temph}[1]{{\sl #1}}

\DeclareSymbolFont{AMSb}{U}{msb}{m}{n}
\DeclareMathSymbol{\N}{\mathbin}{AMSb}{"4E}
\DeclareMathSymbol{\ZZ}{\mathbin}{AMSb}{"5A}
\DeclareMathSymbol{\Y}{\mathbin}{AMSb}{"59}
\DeclareMathSymbol{\U}{\mathbin}{AMSb}{"55}
\DeclareMathSymbol{\RR}{\mathbin}{AMSb}{"52}
\DeclareMathSymbol{\Q}{\mathbin}{AMSb}{"51}
\DeclareMathSymbol{\Prob}{\mathbin}{AMSb}{"50}
\DeclareMathSymbol{\HHH}{\mathbin}{AMSb}{"48}
\DeclareMathSymbol{\I}{\mathbin}{AMSb}{"49}
\DeclareMathSymbol{\C}{\mathbin}{AMSb}{"43}
\DeclareMathSymbol{\E}{\mathbin}{AMSb}{"45}
\DeclareMathSymbol{\A}{\mathbin}{AMSb}{"41}
\DeclareMathSymbol{\C}{\mathbin}{AMSb}{"43}
\DeclareMathSymbol{\D}{\mathbin}{AMSb}{"44}
\DeclareMathSymbol{\OO}{\mathbin}{AMSb}{"4F}
\DeclareMathSymbol{\X}{\mathbin}{AMSb}{"58}
\DeclareMathSymbol{\LL}{\mathbin}{AMSb}{"4C}

\newcommand{\Cl}{\mathcal{C}}

\newcommand{\Yl}{\mathcal{Y}}
\newcommand{\Al}{\mathcal{A}}
\newcommand{\Bl}{\mathcal{B}}
\newcommand{\Xl}{\mathcal{X}}

\newcommand{\Ul}{\mathcal{U}}

\newcommand{\PPl}{\mathcal{P}}
\newcommand{\Pl}{\mathscr{P}}

\newtheorem{theorem}{Theorem}[section]

\newtheorem{proposition}{Proposition}[section]
\newtheorem{definition}{Definition}[section]

\theoremstyle{definition}

\newenvironment{flushitem}{
\begin{itemize}
  \setlength{\leftmargin}{0cm}
     \setlength{\itemsep}{0pt}
    \setlength{\topsep}{0pt}
    \setlength{\partopsep}{0pt}
    \setlength{\parsep}{0pt}
    \setlength{\parskip}{0pt}
}{\end{itemize}}

\usepackage{amssymb}

\journal{Electric Power Systems Research}

\begin{document}

\begin{frontmatter}

\title{A Composable Method for Real-Time Control of Active Distribution Networks with Explicit Power Setpoints. \\ Part I: Framework}

\author[lca2]{Andrey Bernstein \fnref{fn1}}
\ead{andrey.bernstein@epfl.ch}
\author[desl]{Lorenzo Reyes-Chamorro\corref{c1} \fnref{fn1}}
\cortext[c1]{Corresponding Author. Phone number: +41 21 69 37369, Postal address: EPFL STI IEL DESL, ELL037, Station 11, CH-1015 Lausanne}
\fntext[fn1]{Both authors contributed equally in this research work.}
\ead{lorenzo.reyes@epfl.ch}
\author[lca2]{Jean-Yves Le Boudec}
\ead{jean-yves.leboudec@epfl.ch}
\author[desl]{Mario Paolone}
\ead{mario.paolone@epfl.ch}

\address[lca2]{Laboratory for Communications and Applications 2, \'{E}cole Polytechnique F\'{e}d\'{e}rale de Lausanne, CH-1015 Lausanne, Switzerland}
\address[desl]{Distributed Electrical Systems Laboratory, \'{E}cole Polytechnique F\'{e}d\'{e}rale de Lausanne, CH-1015 Lausanne, Switzerland}
\begin{abstract}
The conventional approach for the control of distribution networks, in the presence of active generation and/or controllable loads and storage, involves a combination of both frequency and voltage regulation at different time scales.
With the increased penetration of stochastic resources, distributed generation and demand response, this approach shows severe limitations in both the optimal and feasible operation of these networks, as well as in the aggregation of the network resources for upper-layer power systems.
An alternative approach is to directly control the targeted grid by defining explicit and real-time setpoints for active/reactive power absorptions/injections defined by a solution of a specific optimization problem; but this quickly becomes intractable when systems get large or diverse. In this paper, we address this problem and propose a method for the explicit control of the grid status, based on a common abstract model characterized by the main property of being \emph{composable}. That is to say, subsystems can be aggregated into virtual devices that hide their internal complexity. Thus the proposed method can easily cope with systems of any size or complexity.
The framework is presented in this Part I, whilst in Part II we illustrate its application
to a CIGR\'{E} low voltage benchmark microgrid. In particular, we provide implementation examples with respect to typical devices connected to distribution networks and evaluate of the performance and benefits of the proposed control framework.
\end{abstract}


\end{frontmatter}

\section*{Acronyms}
\begin{tabular}{ll}
GA	& 	Grid Agent\\
PCC &	Point of Common Coupling\\
RA 	& 	Resource Agent
\end{tabular}

\section*{Nomenclature}
\hspace*{-1cm}
\begin{tabular}{rl}
$u=(P_1,Q_1,P_2,Q_2,...,P_n,Q_n)$ & Control (target) setpoints\\
$x = (P'_1,Q'_1,P'_2,Q'_2,...,P'_n,Q'_n)$	& Implemented (actual) setpoints \\
$\hat{x}$ & Current (estimated) setpoints \\
$\Al_i$	& $PQt$ profile of follower $i$ \\&(set of possible target values for $(P_i,Q_i)$)\\
$C_i(P_i, Q_i)$	& Virtual cost of follower $i$  \\
$BF_i(P_i, Q_i)$		& Belief function of follower $i$ \\
& (set of possible $(P'_i,Q'_i)$ when $(P_i,Q_i)$ is requested)\\
$\Al \triangleq \Al_1\times\Al_2\times ...,\times \Al_n $		& Joint $PQt$ profile \\
&(set of possible values for $u$) \\
$\widetilde{\Al}_0$      & Exact aggregated $PQt$ profile \\
&(set of possible values for $(P_0,Q_0)$ power at PCC)	\\
$\widetilde{\Al}^*_0$ & Approximate aggregated $PQt$ profile 	\\
$BF(u)= BF_1(P_1, Q_1)\times ...\times BF_n(P_n, Q_n)$ & Joint belief function\\
&(set of possible $x$ when $u$ is targeted)\\
$\widetilde{BF}_0(P_0, Q_0)$ & Exact aggregated belief function\\
&(set of possible $(P'_0,Q'_0)$ when $(P_0,Q_0)$ is targeted)\\
$\widetilde{BF}_0^*(P_0, Q_0)$ & Approximate aggregated belief function\\
$\Ul$	&  Set of admissible setpoints $u$\\
$J$ & Penalty for electrical state feasibility\\
$J_0$ &	Penalty for power flow deviation at the PCC\\
\end{tabular}

\section{Introduction}

The modern and future electrical infrastructure has to satisfy two main conflicting requirements: (i) provide reliable and secure supply to an increasing number of customers, and (ii) take into account the rational use of energy and the protection of the environment.
This last requirement drives major changes in power systems, where the most evident result is an almost quadratic increase of the connection of renewable energy sources \cite{INV_WORLD_ELE}.
It is generally admitted that these sources need to be massive and distributed, in order to provide a significant part of the consumed electrical energy (e.g. \cite{mackay2008sustainable}).
However, the increased penetration of distributed and/or renewable energy-resources in electrical medium and low-voltage networks is such that, in several countries, operational constraints have already been attained. This calls for a radical re-engineering of the entire electrical infrastructure.
Conventional approaches are unable to scale to such an increase in complexity.

As known, the main controls of an interconnected power system are essentially concerned with
\begin{inparaenum}[(i)]
\item maintaining the power balance and
\item maintaining the voltage levels close to the rated values,
\end{inparaenum}
both performed at various time scales.
These two basic controls are the building blocks used by other more sophisticated regulators responsible for hierarchically superior actions (e.g., angular and voltage stability assessment, congestions in main transmission corridors, etc.).
As well known, the control of (i) is based on the link between the power imbalance and the network frequency (that constitutes the control variable) and it is usually deployed in three main time-frame controls that belongs to primary, secondary and tertiary frequency controls.
There are essentially two main drawbacks to this control philosophy: First, there is a monotonous increasing dependency between the primary/secondary frequency-control reserves and the errors associated with the forecasts of increasing renewable production (especially when distributed in small dispersed units).
Second, the definition of the primary/secondary frequency-control reserves are centralized; hence, distributed control mechanisms, to be deployed in distribution networks with active resources, cannot be easily implemented.
These mechanisms will require increasing reserve scheduling in order to keep acceptable margins and to maintain the grid vulnerability at acceptable levels (e.g. \cite{OptPrimaryReserve}).
An example of such a principle is described in \cite{ENTSO-E_OpHB}.

As for the control of (ii), which requires maintaining the voltage deviations within predetermined limits (e.g., \cite{ENTSO-E_DeConn}), it is implemented at various levels and with different strategies that mainly control reactive-power injections.
However, network voltages fluctuate as a function of various quantities such as the local and overall network load, generation schedule, power system topology changes and contingencies.
The typical approach for voltage-control divides (still into primary, secondary and tertiary controls) the control actions as a function of their dynamics and their area of influence.
The major advantage of such an approach is that it enables a decoupling of the controllers as a function of their area of influence.
However, it is not easily down-scalable to distribution networks because, similarly to the frequency control, it was conceived for interconnected power systems, where the control resources are limited in number, large in size, and centrally controlled.

In general, if we base the equilibrium of the grid in terms of purely power injections, there is always the need to assess adequate reserves in order to guarantee the power balance (both active and reactive) of the system.
In agreement with this approach, the European Network Transmission Systems Operator (ENTSO-E) attempts to extend to distribution the network codes that set up a common framework for network connection agreements between network operators and demand/producers owners \cite{ENTSO-E_DeConn}.
This specific network code requires the distribution networks to provide the same frequency/voltage support provided by other centralized resources (i.e., power plants) directly connected to transmission networks.
Such an approach, however, has many drawbacks in systems characterized by dominant non-dispatchable stochastic renewable energy resources where, to balance the power, the non-desirable use of traditional power plants (usually gas-fired power plants e.g. \cite{MMOp_CCGT} or, when available, hydro power plants) is necessary.
In contrast, if in distribution networks it is possible to expose to a \emph{grid controller} the state of each energy resource (i.e., generation sources, storage systems, and loads) in a scalable way, then it is possible, in principle, to always find an admissible and stable system-equilibrium point with small or negligible power-balancing support from the external grid.
This feature will enable the graceful operation of each local distribution network in both islanded and grid-connected operation modes, thus allowing, for this last one, the possibility of quantifying the amount of the microgrid's ancillary services to the upper power network (i.e., primary and secondary frequency control support, as well as voltage compensation).
Directly controlling every resource however is clearly too complex when resources are numerous and diverse.
This is the challenge we propose to tackle.

Our goal is therefore to define a \emph{scalable} method for the direct and explicit control of real-time nodal power injections/absorptions.
We use software agents, which are responsible for subsystems and resources, and we communicate with other agents in order to define real-time setpoints. To make our method scalable, we use the following features:

\begin{trivlist}
\item[~(a)] \emph{Abstract Framework}. It applies to all electrical subsystems and specifies their capabilities, expected behavior, and a simplified view of their internal state.
    A subsystem advertizes its internal state by using \emph{$PQt$ profiles}, \emph{virtual costs} and \emph{belief functions}, which are expressed using a common device-independent language (Section~\ref{sec:pvcp}).

The existence of a common abstract framework is an essential step for scalability and composability. It was applied, for example, to the control of very large and heterogeneous communication networks in \cite{ForwPHB}.
\item[~(b)] \emph{Composition of Subsystems}. It is possible to aggregate a set of interconnected elements into a single entity.
A local grid with several generation sources, storage facilities and loads can be viewed by the rest of the network as a single resource.
\item[~(c)] \emph{Separation of Concern}. Agents that are responsible for grids (henceforth called ``Grid Agents") manipulate only data expressed by means of the abstract framework and do not need to know the specific nature of the resources in their grid; in particular, there is only one grid agent software for all instances of grid agents. In contrast, resource agents (which are responsible for specific resources)  are specific, but their function is simpler, as it is limited to (i) mapping the internal state of the resource and expressing it in the proposed abstract framework and (ii) implementing the power setpoints received from the grid agent with which they communicate. In other words, agents that need to know details of diverse systems are simple-minded, whereas agents that need to take intelligent decisions have an abstract, simple view of the grid and of their resources.
\end{trivlist}

In view of the complexity of the proposed approach, the paper has been divided into two parts.
In Part I, we give the formal description of the proposed method. In Part II, we present the detailed application with the reference to actual resources connected to active distribution networks and evaluate the performance of the proposed method in a CIGR\'{E} low voltage benchmark microgrid. 
The structure of this first part is the following.
In Section~\ref{sec:soa}, we discuss the state of the art.
In Section \ref{sec:agents}, we present the definition of agents and their interaction and give a global overview of our method. The abstract framework is described in
Section \ref{sec:pvcp} ($PQt$ profiles, virtual cost, and belief functions). In Section~\ref{sec:one_step_opt}, we present the details of the decision process performed in the grid agent.
In Section \ref{sec:comp}, we discuss the composability property and propose methods for aggregation of subsystems.
Finally, we close this part with concluding remarks in Section~\ref{sec:conc}.

\section{State of the Art}
\label{sec:soa}

The literature on the real-time control of microgrids in the presence of stochastic generation tackles the problem by using two main approaches. The first one relies on centralized \emph{stochastic optimisation} control, and the second uses \emph{multi-agent-based control systems} (MAS).

The first approach relies on the possibility of quantifying and using the statistical distributions of both the stochastic generation and the loads in a central controller/dispatcher. In general, the controller/dispatcher is responsible for the solution of an optimal dispatching problem constrained by the grid operation limits. In \cite{stoch1}, a scheduling of microgrid resources is proposed; it accounts for the stochasticity of the wind and plug-in vehicles by means of a brute-force scenario-generation and reduction based on a-priori known statistical distributions of these stochastic variables. In \cite{stoch2}, it is proposed to solve the microgrid dispatch problem, together with its optimal configuration, by representing the stochasticity of renewable resources/loads via a forecasting tool based on the support vector regression technique. The authors of \cite{stoch3} propose a power scheduler aimed at minimizing the microgrid net cost, where the utility of the dispatchable loads accounts for the worst-case transaction cost inferred from the uncertainties in the renewable generation.

In \cite{Sossan_FC} the stochasticity is faced by using a Model Predictive Control strategy when coupling traditional space heating sources with combined heat and power units to achieve \emph{energy replacement}.

The other approach discussed in the literature of microgrids control relies on MAS (e.g., \cite{rehtanz2003autonomous}). In this context, MAS are proposed as a step towards the distribution of control.
Optimization goals in previously proposed methods (e.g., \cite{MG_PM_DA} and \cite{MAS_Opt_MG}) consider the operational costs of the system without accounting for the operational constraints such as voltage magnitudes or line congestions.
More in particular, the MAS approach presented in \cite{MAS_Opt_MG} relies on the availability of droop control that is suitably adjusted by MAS negotiations. However, this method does not express the specific states of the resources in the device-independent advertisements nor does it consider the grid state to ensure an appropriate grid Quality-of-Supply (QoS) and a feasible operation of the grid.
The authors of \cite{MAS_sched} present a centralized control scheme that uses MAS for generation scheduling and demand-side management for secondary frequency regulation, in order to optimize the operational cost of a microgrid in both grid-connected and islanded modes.
However, the method proposed there does not account for the operational constraints associated with the grid and, also, does not take into account the sub-second time constraints associated with the short-term volatility of stochastic resources. Therefore, this method does not appear to be a \emph{real-time} control method.
The case of the post-fault microgrids behaviour is discussed in \cite{MAS_distributed}, and the design principles of a corresponding MAS-based real-time control method are presented. In particular, the proposed method is designed to achieve fast load-shedding strategies in order to maintain the real-time power balance of the microgrid and, as a consequence, avoid its collapse. Their paper does not discuss the use of the MAS with respect to the optimal operation of the grid in normal operating conditions. Additionally, similarly to the other references, the proposed method does not express the internal states of the resources to the other agents and it is not scalable.

Our approach goes several steps beyond.
First, we base our method on a unified, abstract representation of devices and subsystems, which is a central element for simple design and correctness by construction.
Second, our approach is \emph{composable}, i.e., entire subsystems can be abstracted in the same way of a device.
This characteristic makes our approach fully scalable from low-voltage microgrids to medium-voltage distribution networks.
Third, we target stringent real-time control. Specifically, we propose a formal approach capable to close the agent negotiation and deployment of control setpoints in sub-second time scales.

\vspace*{-0.5cm}
\section{Agents and the Interaction between Them}
\label{sec:agents}
\begin{figure}[h!]
\centering
\includegraphics[width=0.8\textwidth]{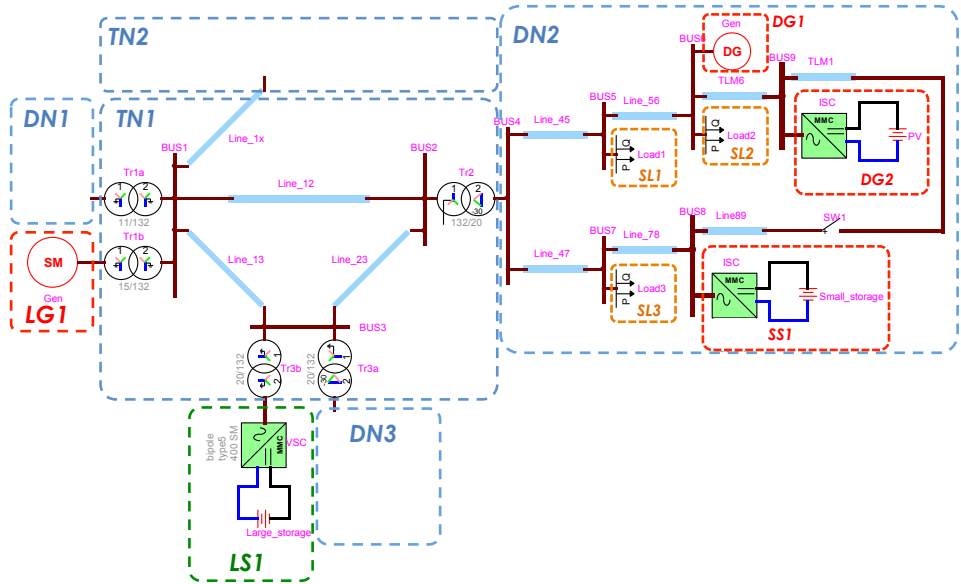}
\caption{Modern interconnected power systems configuration with active distribution networks.\label{fig:real_net}}
\end{figure}

We rely on the current structure of power networks, essentially composed of a number of subsystems interconnected at different voltage levels.
Each subsystem is constituted of \emph{electrical grids} and \emph{resources}:
loads, generators, and storage devices.
For the sake of clarity, we use the example in Fig. \ref{fig:real_net} where a sub-transmission network, with a meshed structure (TN1), interconnects a neighbour transmission network (TN2), a large generator (LG1), a large storage systems (LS1) and distribution networks (DN1, DN2, DN3) that have local generation and storage devices.
The figure also shows details of the distribution network DN2, where we can identify a storage system (SS1), a mini-hydro power plant (DG1), a photovoltaic installation (DG2), as well as secondary substations that represent the local loads (SL1, SL2, SL3).

We use software \emph{agents}, i.e., pieces of software that are able to speak for, and control, a set of electrical systems.
An agent can be associated with a resource, or an entire system including a grid and/or a number of devices. An agent can be implemented in a stand-alone processor, as a process on a control computer, or as an embedded system.
Small systems such as appliances, boilers or small photovoltaic roofs do not necessarily need to have a specific agent. Instead, they can be controlled and represented by one single group-aggregating agent that uses a broadcast protocol such as GECN \cite{GECN, GECN2}.
An agent that controls an entire grid is called ``Grid Agent" (GA); other agents are called ``Resource Agents" (RAs). Each agent is assigned a role of a \emph{leader} of one or more other agents that we term the \emph{followers} of that leader. The roles follow the hierarchy of distribution and transmission networks. For example, in Fig. \ref{fig:real_net}, the grid agent of DN2 is a leader of the resource agents of SS1, DG1, DG2, and SLx. Also, DN2's Grid Agent is a follower of TN1's Grid Agent. Resource Agents are always followers, but grid agents can assume both roles.
Agents communicate using a simple Advertisement/Request protocol, as follows.
\begin{trivlist}
\item[~~1.] A follower agent (GA or RA) periodically advertises an abstract view of the internal state of the resource or grid to its leader in the form defined in Section \ref{sec:pvcp}.
\item[~~2.] A leader agent (GA) has knowledge of the state of its electrical grid and uses the information received from its followers, together with the requested setpoints from its own leader, in order to compute the requested power setpoints of its followers (Section~\ref{sec:one_step_opt}); the setpoints are then sent to all the followers.
\item[~~3.] On receiving the requested setpoints, the followers set, if possible, their operation according to the requested setpoints and respond with a new advertisement that also serves as a confirmation to the leader that the setpoints were set.
\end{trivlist}
The process is repeated periodically every $T$ time units.
In this paper we take $T=100ms$, a value short enough to cope with the fastest possible volatility of distributed resources and large enough to be compatible with the need to estimate the electrical state of the grid. However, this time can be shortened down to the limit of the adopted telecom infrastructure.
We also assume that a GA receives measurements from the grid under its responsibility; they are used to estimate the electrical state. The measurement messages are also sent periodically, with a period $\tilde{T}=20ms$.
Such a time period is compatible with the data frames of modern monitoring systems equipped, for instance, with phasor measurement units (PMUs). These devices typically provide synchrophasor measurements ranging from 10 to 60 frames-per-second, as required by the IEEE Std.~C37.118 \cite{IEEE_Std_Synch, IEEE_Std_Synch_DT}.
Examples of time latencies and accuracy of real-time state estimation processes fully based on PMU data are discussed in \cite{TPh_SE, RT_SE_ADN}.
Typical time latencies of less than $200ms$ can be achieved in order to determine the system state with relevant refresh rates of some tens of milliseconds.

\section{Virtual Cost, $PQt$ profile and Belief Function}
\label{sec:pvcp}
In this section, we describe the elements of the advertisement messages that are sent by the followers to their leader.
\subsection{Virtual Cost and $PQt$ Profile} An agent advertises to its leader a region in the $(P,Q)$ plane (for active and reactive power) that the subsystem, under the control of this agent, can deploy (\emph{negative power means consumption}). More precisely, the\emph{ $PQt$ profile} advertised by the agent is the set of $(P,Q)$ values that this subsystem is willing to implement in the next iteration. In addition, the agent also advertises a \emph{virtual cost} function, defined for every $(P,Q)$ in the $PQt$ profile. The virtual cost is interpreted as the cost to this subsystem of applying a requested power setpoint.
It is worth observing that this virtual cost does not make reference to a real electricity cost. On the contrary, its role is to quantify the propensity 
of this subsystem to deploy $(P, Q)$ setpoints within particular zones of the $PQt$ profile.
For instance, if the subsystem is a storage system, when the state of charge is close to $100\%$, its agent advertises a negative cost for positive $P$ and positive cost for negative $P$, thus signalling to the grid agent that the storage system would prefer to be discharged. Note that agents do \emph{not} advertise device specific information such as state of charge; this is an intentional feature of our approach, because keeping the advertised information generic enables aggregation and composition of systems.

\subsection{Belief Function}
A follower agent also advertises its \emph{belief function}, which returns the set of all possible (actual) setpoints that this subsystem might implement. Formally, we call $BF$ the belief function of an agent and assume that it receives from its leader GA a request to implement a setpoint $(P,Q)$; then the \emph{actual} setpoint $(P',Q')$ that this subsystem does implement lies in the set $BF(P,Q)$ with overwhelming probability.

The belief function accounts for the uncertainty in subsystem operation. In particular, highly controllable subsystems, such as batteries and generators, are expected to have (almost) ideal beliefs, namely $BF(P,Q) = \{(P,Q)\}$. For subsystems such as PV/wind farms, or loads, the belief function will return larger sets, to account for their volatility.

It is important to underline the difference between a $PQt$ profile and a belief function: the former indicates the deployable setpoints that this subsystem is willing to receive, whereas the latter indicates all the possible operating conditions that might result from applying a requested setpoint due to the stochasticity of the process controlled by the resource agent. In particular, the $PQt$ profile represents \emph{the domain of definition} of both the virtual cost function and the belief function.

Recall that the advertisement messages (with
$PQt$ profiles, virtual costs and belief functions) are sent not only by RAs to their GA but also by a GA to its own leader, by means of the \emph{composability} property described in Section~\ref{sec:comp}.

\subsection{Valid Approximations of $PQt$ Profiles and Belief Functions}
\label{sec:simp}
As our goal is to develop a \emph{real-time} system that completes its cycle in approximately $100ms$, we often use simplifications.
To be valid (i.e., to keep the system in a feasible electrical state), any simplification must satisfy the following property:
\begin{definition}[Validity Property for $PQt$ profiles and Belief Functions] \label{def:validity}
$(\Al_i, BF_i)$ is a valid pair of $PQt$ profile and belief function for a given subsystem $i$ if, whenever this subsystem receives a target setpoint $(P,Q)\in\Al_i$, the actual power injected by this subsystem lies in the set $BF_i(P,Q)$.
\end{definition}

As will become clear below, a grid agent can simplify its computation by using approximations of the advertised $PQt$ profiles and belief functions, instead of those sent by the followers, as long as the approximation satisfies the validity property. In particular, if we replace an original $PQt$ profile and belief function by an approximating \emph{subset} for the $PQt$ profile and approximating \emph{supersets} for the belief function, then the approximation is valid. We present concrete examples of such simplifications in Sections \ref{sec:simp_adm} and \ref{sec:pqt_simpl}.

\section{The Grid Agent's Decision Process}
\label{sec:one_step_opt}
\begin{figure}[h!]
\centering
\includegraphics[width=0.7\textwidth]{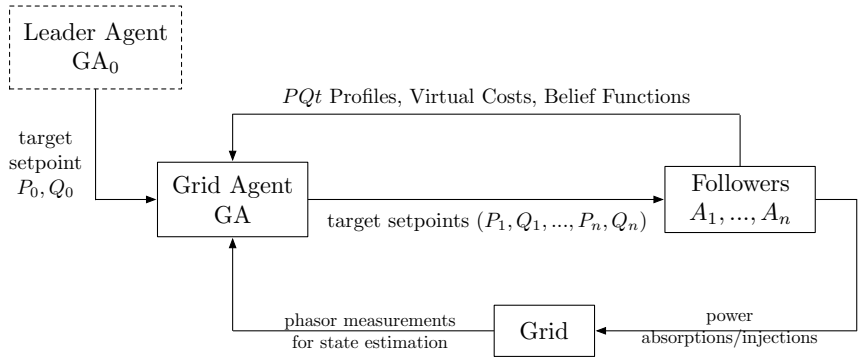}
\caption{Illustration of the decision process made by GA.
 \label{fig:decision}}
\end{figure}

In this section, we describe the decision process performed by the Grid Agent (GA), as illustrated in Figure~\ref{fig:decision}.
Consider a GA responsible for a grid in which there are $n$ subsystems, each with
their own agent $A_1...A_n$ (recall that some of these subsystems may be entire grids). This GA can also be following a leader grid agent, say GA$_0$.

At every time step, this GA receives the $PQt$ profiles, virtual costs and belief functions of followers $A_1,...,A_n$, the target setpoint $(P_0,Q_0)$ for the power flow to/from its leader grid (requested by the leader GA$_0$),
and an estimation of the electrical state of the grid.
The goal of the GA is then to \emph{steer}, using frequent updates, the electrical state of its grid by explicitly setting the power setpoints so that (i) the virtual costs of the followers are minimized, (ii) the setpoint $(P_0,Q_0)$ is satisfied as much as possible and that (iii) the grid is in a \emph{feasible state of operation}, as defined in Section~\ref{sec:feasible}.

To reach its goal, the GA applies its \emph{decision process}, as described in Section~\ref{sec:decision}, and obtains target power setpoints $(P_1,Q_1,...,P_n,Q_n)$ that are then communicated to the follower agents $A_1,...,A_n$. All subsystems do their best to implement the target setpoints, but subsystem $i$ might end up applying any actual setpoint $(P'_i, Q'_i)\in BF_i(P_i,Q_i)$ where $BF_i$ is the belief function previously advertised by $A_i$. The process is then repeated at the next time step.

Observe that in our hierarchy of agents, there can be two
possible types of GAs: (i) a GA that is connected to a higher level network with a higher network's GA being its leader, and (ii) a GA that has no leader and works in an \emph{islanded mode} (e.g., GA$_0$ in the above example). We term the former \emph{internal GA} (representing the internal node in the tree hierarchy) and the latter \emph{root GA} (representing the root of the tree hierarchy). A root GA naturally relies on \emph{a slack resource} (such as a storage system or synchronous generator working in the voltage control mode) in order to operate its grid. Thus, in this case, one of the follower agents becomes a \emph{slack agent}.

In the following subsections, we describe the details of the GA's decision process, emphasizing the difference between these two types of GAs.

\subsection{Feasible State of Operation and Admissible Target Setpoints}\label{sec:feasible}

In the case of the internal GA, we consider that the grid under the control of the GA is connected to the grid of the leader grid agent, say GA$_0$, at a point of common coupling (PCC).
We assume that GA knows the equivalent Thevenin impedance between the PCC and the slack bus of the overall system. This impedance can be computed using a given methodology such as \cite{ImpEst_KF, ImpEst_RT_WB, ImpEst_AdapControl}. Thus, for the computation of the load-flow, we consider as a slack bus the node after this impedance that is included in the admittance matrix of the grid.

In the case of the root GA, the slack bus is naturally the bus where the slack resource is connected.

The \emph{electrical state} of this grid is given by the set $\{\delta_k, V_k\}$ of the voltage angles and magnitudes at different buses $k$, with $\delta_0 = 0^{\circ}$ and $V_0$ is fixed (given) by convention of the slack bus.
We assume that the GA has the means to estimate the electrical state of its grid with a sufficient refresh rate compatible with the frequency of power setpoints updates.

We say that an electrical state is \emph{feasible} if
\begin{enumerate}
\item[(i)] It satisfies static conditions on voltage and currents, of the form
\begin{equation} \label{eqn:elec_safe}
V_k \in [V_k^{nom} - \beta_k, V_k^{nom} + \beta_k], \quad I_{\ell} \leq I_{\ell}^{\max}
\end{equation}
where $V_k^{nom}$ is the nominal voltage value of $V_k$ (which depends on the voltage rating of the GA's grid), $I_{\ell}$ is the current magnitude at a line\footnote{If the current getting into line $(k,k')$ is different from the one getting into line $(k',k)$, $I_{\ell}$ is the maximum one among them.} $\ell = (k, k')$, and $\beta_k$ and $I_{\ell}^{\max}$ are given threshold variables.
\item[(ii)] The power injection at the slack bus is within a specified region $R_0$:
\begin{equation} \label{eqn:elec_safe-slack}
(P_0,Q_0) \in R_0.
\end{equation}
\end{enumerate}
Note that, for the sake of clarity, the concept of \emph{feasible electrical state} defined by \eqref{eqn:elec_safe} and \eqref{eqn:elec_safe-slack} has been intentionally simplified to include only these steady-state feasibility conditions. This concept can be further extended to take into account other conditions that formalize the \emph{dynamic stability} of a grid such as voltage and angular stability. In this respect, recent literature has discussed the stability aspects of microgrids in the case of constant-power injections/absorptions actuated by modern power electronics (e.g., \cite{molinas_2011} and \cite{molinas_2012}). Although this aspect goes beyond the scope of this paper, it is worth observing that for this specific case, as the GA is assumed to know the state of the grid, it also knows the admittance matrix of the system. As a consequence, it can formally compute the frequency-domain input impedances in the correspondence of each node where the resources are connected. Therefore, the GA can potentially use this information to augment the computation of the feasible electrical state of the grid and, thus, include further stability requirements.

We are interested in radial distribution networks, in which it is known that the load-flow problem has a unique solution if voltage magnitudes are close to nominal values \cite{LFUniqDistr, LFMonoDistr}. We consider a grid operating in this regime. In particular, we assume that the voltage magnitude bounds $\beta_k$ are small enough to guarantee the uniqueness of a solution to the load flow equation and that it contains a margin compatible with the accuracy of state estimation. This implies that the electrical state of the grid and the power flow at the PCC are uniquely determined by the injections/absorptions at all subsystems $(P_1,Q_1, ...,P_n, Q_n)$ and the voltage $(\delta_0, V_0)$ at the slack bus.

We aim at controlling the power setpoints at all subsystems, which therefore allows to control the electrical state of the grid. However, the actual power setpoints implemented by subsystems can differ from targets, and this is captured by the belief functions. This suggests the following definition.
\begin{definition}[Admissible Target Setpoints] \label{def:adm}
We say that a collection of target setpoints $(P_1, Q_1,...,P_n,Q_n)$ is \temph{admissible} if for any actual implementation that is compatible with the belief function, the resulting electrical state is feasible. We denote the admissible set by $\Ul$.
\end{definition}

\subsection{Formulation of the Objective Function and Overview of The Decision Process}
\label{sec:decision}
The goal of the grid agent GA is to compute the collection
\begin{equation} \label{eqn:sp}
u=(P_1,Q_1,...,P_n,Q_n)
\end{equation}
of target setpoints that it will send to its followers.
In order to do so, the virtual costs of the followers are aggregated by means of
the weighted total cost
\begin{equation} \label{eqn:total_cost}
C(u) \triangleq \sum_{i = 1}^n w_i C_i(P_i,Q_i),
\end{equation}
where $C_i$ is the virtual cost function of follower $i$, and the weights $\{w_i\}$ express the preference of GA of one follower over another.

In particular, GA would put a higher weight on the cost function of a device that \emph{provides more service} to the grid and is more \emph{controllable}. For example, more weight can be put on storage systems, whereas less weight on loads and semi-controllable generators (such as PV or wind farms).
The optimal computation of the weights is out of the scope of this paper. We just mention that, in general, some adaptive learning procedures can be applied to adjust the weights based on the observed history of the followers.
In this paper, we assume that the weights are pre-computed based on the prior knowledge on the followers and that they are fixed during the system's operation.

Further, we add a penalty term that represents the constraints on the power at the PCC/slack bus. Here and below, we use a \emph{superscript} $(I)$ to denote the variables related to the internal GA, and $(R)$ to denote these of the root GA.
In the case of the internal GA, it needs to accommodate the request $(P_0,Q_0)$ sent by its leader agent GA$_0$; this is captured by adding to the virtual cost a penalty term
$$J_0^{(I)}(P_0,Q_0; P_0(u), Q_0(u))$$
where $J_0(\cdot)$ is some measure of the distance between $(P_0,Q_0)$ and $(P_0(u), Q_0(u))$. Here,  $(P_0(u), Q_0(u))$ is the resulting power flow through the PCC when the collection of power flows at other systems is given by $u$. In this paper, we use the function $J_0$ defined by
\[
J_0^{(I)}((P_0, Q_0), (P'_0, Q'_0)) = w_0((P_0 - P'_0)^2 + (Q_0 - Q'_0)^2)
\]
for $w_0 > 0$. On the other hand, as the root GA works in an islanded mode, there is no request for the power at the connection point. Instead, we use the advertised cost of the slack agent as a cost for power at the slack bus, namely
\vspace*{-0.5cm}
\[
J_0^{(R)}((P_0, Q_0), (P'_0, Q'_0)) = w_0 C_0(P'_0, Q'_0),
\]
where $C_0$ is the cost advertised by the slack agent\footnote{It is worth observing that, as defined by the proposed control framework, the slack agent sends the advertisement messages to the root GA (namely, $PQt$ profile, cost, and belief functions); however, instead of implementing setpoints, the converter of the slack device works in the voltage control mode, satisfying any instantaneous $(P, Q)$ request within its capability limits.}.

Finally, a penalty term $J(u)$ is added to capture the distance of the electrical state to the boundary of the feasible region. This function is defined in terms of both threshold variables $\beta_k$ and $I_{\ell}^{\max}$.
In this paper, we consider that the first is constant (e.g., 10\% of the nominal voltage), whereas the second might change dynamically. 
In particular, we allow a line to transfer current over its ampacity limit by considering the \emph{specific energy} of its conductor, normally available from the manufacturer.
With this, $I_{\ell}^{max}$ is computed as the current that makes the \emph{Joule Integral} $\int I_{\ell}(t)^2 dt$ reach the \emph{specific energy} characteristic.

In this paper we use the function $J$ defined by
\begin{eqnarray}
\label{eqn:objective} J(u) &\triangleq & \sum_k \frac{(V_k(u) - V_k^{nom})^2}{\beta_k^2 - (V_k(u) - V_k^{nom})^2} + \sum_{\ell} \frac{I_{\ell}(u)^2}{(I_{\ell}^{max})^2 -I_{\ell}(u)^2} \\
&&\mbox{ if $\{V_k(u), I_{\ell}(u)\}$ satisfy \eqref{eqn:elec_safe}}\notag \\
&=&\infty \mbox{ otherwise.} \notag
\end{eqnarray}

In the above $V_k(u)$, $I_{\ell}(u)$ are the voltage and current magnitudes that result from the load-flow solution when the collection of power injections/absorptions is given by $u$.

Ideally, GA would like to find a collection of target setpoints $u$ that (i) is admissible (as per Definition \ref{def:adm}) and (ii) minimizes
 \begin{equation}\label{eq:opt_consr}
 C(u)+J_0(P_0,Q_0;P_0(u),Q_0(u))+J(u)
 \end{equation} over all admissible $u$.

However this is not easily computable. Instead, we propose to steer the power injections in the direction of the optimum of (\ref{eq:opt_consr}), using a \emph{gradient descent approximation}.
More precisely, the decision process at the grid agent computes

\begin{equation} \label{eqn:opt_grad}
\begin{array}{l}
u = \Pl_{\Ul} \Big\{\hat{x}
- \alpha \nabla_u \big[C(u)+J_0(P_0,Q_0;P_0(u),Q_0(u))+J(u)\big]\Big|_{u = \hat{x}}  \Big\},
\end{array}
\end{equation}
where $\Pl_{\Ul}$ is the Euclidean ``projection''\footnote{In the general case, the set of admissible setpoint $\Ul$ is non-convex. Hence, in the practical implementation, the projection is replaced with finding a closest point in $\Ul$ as described in \ref{sec:algos}.} to the admissible set $\Ul$, $\hat{x}=(\hat{P}_1, \hat{Q}_1,...,\hat{P}_n, \hat{Q}_n)$ is the last estimated power setpoint (obtained from the GA's state estimation process), and $\alpha$ is a step size parameter.

Note that the algorithm defined in \eqref{eqn:opt_grad} requires two major computations: (i) that of the gradient of the objective function, and (ii) the projection to $\Ul$. We next describe the methods that are used in this paper in order to perform these computations in real time.

\subsubsection{Gradient of the Objective Function}
By using the definition of $J$ in \ref{eqn:objective}, it can be easily verified that
\vspace*{-0.5cm}
\begin{multline}  \label{eqn:grad_objective}
\nabla_x J(x) = \sum_k \frac{2\beta^2 (V_k(x) - V^{nom})}{(\beta^2 - (V_k(x) - V^{nom})^2)^2} \nabla_x V_k(x) \\ + \sum_{\ell} \frac{2 (I_{\ell}^{max})^2 ((I_{\ell}^{max})^2 - I_{\ell}(u)^2)}{\l((I_{\ell}^{max})^2 - I_{\ell}(u)^2\r)^2} \nabla_x I_{\ell}(x).
\end{multline}

This requires the knowledge of $V_k(x)$ and $I_{\ell}(x)$ and, in particular, its dependence on the setpoint $x$. The exact dependence is complicated, as it follows from the solution of the power flow equations. We use, instead,  a \emph{linear approximation} of this dependence. In particular, given the current state $\hat{V} = \{\hat{V}_k\}$ and $\hat{I} = \{\hat{I}_{\ell}\}$ (obtained from the state estimation procedure), we let
$
\widetilde{V}(x) = \hat{V} + K_V (x - \hat{x}), \text{ and } \widetilde{I}(x) = \hat{I} + K_I (x - \hat{x}),
$
where $K_V$ and $K_I$ are the voltage and current \emph{sensitivity coefficients} computed using methods as in \cite{SensCoeffNadia, Zhou_Bialek}. By using this approximation, we have that $\nabla_x \widetilde{V}_k(x) = (K_V)_k$ and $\nabla_x \widetilde{I}_k(x) = (K_I)_k$. Moreover, as the gradient $\nabla_x J(x)$ is computed at $x = \hat{x}$, we have that $\widetilde{V}(x) = \hat{V}$ and $\widetilde{I}(x) = \hat{I}$. Therefore, Equation \eqref{eqn:grad_objective} and the approximated values above provide us with an approximation of the gradient of the objective function.

A similar approach is taken in order to compute the gradient of $J_0(u_0, X_0(x))$, where $X_0(x) = (P_0(x),Q_0(x))$: the exact dependence of $X_0(x)$ is replaced by an approximated linear one, and the corresponding gradient is computed. Finally, the gradient of the cost function $C(x)$ is computed either by using the analytical form of the cost function advertised to GA, or by numerical approximation.

\subsubsection{Computation of the Beliefs and Projection to the Admissible Set} \label{sec:simp_adm}
We next give explicit expressions for the admissible sets used in  both types of grid agents.
For clarity, we now need to introduce some more notations. We denote with $\Al$ the \emph{joint} $PQt$ profile, i.e. set of all collections of setpoints that are in the advertized $PQt$ profiles. Namely, $\Al \triangleq \Pi_{i = 1}^n \Al_i$, where $\Al_i$ is the $PQt$ profile advertised by follower $i$ and $\Pi_{i = 1}^n$ stands for the Cartesian product of sets. Similarly, we denote with $BF$ the \emph{joint} belief function, defined for all $u=(P_1,P_2,...,P_n,Q_n)\in\Al$ by $BF(u) \triangleq \Pi_{i = 1}^n BF_i(P_i, Q_i)$.
Note that $\Al$ represents the \emph{domain of definition} of the two functions $C(u)$ and $BF(u)$.

Since the internal GA works in grid-connected mode, we assume that there is no explicit constraint on the power flow at the PCC (namely, constraint \eqref{eqn:elec_safe-slack} is inactive). Hence, the admissible set can be written as
\vspace*{-0.5cm}
\begin{equation} \label{eqn:adm_mod_LVGA}
\Ul^{(I)} \triangleq \l\{u \in \Al^{(I)}: \forall x \in BF^{(I)}(u), \, J^{(I)}(x) < \infty \r\},
\end{equation}
where $x = (P'_1, Q'_1, ..., P'_n, Q'_n)$ is any \emph{actual} implementation of the setpoints. Observe that, in this case, the feasibility condition \eqref{eqn:elec_safe} is equivalent to $J^{(I)}(x) < \infty$ (see \eqref{eqn:objective}).
In the admissible set of the root GA, on the contrary, the feasibility condition \eqref{eqn:elec_safe-slack} is active, as it takes into account the capability limits of the slack resource. In particular,
\vspace*{-0.5cm}
\begin{equation} \label{eqn:adm_mod_MVGA}
\Ul^{(R)} \triangleq \l\{u \in \Al^{(R)}: \forall x \in BF^{(R)}(u), \, J^{(R)}(x) < \infty \text{ and } X^{(R)}_0(x) \in \Al^{(R)}_0  \r\},
\end{equation}
where $\Al^{(R)}_0$ is the $PQt$ profile of the slack resource, and $X^{(R)}_0(x)$ is the power injected/absorbed\footnote{Note that the setpoint computed by the root GA does not include that of the slack resource.} by it when the power setpoint of the other followers is $x$.

We next relax the exact computation of the projection to the set of admissible target setpoints $\Ul$, which is required in algorithm (\ref{eqn:opt_grad}). Note that belief functions are used to ensure a feasible operation of the grid, hence we need to guarantee that the relaxation maintains the feasibility property.

For brevity, we focus in the rest of this subsection on the internal GA; similar computations are preformed in the root GA. First, consider the subproblem of testing whether a given control $u$ is in $\Ul$. We refer to this process as the \emph{admissibility test}. As follows from \eqref{eqn:adm_mod_LVGA}, in order to carry out this test, we should solve $\max_{x \in BF(u)} J(x)$ and verify whether the result is finite.

The above optimization is hard in general, hence we propose to relax it as follows.
First, observe that using Definition \ref{def:validity}, we can replace the exact belief functions with \emph{supersets}.
We thus assume that the grid agent has access to functions $\overline{BF}_i(P_i, Q_i)$ with the following two properties:
\begin{inparaenum}[(i)]
\item[(i)] $BF_i(P_i, Q_i) \subseteq \overline{BF}_i(P_i, Q_i)$, and
\item[(ii)] $\overline{BF}_i(P_i, Q_i)$ is a \emph{rectangle} in $\RR^2$.
\end{inparaenum}
We note that the rectangular \emph{super beliefs} can be either sent directly by the follower agents, or computed by the grid agent from the advertised exact beliefs.

In addition, we use the following property of the load-flow solution that holds true in the networks considered in this paper.
It was shown in \cite{LFMonoDistr} that the solution is \emph{monotonic} in radial distribution networks, whenever the shunt elements of the lines are neglected. Specifically, every voltage magnitude can either increase or decrease monotonically\footnote{This is true for actual low voltage grids and, in particular, for the one considered in the case study used in this paper in Part~II. Regarding the MV network, for which the transverse line parameters cannot be neglected, we have numerically validated this property.} as a function a single power injection (while the other injections are kept fixed). Similarly, the extreme values of the current magnitudes are obtained at the extreme values of a power injection.
By using the definition of $J$ in \eqref{eqn:objective}, it follows that only a small finite number of simple computations is required in order to perform the admissibility test of a control $u$.
In particular, for each \emph{vertex} $v$ of $\overline{BF}(u)$, we should test whether (i) there exists a solution to the load-flow equations, and (ii) $J(v) < \infty$.

Given this simplified admissibility test, we can devise an efficient method for projection to $\Ul$. As the projection is needed only in a local vicinity of the current setpoint $\hat{x}$,
it can be efficiently computed by doing a search of the closest point in $\Ul$, and using a relatively small number of the (simplified) admissibility tests. We present the details of the related algorithms in \ref{sec:algos}.

\section{Composition of Subsystems}
\label{sec:comp}

A key aspect of our framework is \emph{composability}: subsystems can be aggregated and viewed by others as a single entity that exhibits the same properties of a single subsystem (i.e., $PQt$ profile, virtual cost, and belief function). This is essential for the application of the method to systems of any size and complexity.

\begin{figure}[h!]
\centering
\begin{minipage}{0.5\textwidth}
\centering
\includegraphics[width=0.83\textwidth]{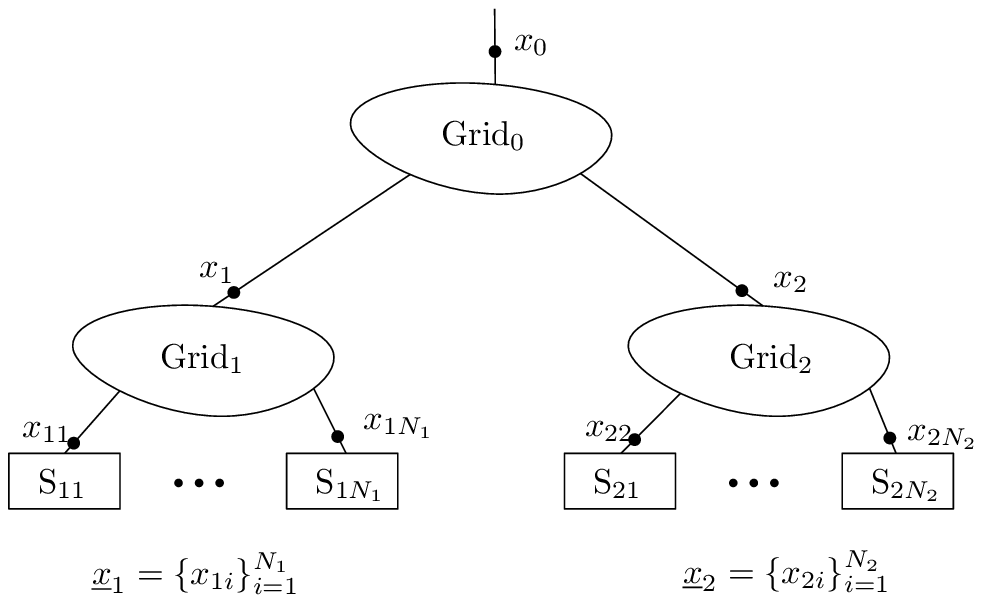}
\captionof{figure}{Schematic representation of a radial distribution network. \label{fig:grid_aggr}}
\end{minipage}%
\begin{minipage}{0.5\textwidth}
\centering
\includegraphics[width=1\textwidth]{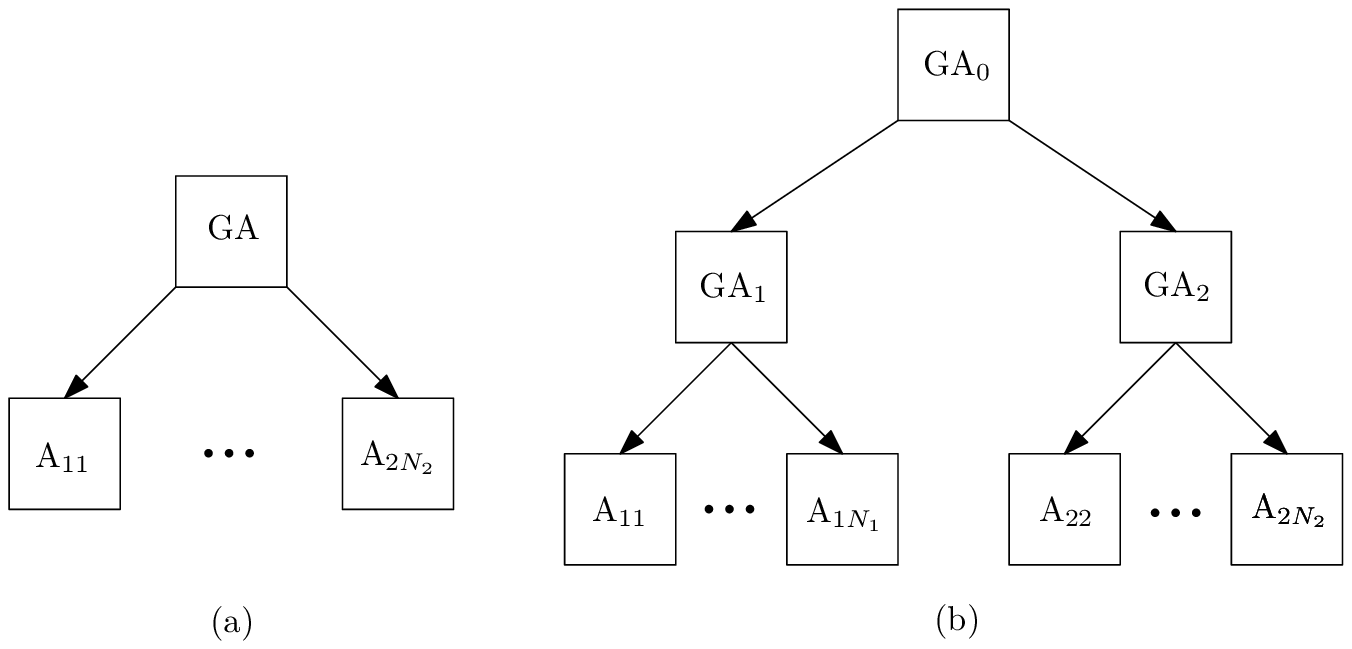}
\captionof{figure}{(a) A flat architecture, with one centralized grid agent. (b) A hierarchical architecture, with three grid agents. \label{fig:agents_aggr}}
\end{minipage}
\end{figure}

To illustrate the idea, we consider a radial power network shown schematically in Fig.~\ref{fig:grid_aggr} and two settings of agents depicted in Fig.~\ref{fig:agents_aggr}. In the \emph{flat setting} of Fig.~\ref{fig:agents_aggr} (a), there is a single grid agent (which is a root GA) that is responsible for the whole grid (Grid$_0$, Grid$_1$, and Grid$_2$) and is a leader of $N_1 + N_2$ agents $A_{11}, ..., A_{1, N_1}, A_{21}, ..., A_{2N_2}$. In the \emph{hierarchical setting} of Fig.~\ref{fig:agents_aggr} (b), there are three grid agents GA$_0$, GA$_1$, and GA$_2$, each responsible for Grid$_0$, Grid$_1$, and Grid$_2$, respectively. Consequently, in the hierarchical setting, GA$_1$ is an internal GA that is the leader of $N_1$ agents $A_{11}, ..., A_{1N_1}$, GA$_2$ is an internal GA that is the leader of $N_2$ agents $A_{21}, ..., A_{2N_2}$, and GA$_0$ is a root GA that is the leader of GA$_1$ and GA$_2$.

In this case, GA$_1$ and GA$_2$ represent their internal state to GA$_0$ by advertising \emph{aggregated} $PQt$ profiles, virtual costs, and belief functions.
To define these elements, consider the follower grid agent GA$_1$, aggregating and sending advertisements to its leader grid agent GA$_0$.
First, GA$_1$ computes the \emph{aggregated PQt profile} as the set of all possible power flows $(P^1_0,Q^1_0)$ at the PCC of Grid$_1$ and Grid$_0$, given that the powers injected by the followers of GA$_1$ belong to their respective $PQt$ profiles, advertised to GA$_1$; this is computed by solving the load-flow for Grid$_1$.
Second, GA$_1$ computes the value of the \emph{aggregated virtual cost} function for every $(P^1_0,Q^1_0)$ that is in the aggregated $PQt$ profile as follows: GA$_1$ applies its decision process in order to obtain a collection of power setpoints for its set of followers and returns the corresponding value of the objective function \eqref{eq:opt_consr}.
Last, the value of the \emph{aggregated belief function} at $(P^1_0,Q^1_0)$ is the set equal to the union of all possible \emph{actual} power flows at the PCC of Grid$_1$ and Grid$_0$ over all possible actual power injections at the followers, given by the belief functions advertised to GA$_1$.

In theory, such an aggregation is transparent, i.e., the operation of the grid is the same in both the flat and the hierarchical settings. More precisely, assume that

\begin{enumerate}
\item[(i)] The two systems in Fig.~\ref{fig:agents_aggr} (a) and (b) are synchronized in communication exchange;
\item[(ii)] There is no delay in the transmission of information over the channels;
\item[(iii)] Both GA$_1$ and GA$_2$ know the exact equivalent Thevenin impedance between their PCC and the slack bus of the overall system;
\item[(iv)] All the grid agents implement the \emph{optimal} control defined by the optimization (\ref{eq:opt_consr}), with the penalty function $J_0$ given by
\vspace{-0.4cm}
\[
J_0(P_0, Q_0; u_0) =
\begin{cases}
0, & \text{ if } (P_0, Q_0) = u_0, \\
\infty, & \text{ otherwise.}
\end{cases}
\]
(Namely, we have a strict constraint $(P_0, Q_0) = u_0$ at the PCC.)
\end{enumerate}

\begin{proposition} \label{prop:equiv}
In the ideal case defined above, the target power setpoints computed at one step of the decision process are the same in both settings of Fig.~\ref{fig:agents_aggr} (a) and (b).
\end{proposition}

The proof of this proposition can be found in the Appendix.

In Part~II of this paper, we propose practical methods for computing the aggregated elements described above and we show their performance.

\subsection{Aggregated $PQt$ Profile and Belief Function} \label{sec:pqt_simpl}
In order to aggregate in practice the $PQt$ profiles and belief functions of an internal GA,
we use the validity property formulated in Deﬁnition \ref{def:validity}.

To this end, we first write the load-flow constraints more explicitly, in terms of the power injections in the grid and the powers at the slack bus:
\begin{equation} \label{eqn:slack_pfe}
P_0 = \sum_{i = 1}^N P_i - L_{P}(\{P_i, Q_i \}), \quad Q_0 = \sum_{i = 1}^N Q_i - L_{Q}(\{P_i, Q_i \}),
\end{equation}
where $L_{P}(\{P_i, Q_i \}) \geq 0$ and $L_{Q}(\{P_i, Q_i \})$ is the active and reactive total power loss. Alternatively, \eqref{eqn:slack_pfe} can be written as
\[
X_0(u) = \sum_i u_i - L(u), \quad L(u) \triangleq (L_{P}(\{P_i, Q_i \}), L_{Q}(\{P_i, Q_i \})),
\]
and the exact aggregated $PQt$ profile reads
\begin{equation} \label{eqn:pqt_exact}
\widetilde{\Al}_0 = \bigcup_{u \in  \Ul} \l\{u_0 = \sum_i u_i - L(u) \r\}.
\end{equation}
Our method for approximation is based on (i) omitting the loss term when computing the aggregated $PQt$ profile, and (ii) accounting for the resulting error in the aggregated belief function. Next, we describe these two procedures in detail.

\noindent \textbf{Computing the aggregated $PQt$ profile:}
\vspace*{-0.25cm}
\begin{trivlist}
\item[~(i)] Given the current setpoint $\hat{x}$ (again, assumed as known via a state-estimation process), randomly and uniformly generate $M$ setpoints $u^{k} \in \Al$, $k = 1, ..., M$ with the following two properties:
    \begin{itemize}
    \item \textbf{Locality:} $\l\|u^{k} -  \hat{x}\r\| \leq \alpha G_{\max}$, where $\alpha$  is the step size of the gradient descent algorithm \eqref{eqn:opt_grad} and $G_{\max}$ is an upper bound for the gradient.
    \item \textbf{Admissibility:}
    \begin{equation} \label{eqn:safety}
    \forall x \in \conv\l\{BF(u^{k})\r\}, \quad J(x) < \infty,
    \end{equation}
    where $\conv\l\{BF(u^{k})\r\}$ is the \emph{convex hull} of the sets defined by the belief functions $BF(u^{k}), k = 1, ..., M$.
        Observe that \eqref{eqn:safety} is a \emph{stronger} requirement than just $u^{k} \in \Ul, k = 1,...,M$. This step can be performed efficiently by using local methods for projection described in \ref{sec:algos}. In particular, similarly to the methods described in Section \ref{sec:simp_adm}, $\conv\l\{BF(u^{k})\r\}$ can be overapproximated by a \emph{rectangular} set, and the feasibility property is then trivially tested only on the vertices of the rectangle.
    \end{itemize}

\item[~(ii)] Compute the corresponding \emph{ideal} powers at the slack bus $u_0^k = \sum_{i = 1}^N u_i^k$, and advertize the following approximation for the aggregated $PQt$ profile:
\begin{equation} \label{eqn:pqt_approx}
\widetilde{\Al}^*_0 = \conv (\{u_0^k \}_{k = 1}^M),
\end{equation}
namely the \emph{convex hull} of $\{u_0^k \}_{k = 1}^M$.
\end{trivlist}

The belief functions can be aggregated by solving the following four optimal power flow problems (OPFs) for a each $u_0 \in \widetilde{\Al}_0^*$:
\begin{equation} \label{eqn:belief_comp}
\begin{array}{l}
\max / \min P_0 \\
\text{s.t.} \begin{cases}
x \in BF(F(u_0)), & \\
(P_0, Q_0) = X_0(x),& \\
\end{cases}
\end{array}
\text{ }
\begin{array}{l}
\max / \min Q_0 \\
\text{s.t.} \begin{cases}
x \in BF(F(u_0)), & \\
(P_0, Q_0) = X_0(x),& \\
\end{cases}
\end{array}
\end{equation}
where $u = F(u_0)$ represents the algorithm of (\ref{eqn:opt_grad}).
This will yield a \emph{rectangular} belief function that represents a superset of the true aggregated belief. In this paper, due to timing constraints, we avoid solving these exact OPFs; instead we use \eqref{eqn:slack_pfe} and \emph{bounds on the losses}.
As a preliminary step, these bounds are estimated \emph{offline}:
\[
\overline{L}_P = \max L_{P}(\{P_i, Q_i \}), \quad \underline{L}_P = \min L_{P}(\{P_i, Q_i \}),
\]
\[
\overline{L}_Q = \max L_{Q}(\{P_i, Q_i \}), \quad \underline{L}_Q = \min L_{Q}(\{P_i, Q_i \}),
\]
where the optimization is done over all possible setpoints.

\noindent \textbf{Computing the aggregated belief function:}
\vspace*{-0.25cm}
\begin{trivlist}
\item[~(i)] Generate a uniform partition $\PPl_0$  over $\widetilde{\Al}^*_0$. A given requested setpoint $u_0 \in \widetilde{\Al}^*_0$ is mapped into a \emph{representative} request $u_0^{\PPl} \in \PPl_0$ (e.g., the \emph{closest point} to $u_0$ in $\PPl_0$).
\item[~(ii)] For each $u_0^{\PPl} \in \PPl_0$, compute
\begin{enumerate}
\item[(a)] The corresponding setpoints for the followers $u = \{u_i\} =  F\l(u_0^{\PPl}\r)$.
\item[(b)] The bounds for the power at the connection point, using the bounds on the losses:
\begin{equation} \label{eqn:belief_bounds_approx}
P_0^{\max}(u_0^{\PPl}) = \max_{(P_i, Q_i) \in BF_i(u_i)} \sum_i P_i - \underline{L}_P, \quad P_0^{\min}(u_0^{\PPl}) = \min_{(P_i, Q_i) \in BF_i(u_i)} \sum_i P_i  - \overline{L}_P,
\end{equation}
and similarly for $Q_0^{\max}(u_0^{\PPl})$ and $Q_0^{\min}(u_0^{\PPl})$. Observe that if $BF_i$ are \emph{rectangular}, \eqref{eqn:belief_bounds_approx} is just the summation of the corresponding individual upper/lower bounds.
\end{enumerate}
\item[~(iii)] Advertise the resulting belief function over $\PPl_0$ with the interpretation that for each $u_0 \in \widetilde{\Al}^*_0$,
\begin{equation} \label{eqn:belief_approx}
\widetilde{BF}^*_0 (u_0) = \l[P_0^{\min}(u_0^{\PPl}), P_0^{\max}(u_0^{\PPl})  \r] \times \l[Q_0^{\min}(u_0^{\PPl}), Q_0^{\max}(u_0^{\PPl}) \r],
\end{equation}
where $u_0^{\PPl}$ is the representative element for $u_0$ in $\PPl_0$.
\end{trivlist}

It follows from this construction that for any request $u_0 \in \widetilde{\Al}^*_0$, the actual power at the connection point $x_0$ satisfies that $x_0 \in \widetilde{BF}^*_0 (u_0)$. In other words, the pair  $(\widetilde{\Al}^*_0, \widetilde{BF}^*_0)$ is a valid pair of a $PQt$ profile and a belief function as per Definition \ref{def:validity}.

We note that the previous statement does not pose any requirements on the accuracy of the aggregated $PQt$ profile $\widetilde{\Al}^*_0$. In fact, it is valid for \emph{any} $\widetilde{\Al}^*_0$ provided that the belief function is constructed as above. The next result shows that when the losses are small and the belief functions satisfy a certain \emph{concavity} property, the proposed construction for $\widetilde{\Al}^*_0$ provides us with a good approximation of the true aggregated $PQt$ profile. The proof can be found in \ref{sec:proofs}.

\begin{theorem} \label{theo:approx_PQt}
Suppose that $BF(u)$ is a \temph{concave} set-valued function, namely for all $u_1, u_2 \in \Al$ and $\alpha \in [0, 1]$
\[
BF(\alpha u_1 + (1 - \alpha) u_2) \subseteq \alpha BF(u_1) + (1 - \alpha) BF(u_2)
\]
(where the second plus sign stands for the Minkowski sum).
Then $\widetilde{\Al}^*_0$ is a convex $\delta$-approximation of the exact $PQt$ profile
$\widetilde{\Al}_0$ \eqref{eqn:pqt_exact}, with $\delta \triangleq \max_{x} \l\| L(x)\r\|$. Namely, for any $x_0' \in \widetilde{\Al}^*_0$ there exists $x_0 \in \widetilde{\Al}_0$ such that $\l\|x_0 - x_0'\r\| \leq \delta$.
\end{theorem}

It can be verified that the concavity requirement of $BF(u)$ holds for the resources considered in the case study in Part II of this paper.

\noindent \textbf{Remark.} As follows from Theorem \ref{theo:approx_PQt}, the proposed construction for the approximate aggregated $PQt$ profile is a good approximation of the exact aggregated $PQt$ profile whenever the bound on the losses $\delta \triangleq \max_{x} \l\| L(x)\r\|$ is small. Hence, this approximation can be successfully used in the LV networks, such as the microgrid case in our paper. However, it is possible to compute an approximation to the aggregated $PQt$ profile also in other cases where the losses are not negligible (such as the MV network in the case study). For example, instead of omitting the losses when computing \eqref{eqn:pqt_approx}, we can solve the exact load-flow problem for each $u^k$ in order to obtain the exact power at the PCC $u_0^k$. Furthermore, as mentioned above, the computation of the aggregated belief function can be made more accurate as well, by solving the OPFs \eqref{eqn:belief_comp}. This will certainly provide us with better approximations in the case of networks where losses are high. However, the resulting procedures will be more demanding computationally. In particular, a special procedure should be devised in order to choose \emph{a small} set of \emph{representative} candidate setpoints $u^k$, hence reducing the overall computational complexity. This is out of the scope of this paper.

\subsection{Aggregated Cost Function}
As was already mentioned, for computing the aggregated virtual cost function at a given requested setpoint at the PCC $u_0 = (P_0, Q_0)$, the internal GA applies its gradient descent algorithm \eqref{eqn:opt_grad} in order to obtain a collection of power setpoints $u$ for its set of followers, and it returns the corresponding value of the objective function \eqref{eq:opt_consr}. In this paper, in order to advertize the virtual cost for \emph{every} $u_0 \in \widetilde{\Al}^*_0$, we compute it on a sparse partition of the aggregated $PQt$ profile $\widetilde{\Al}^*_0$ and advertize a linear interpolation thereof.

\section{Conclusion}
\label{sec:conc}
We have described the elements of a method that uses explicit power setpoints in order to control electrical grids in a scalable and reliable way. The proposed approach enables the behaviour of a complex electrical system to emerge as a property of a combination of agents, irrespective of the stochastic or deterministic nature of the energy resources.
In this respect, a first feature of this Part I is to guarantee that any system that implements the proposed framework must be correctly controllable by construction.
The correctness of the control is such that it guarantees not only a feasible operation point but, also, some form of optimality. This is achieved by combining the cost of the grid and the virtual costs of the resources.
The proposed framework has been designed to manage, by leveraging on the abstraction of the devices state, systems characterized by high volatility of energy resources. This property guarantees the inherent minimization of the required reserve usually needed in traditional control schemes for power systems.

A second feature of the proposed framework is composability. The same abstraction and protocol is used uniformly, regardless of the specifics of the resources or sets of resources and of system size. The rules for the abstraction of devices and subsystems have been provided, together with the proof of the main aggregation property. This allows an entire network and its resources to be viewed and handled as a single resource: this is a key characteristic, as it enables the method to be scaled to systems of any size.

In Part II, we use a practical case study to demonstrate a detailed implementation of the method and we evaluate its performance benefits.

\appendix
{ 
\setstretch{1.7}
\section{Algorithms} \label{sec:algos}
\noindent \textbf{\emph{Algorithm 1: Admissibility Test}}
\label{algo:admiss}
\begin{trivlist}
\item \textbf{Input:} Control $u = (u_j)$ to be tested.
\item \textbf{Parameters:}  Belief functions of the resources, given in terms of $\Bl_j(u_j)$ -- finite sets of representative ``worst-case'' setpoints that $u$ can give rise to (e.g., vertices of a rectangular belief).
\item \textbf{Do:} Obtain worst-case setpoints of a resource $j$ by using the belief function,
$
B_j = \Bl_j(u_j),
$
and test all possible combinations of the setpoints in $B_j$. That is, for each $x_j \in B_j$,
compute $d\l(x, \Yl\r)$, where $d(y, \Yl)$ is a certain ``distance'' of $x$ from the set  $\Yl \triangleq \{x: \, J(x) < \infty \}$. This distance can be computed using the definition of $J$ in Part I, Equation (5).
\item \textbf{Output:} Maximum violation $\Delta_{\max} = \max_{x: \, x_j \in B_j} d\l(x, \Yl\r)$.
\end{trivlist}

\noindent \textbf{\emph{Algorithm 2: Projection onto $\Ul$}}
\begin{trivlist}
\item \textbf{Input:} Control $u = (u_j)$ to be projected.
\item \textbf{Parameters:}  Search step $\Delta u$, number of search directions $n$.
\item \textbf{Initialization:}  The min-max violation $\Delta_{\min\max} = C > 0$.
\item \textbf{While $\Delta_{\min\max} > 0$}:
\begin{flushitem}
\item Generate $n$ test point $\{x_m, m = 1,...,n\}$ uniformly spread on a sphere with radius $\Delta u$ around $u$, so that $\l\|x_m - u \r\| = \Delta u$.
\item \textbf{For $m = 1,...,n$:}
\begin{flushitem}
\item Project $x_m$ to $\Al$ (using, e.g., the alternating projections method \cite{AltProj}):
$
x_m := \Pl_{\Al} \l\{ x_m\r\}.
$
\item Use Algorithm 1
to test admissibility of $x_m$, save the output to $\Delta_{m, \max}$.
\end{flushitem}
\item Compute the direction of the minimum violation:
$
m^* \in \argmin_{m = 1,...,n} \Delta_{m, \max},
$
and the corresponding violation:
$
\Delta_{\min\max} = \min_{m = 1,...,n} \Delta_{m, \max}.
$
\item Update $u := x_{m^*}$.
\end{flushitem}
\item \textbf{Output:} The projected control $u$.
\end{trivlist}

\section{Proof of Proposition \ref{prop:equiv}}
Consider first the flat setting of Figure \ref{fig:agents_aggr} (a).
Let $\textbf{u} = (\textbf{u}_1, \textbf{u}_2)$ denote the collection of setpoints in the overall grid of Figure \ref{fig:grid_aggr}.
The goal of GA is to minimize\footnote{In this proof, we omit the constraint $(P_0, Q_0) = u_0$ at the PCC of GA to its own leader, as it is fixed throughout.} (subject to constraints) $C(\textbf{u}) + J(\textbf{u})$, where, as before,
$
C(\textbf{u}) = \sum_{j = 1}^{N_1}w_{1j} C_{1j}(u_{1j}) + \sum_{j = 1}^{N_2} w_{2j} C_{2j}(u_{2j})
$
is the total cost advertised by the followers and $J(\textbf{u})$ is the internal objective function of GA, which captures the distance of the electrical state to the boundary of the feasible region, as in \eqref{eqn:objective}.
Naturally, the virtual cost is separable, namely
$
C(\textbf{u}) = C_1(\textbf{u}_1) + C_2(\textbf{u}_2),
$
where $C_i(\textbf{u}_i)$ is the advertised cost of the followers in grid $i = 1, 2$. The penalty function $J(\textbf{u})$ is also separable in the sense:
$
J(\textbf{u}) = J_1(\textbf{u}_1) + J_2(\textbf{u}_2) + J_0(X_1(\textbf{u}_1), X_2(\textbf{u}_2)),
$
where $J_i$ is the penalty function in grid $i = 0, 1, 2$, and $X_1(\textbf{u}_1)$ (respectively, $X_2(\textbf{u}_2)$) is the power at the PCC between grid $1$ (respectively, $2$) and grid $0$ at the given setpoint $\textbf{u}_1$ (respectively, $\textbf{u}_2$), when the perfect knowledge of the corresponding equivalent Thevenin impedance is assumed.

Now, \emph{without constraining} $\textbf{u}$ to the admissible setpoints dictated by the belief functions, the optimization problem of GA in the flat setting is

\vspace{-0.5cm}
\begin{eqnarray*}
&&\min_{\textbf{u}} \l(C(\textbf{u}) + J(\textbf{u})\r) \\
&& = \min_{u_1, u_2} \l\{ \min_{\textbf{u}_i, X_i(\textbf{u}_i) = u_i, i = 1, 2} \l[ C_1(\textbf{u}_1) + C_2(\textbf{u}_2) + J_1(\textbf{u}_1) + J_2(\textbf{u}_2) + J_0(X_1(\textbf{u}_1), X_2(\textbf{u}_2)) \r] \r\} \\
&& = \min_{u_1, u_2} \bigg \{ J_0(u_1, u_2)
+ \min_{\textbf{u}_1, X_1(\textbf{u}_1) = u_1} \l[ C_1(\textbf{u}_1) + J_1(\textbf{u}_1) \r] + \min_{\textbf{u}_2, X_2(\textbf{u}_2) = u_2} \l[ C_2(\textbf{u}_2) + J_2(\textbf{u}_2) \r] \bigg \}.
\end{eqnarray*}
Observe that the outer optimization problem is the problem solved by the grid agent GA$_0$ in the hierarchical setting of Figure \ref{fig:agents_aggr} (b), not considering the admissibility constraints.

We next take the constraints into account. Let $\Ul$ denote the set of admissible setpoints of grid agent GA in the flat setting of Figure \ref{fig:agents_aggr} (a). Similarly, let $\Ul_1$ and $\Ul_2$ denote the set of admissible setpoints of grid agents GA$_1$ and GA$_2$ in the hierarchical setting of Figure \ref{fig:agents_aggr} (b); these sets are computed using the perfect knowledge of the corresponding equivalent Thevenin impedances. Finally, let $\Ul_0$
denote the set of admissible setpoints of grid agent GA$_0$; this set is computed using the aggregated belief functions advertised by GA$_1$ and GA$_2$ (as explained in Section \ref{sec:comp}). Now, it is easily seen that, under the assumed ideal conditions, the target setpoint $\textbf{u}$ is in $\Ul$ if and only if $\textbf{u}_1 \in \Ul_1$, $\textbf{u}_2 \in \Ul_2$, and $\l(X_1(\textbf{u}_1), X_2(\textbf{u}_2)\r) \in \Ul_0$, completing the proof of the Proposition.

\section{Proof of Theorem \ref{theo:approx_PQt}} \label{sec:proofs}
First, note that, by the concavity of $BF(u)$, we have that
\[
\conv\l\{BF(u^{k})\r\} \supseteq BF\l(  \conv\l\{ u^{k} \r\} \r).
\]
Thus, by \eqref{eqn:safety}, any $u \in \Cl \triangleq \conv (\{u^k \}_{k = 1}^M)$ satisfies
\[
\forall x \in BF(u), \quad J(x) < \infty,
\]
implying that $\Cl \subseteq \Ul$.

Let
\[
\widetilde{\Al}'_0 = \bigcup_{u \in \Ul}\l\{u_0 = \sum_i u_i \r\}.
\]
Now, $\widetilde{\Al}'_0$ is a $\delta$-approximation of $\widetilde{\Al}_0$ where $\delta = \max_{x \in \Xl} \l\| L(x)\r\|$.
We thus show below that $\widetilde{\Al}^*_0 \subseteq \widetilde{\Al}'_0$, implying that $\widetilde{\Al}^*_0$ is a $\delta$-approximation of $\widetilde{\Al}_0$ as well.
Indeed, let $u_0 \in \widetilde{\Al}^*_0$. Hence, there exist $\{\gamma_k\}_{k = 1}^M$, $\gamma_k \geq 0$, $\sum_k \gamma_k = 1$, so that
\[
u_0 = \sum_{k = 1}^M \gamma_k u_0^k
= \sum_{k = 1}^M \gamma_k \sum_i u_i^k
= \sum_i \sum_{k = 1}^M \gamma_k u_i^k
= \sum_i u^*_i,
\]
with
\[
u^* \triangleq \sum_k \gamma_k u^k \in \Cl \subseteq \Ul
\]
Therefore, $u_0 \in \widetilde{\Al}'_0$.

\singlespacing
\bibliographystyle{elsarticle-num}
\bibliography{Bib_Commelec_Paper}
} 

\end{document}


\begin{frontmatter}

\title{A Composable Method for Real-Time Control of Active Distribution Networks with Explicit Power Setpoints. \\ Part II: Implementation and Validation}

\author[desl]{Lorenzo Reyes-Chamorro\corref{c1} \fnref{fn1}}
\cortext[c1]{Corresponding Author. Phone number: +41 21 69 37369, Postal address: EPFL STI IEL DESL, ELL037, Station 11, CH-1015 Lausanne}
\ead{lorenzo.reyes@epfl.ch}
\author[lca2]{Andrey Bernstein \fnref{fn1}}
\fntext[fn1]{Both authors contributed equally in this research work.}
\ead{andrey.bernstein@epfl.ch}
\author[lca2]{Jean-Yves Le Boudec}
\ead{jean-yves.leboudec@epfl.ch}
\author[desl]{Mario Paolone}
\ead{mario.paolone@epfl.ch}

\address[lca2]{Laboratory for Communications and Applications 2, \'{E}cole Polytechnique F\'{e}d\'{e}rale de Lausanne, CH-1015 Lausanne, Switzerland}
\address[desl]{Distributed Electrical Systems Laboratory, \'{E}cole Polytechnique F\'{e}d\'{e}rale de Lausanne, CH-1015 Lausanne, Switzerland}

\begin{abstract}
In this second part of the paper, we evaluate the performances of our control framework by applying it to a case study that contains a minimum set of elements that allows to show its applicability and potentials.
We show how the computation of the \textit{PQt} profiles, belief functions, and virtual costs can be synthesized for generic network resources (i.e., dispatchable and stochastic generation systems, storage units, loads).
The metrics of interest are: quality-of-service of the network represented by voltages magnitudes together with current line magnitudes in comparison with their operational boundaries;
state-of-charge of electric and thermal storage devices; proportion of curtailed renewables;  and propensity of microgrid collapse in the case of renewables overproduction.
We compare our method to two classic ones relying on droop control: the first one with only primary control on both frequency and voltage and the second one with an additional secondary frequency control operated by the slack device.
We find that our method is able to indirectly control the reserve of the storage systems connected to the microgrid, thus maximizing the autonomy in the islanded operation and, at the same time, reducing renewables curtailment.
Moreover, the proposed control framework keeps the system in feasible operation conditions, better explores the various degrees of freedom of the whole system and connected devices, and prevents its collapse in case of extreme operation of stochastic resources.
All of these properties are obtained with a simple and generic control framework that supports aggregation and composability.
\end{abstract}

\begin{keyword}
Decentralized control, explicit distributed optimization, power and voltage control, software agents.
\end{keyword}

\end{frontmatter}

\section*{Nomenclature}
\hspace*{-1cm}
\begin{tabular}{rl}
$u=(P_1,Q_1,P_2,Q_2,...,P_n,Q_n)$ & Control (target) setpoints\\
$x = (P'_1,Q'_1,P'_2,Q'_2,...,P'_n,Q'_n)$	& Implemented (actual) setpoints \\
$\hat{x}$ & Current (estimated) setpoints \\
$\Al_i$	& $PQt$ profile of follower $i$ \\&(set of possible target values for $(P_i,Q_i)$)\\
$C_i(P_i, Q_i)$	& Virtual cost of follower $i$  \\
$BF_i(P_i, Q_i)$		& Belief function of follower $i$ \\
& (set of possible $(P'_i,Q'_i)$ when $(P_i,Q_i)$ is targeted)\\
$\Al \triangleq \Al_1\times\Al_2\times ...,\times \Al_n $		& Joint $PQt$ profile \\
&(set of possible values for $u$) \\
$\widetilde{\Al}_0$      & Exact aggregated $PQt$ profile \\
&(set of possible values for $(P_0,Q_0)$ power at PCC)	\\
$\widetilde{\Al}^*_0$ & Approximate aggregated $PQt$ profile 	\\
$BF(u)= BF_1(P_1, Q_1)\times ...\times BF_n(P_n, Q_n)$ & Joint belief function\\
&(set of possible $x$ when $u$ is targeted)\\
$\widetilde{BF}_0(P_0, Q_0)$ & Exact aggregated belief function\\
&(set of possible $(P'_0,Q'_0)$ when $(P_0,Q_0)$ is targeted)\\
$\widetilde{BF}_0^*(P_0, Q_0)$ & Approximate aggregated belief function\\
$\Ul$	&  Set of admissible setpoints $u$\\
$J$ & Penalty for electrical state feasibility\\
$J_0$ &	Penalty for power flow deviation at the PCC\\
\end{tabular}

\section{Introduction}
In this second part of the paper, we discuss the implementation aspects and evaluate the performance of the control framework, which we henceforth refer to as \emph{Commelec} (which stands for the joint-operation of Communication and Electricity systems). This assessment is done by using a suitably developed simulation environment.
We consider a case study that makes reference to the low voltage microgrid benchmark defined by the CIGR\'{E} Task Force C6.04.02 \cite{LVBenchmark}, connected to a generic medium voltage feeder that contains the minimum number of elements that allow us to show the applicability and potentials of the proposed control framework.
In particular, while the formal description of the framework for controlling the grid using power setpoints is presented in Part I, here we show how to specifically implement the request/advertise messages between agents, how we can derive the $PQt$ profiles, belief and cost functions of the resources, and how the grid agent computes the resources setpoints and aggregates their internal elements.

The considered case study exhibits the following characteristics:
\begin{inparaenum}[(i)]
	\item the system is in islanded conditions,
	\item the slack bus is provided by the storage system connected to the medium voltage network (ESS),
	\item storage is distributed in both low and medium voltage,
	\item thermal loads (water boilers) are used as virtual storage, and that
	\item the randomness comes from the loads absorption patterns and solar irradiation. For the latter, we used a high time-resolution profile (sampled each $50$ ms) obtained from the measurements on solar panels in the authors' laboratory.
\end{inparaenum}

A challenge in such a system is that most of the inertia comes from storage and thermal loads rather than rotating machines; it is precisely the goal of our real-time control method, to overcome this difficulty in the presence of extremely volatile resources (e.g., PVs).

In order to assess its performance, we used the following metrics: the distances of node voltages and line currents to their operational limits, the state-of-charge of electric and thermal storage devices, the proportion of curtailed renewables, and the robustness against system collapse in case of overproduction from renewables..

We compare our method to two classic ones that rely on droop control: the first one with only \emph{primary control} on both frequency and voltage and the second one with an additional \emph{secondary frequency control} at the slack device (see, e.g., \cite{MG_Nikos}).
We find that our method is able to indirectly control the reserve of the storage systems, thus maximizing the autonomy of the islanded operation.
It reduces the curtailment of renewables, compared to the droop based methods, and it is able to implicitly identify local power compensation.
Further, it keeps the system in feasible operation conditions and better explores the various degrees of freedom of both network and energy resources.
Most importantly, it prevents system collapse in case of overproduction of renewables, in contrast to the droop control strategies.

Further, we show that the properties of Commelec are fundamental in the case of \emph{inertia-less grids} associated with the penetration of energy conversion systems that do not have any rotating mass (e.g., photovoltaic plants) or other conversion systems interfaced with the grid by means of power electronic converters. Indeed, in the cases where these energy conversion systems represent the majority of the electricity supplying means, the control strategies have to be re-thought (e.g., \cite{Droop_virtual_impedance}).
In this respect, as the proposed method does not rely on any shared signals (i.e., frequency), it can inherently account for the control of inertia-less grids.

All of these characteristics are obtained in real-time with a simple and generic framework; the specific properties of electric and thermal resources are known only by their local agents, whereas grid agents are generic and independent of the specific resources they control. As introduced in Part I, a key property is composability: an entire grid can be viewed as a single generic resource, the details of which need not be known by the higher-level grid agent.
In this part of the paper, we also evaluate the effect of the simplifications resulting from the aggregation process, and we find that it is essentially negligible.

The structure of this second part is the following. In Section \ref{sec:case}, we present the case to be studied, the simulation environment, the related control algorithms, the profiles' data, and the performance metrics. In Section \ref{sec:RA}, we define the different resource agents and how they manage their exchanged messages.
In Section \ref{sec:results}, we present the simulation results. A discussion section and a conclusion follow.

\section{Case Study} \label{sec:case}
In this section, we present a case study where the proposed control framework is implemented.
To show the applicability of the proposed framework, we have selected a closed system that contains all types of agents described in Part~I. In order to evaluate its performance, we implemented a generic event-driven simulation environment in Matlab \textregistered .

\subsection{System Details}
We consider a $0.4[kV]$ LV network that includes
\begin{inparaenum}[(i)]
	\item distributed generation composed of photovoltaic plants (PV$i$) and a hydraulic microturbine ($\mu$H),
	\item a storage system represented by a battery (ESS1),
	\item uncontrollable loads (UL$i$) and
	\item controllable loads (WB$i$) modelled as water boilers all capable of deploying explicit control setpoints.
\end{inparaenum}
The topology and parameters of this LV grid are taken from \cite{LVBenchmark}.
As typically used in a microgrid (MG) setup, we assume that all the generation/storage units connected to the LV MG are interfaced with the grid through power electronic devices \cite{MG_ElecInterf}.

\begin{figure}[h!]
\centering
\subfloat[]{
\includegraphics[width=0.4\textwidth]{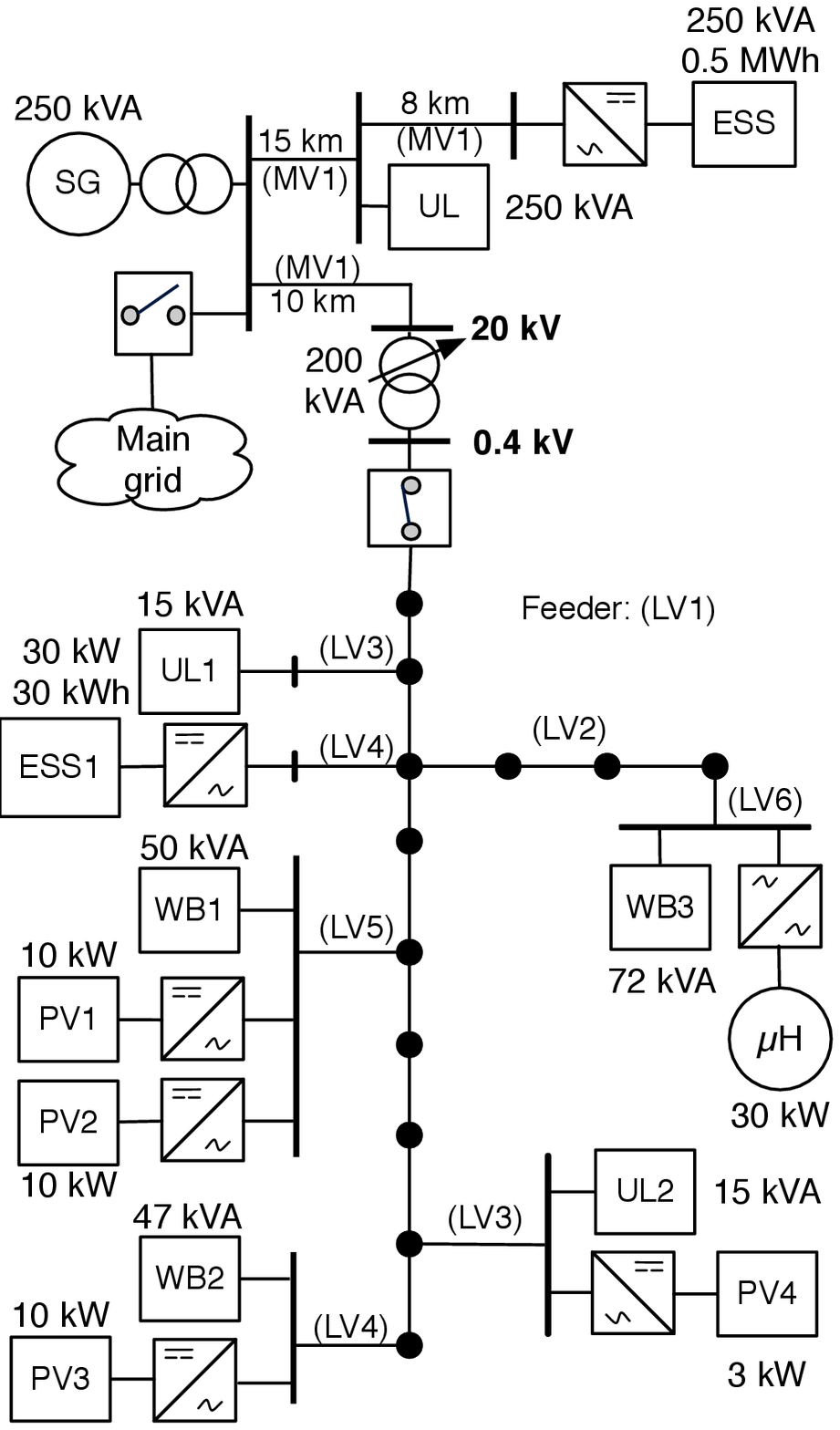}}
\subfloat[]{
\includegraphics[width=0.4\textwidth]{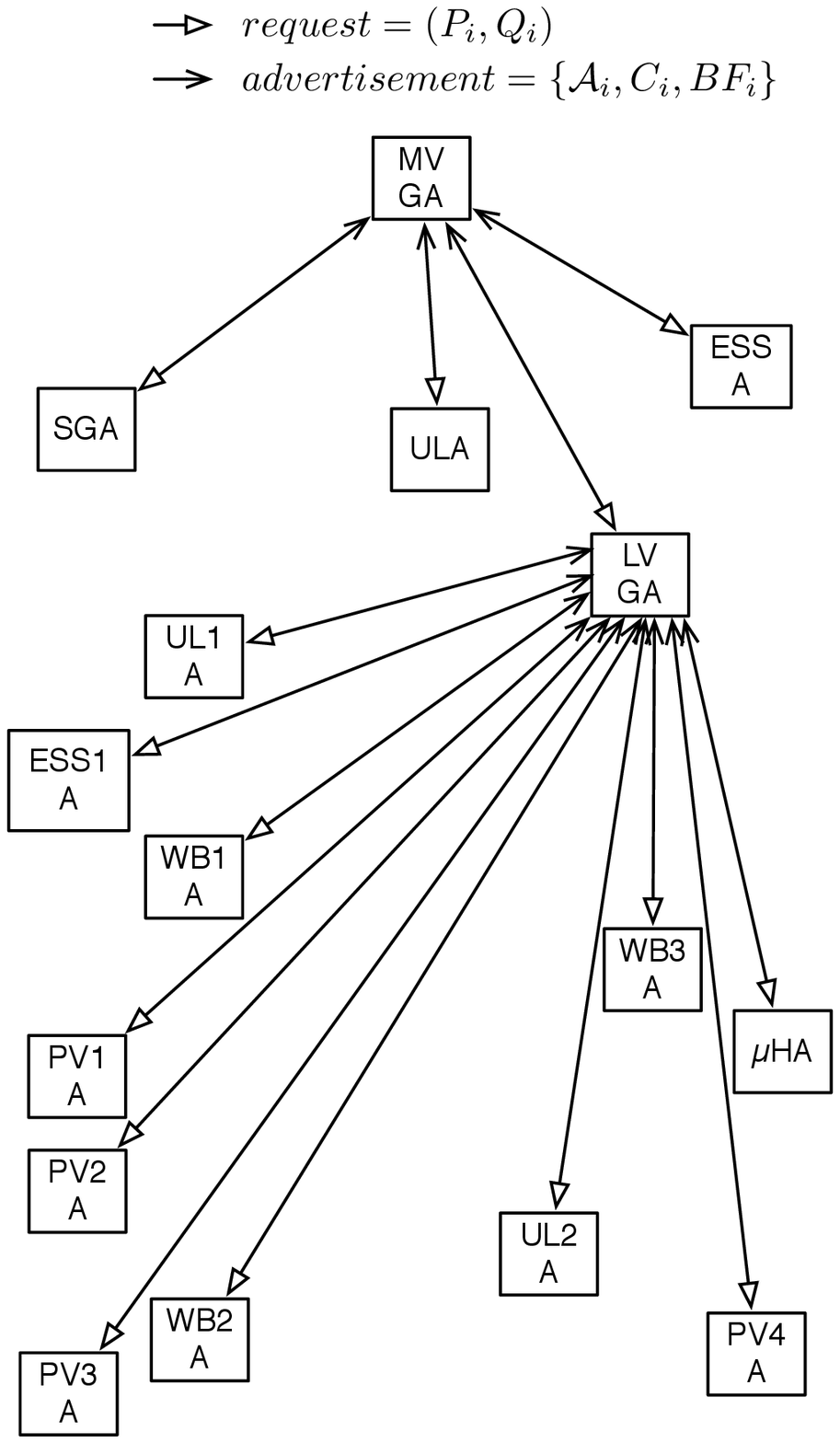}}
\caption{The electrical network and agents for the this case study. (a) Microgrid. (b) Agents.\label{fig:Grid&AgentsCaseStudy}}
\end{figure}

To show the interaction between different grids, the MG is connected to a $20[kV]$ MV distribution system that interconnects
\begin{inparaenum}[(i)]
	\item a large battery storage system (ESS),
	\item a combined heat and power generator interfaced with the MV grid by means of a synchronous generator (SG) and
	\item an industrial uncontrollable load (UL).
\end{inparaenum}
The corresponding electrical diagram for the case study is presented in Fig. \ref{fig:Grid&AgentsCaseStudy}(a).

To illustrate the mapping between physical subsystems and agents, we consider the hierarchical agents setting shown in Fig.~\ref{fig:Grid&AgentsCaseStudy}(b) where the microgrid agent (LVGA) is in charge of the resources in the LV network, whereas the medium voltage grid agent (MVGA) is in charge of the ones in the MV network and the LVGA. In the terminology of Part I of the paper, the LVGA is an internal GA, while the MVGA is a root GA.

The line parameters used for the network are presented in the Table \ref{table:line_param}.

\begin{table}[h!]
\caption{MV and LV power lines parameters \label{table:line_param}}
\centering
\begin{tabular}{ccccc}
\hline
Type		& Resistance		& Reactance 		& 	Susceptance		& Ampacity	\\
		& $R[\Omega/km]$	& $X[\Omega/km]$	&	$B[\mu S/km]$	& $[A]$		\\
\hline
MV1		&	3.9378		&	1.9689		&	2.7798			&	25	\\
LV1		&	0.284		&	0.083		&	0				&	170	\\
LV2		&	0.497		&	0.086		&	0				&	120	\\
LV3		&	3.690		&	0.094		&	0				&	31	\\
LV4		&	1.380		&	0.082		&	0				&	60	\\
LV5		&	0.871		&	0.081		&	0				&	73	\\
LV6		&	0.822		&	0.077		&	0				&	140	\\
\hline
\end{tabular}
\end{table}

We use the base system and the voltage bounds presented in Table \ref{table:base_trafo}(a), while the parameters of the MV/LV transformer used in our case study are shown in Table \ref{table:base_trafo}(b).
We use a conventional transformer model as in \cite{Grainger_Stevenson}.

\begin{table}[ht!]
\caption{System Parameters\label{table:base_trafo}}
\subfloat[Base systems and voltage bounds]{
\begin{tabular}{l|c|c}
\hline
Parameter	&	Value	 	&	Unit		\\
\hline
Base voltage in MV	& 	20	&	$[kV]$	\\
Base voltage in LV	& 	0.4	&	$[kV]$	\\	
Base power 			& 	1	&	$[MVA]$ 	\\
Voltage lower bound	&	0.9	&	$[pu]$	\\
Voltage upper bound	&	1.1	&	$[pu]$	\\
\hline
\end{tabular}}
\subfloat[MV/LV Transformer]{
\begin{tabular}{l|c|c}
\hline
Parameter	&	Value	 	&	Unit		\\
\hline
Primary voltage 			&	20	&	$[kV]$	\\
Secondary voltage 		& 	0.4	& 	$[kV]$	\\
Rated power				& 	400	&	$[kVA]$	\\
Short-circuit voltage	&	4	&	$[\%]$	\\
Short-circuit resistance	&	1	&	$[\%]$	\\
\hline
\end{tabular}}
\end{table}

\subsection{Control Methods}
We performed a comparison between the following control methods applied to our case study.
\begin{trivlist}
\item[~~(i)] The Commelec architecture shown in Figure \ref{fig:AgentsCaseStudy}(a). We show in the following sections how we implement our framework in this case study.
In addition, in order to validate the composability property we performed a simulation of the ``flat'' setting of agents shown in Figure \ref{fig:AgentsCaseStudy}(b).

\begin{figure}[h!]
\centering
\subfloat[]{
\includegraphics[width=0.45\textwidth]{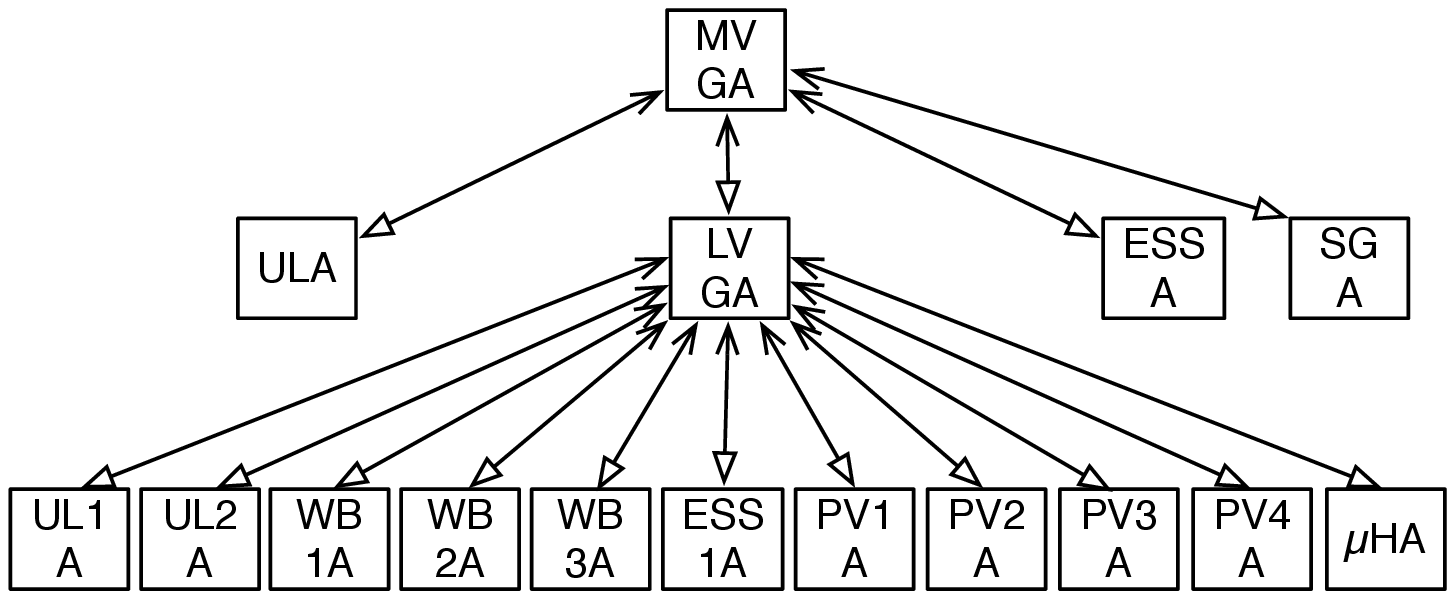}}
\subfloat[]{
\includegraphics[width=0.5\textwidth]{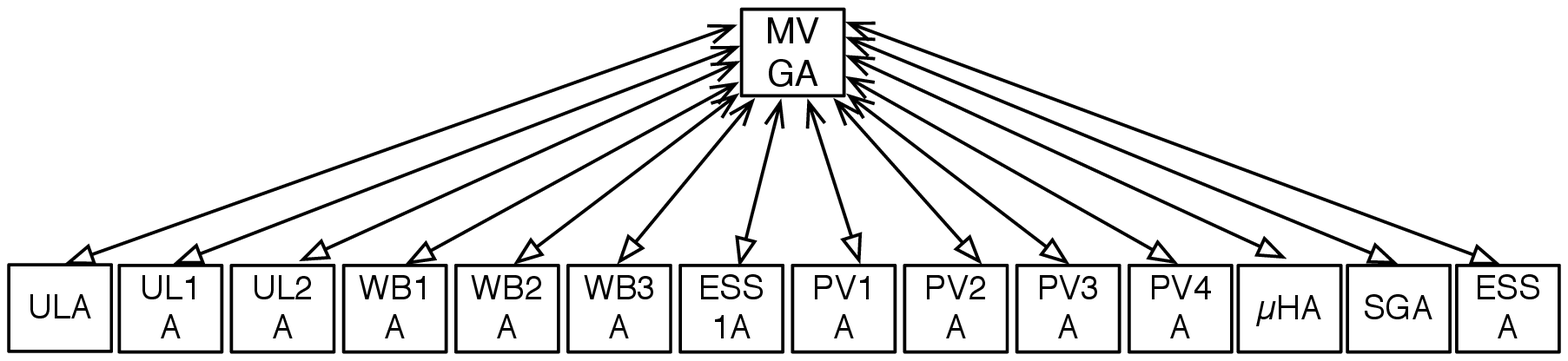}}
\caption{Agents architecture for the case study. (a) Hierarchical: Two grid agents, LVGA in charge of sending requests to the LV microgrid’s resources and MVGA in charge of the MV grid’s resources and the low-voltage grid agent, (b) Flat: A solely grid agent in charge of all the resources in MV and LV grids.\label{fig:AgentsCaseStudy}}
\end{figure}

\item[~~(ii)] The droop control method, with only a \emph{primary control} at each device capable of modifying power setpoints (i.e., all with the exception of the uncontrollable loads).
In the slack resource, the output frequency is calculated using the conventional droop control strategy, assuming a null inertia (as it is the case of ESS).
This is the signal that will be used for all the other resources to compute their power production.
As a result, the frequency is given by
\[
f = f_0 - m_f(P - P_0),
\]
where $f_0$ is the rated frequency (in our case $50$ Hz), $m_f$ is the curve slope, and $P_0$ is the active power when $f=f_0$.
The corresponding frequency and voltage droop curves in other resources were set to
\begin{eqnarray}
P = P_0 - (f - f_0)/m_f,\nonumber\\
Q = Q_0 - (V - V_0)/m_V,\nonumber
\end{eqnarray}
where $V$ is the measured voltage magnitude and $f$ the measured frequency, $V_0 = 1$ pu is the rated voltage, $m_V$ is the curve slope, and $Q_0$ is the reactive power when $V=V_0$.
It is worth noting that the droop parameters are in general different for each resource.
We have selected the droop parameters for the resources using typical values adopted in the literature (e.g. \cite{MG_Nikos}).
The selected droop parameters are shown in Table~\ref{table:droop_param}.

\item[~~(iii)] The droop control method as above, with additional \emph{secondary frequency control} at the slack device, using local frequency-error integrator.
In particular, the frequency droop curve in the slack resource was set to
\[
f(t) = f_0 - m_f(P - P_0) + (1/T_i)\int_{t_0}^{t^{-}}(f_0 - f(\tau))d\tau,
\]
where $T_i = 50$ sec is the chosen integration constant.

\end{trivlist}

\begin{table}[h!]
\centering
\caption{Droop parameters\label{table:droop_param}}
\begin{tabular}{l|cccccc}
\hline
Resource 	&$f_0[Hz]$	&$P_0[pu]$	&$m_f[Hz/pu]$&$V_0[pu]$&$Q_0[pu]$&$m_V[pu]$	\\
\hline
ESS/ESS$1$	&	50	& 	0	&	-0.5		&	1	&	0	& 	-0.04	\\
PV$i$		&	50	&	0.5	&	-1		&	1	&	0.5	&	-0.08	\\
SG/$\mu$H	&	50	&	0.5	&	-0.8		&	1	&	0.5	&	-0.08	\\
\hline
\end{tabular}
\end{table}

\subsection{Profiles Data} \label{sec:scen}
We chose a simulation scenario for all three control methods (including initial conditions), where we could simulate the case of overall overproduction in the grid from renewables (essentially PV) with minimum load consumption.
This scenario is adopted to challenge both control methods to deal with a system characterized by a low margin of controllability.
For this purpose, the scenario has the following initial conditions:
\begin{flushitem}
\item All batteries are close to their maximum stored energy capacity. In particular, the initial state of charge (\emph{SoC}) of both MV and LV battery was set to $0.9$.
\item The boilers are undercharged, with initial state of $2.5$ kWh.
\item There is a high production from PVs, at a partially sunny day, thus representing high irradiation variability.
\item The loads in the LV grid have zero-consumption profiles, whereas the MV load uses a dynamic profile representing changes with time resolution of $1$ min.
\end{flushitem}

\begin{figure}[h!]
\centering
\subfloat{
\includegraphics[width=0.8\textwidth]{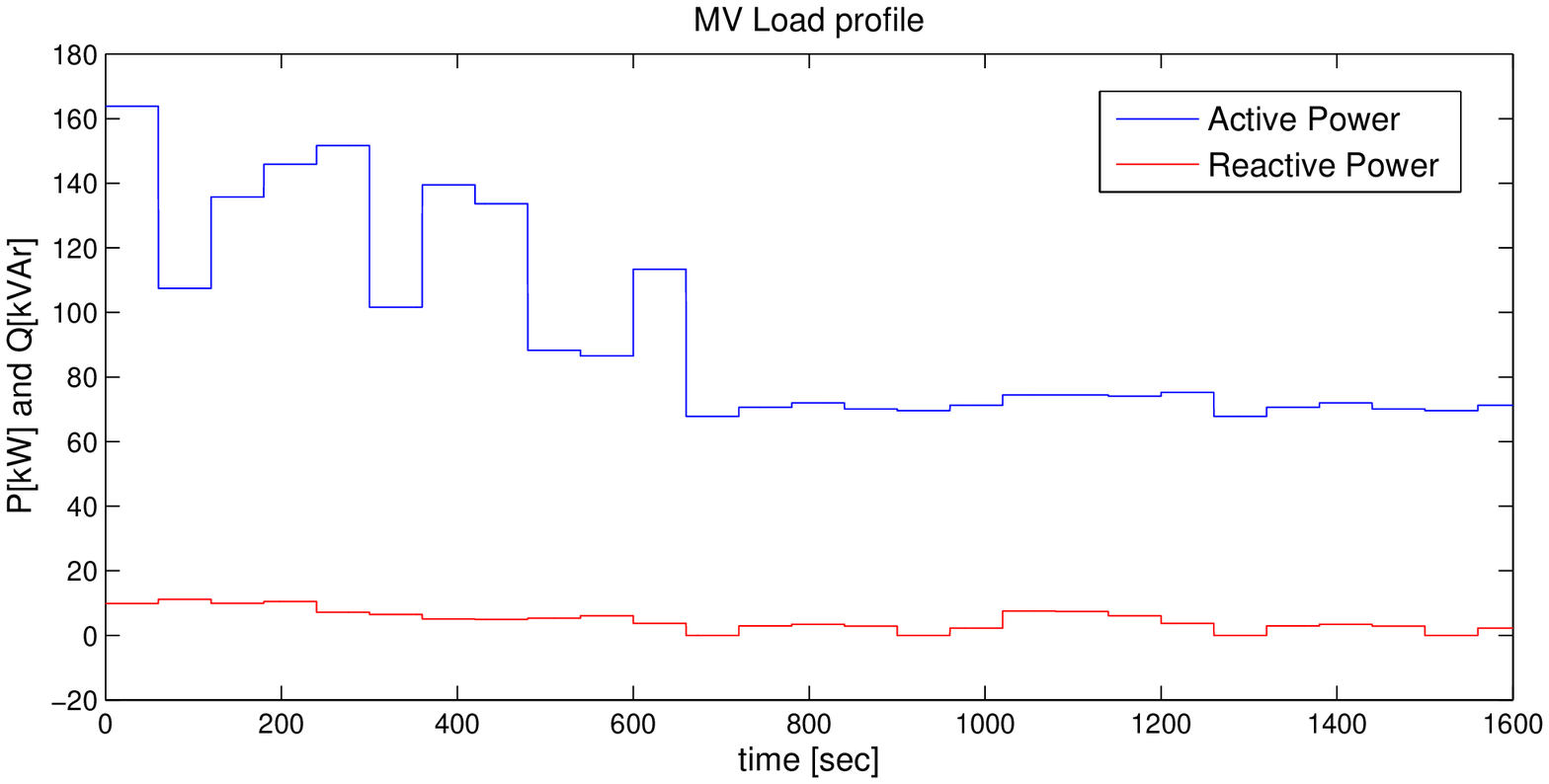}}\\
\subfloat{
\includegraphics[width=0.8\textwidth]{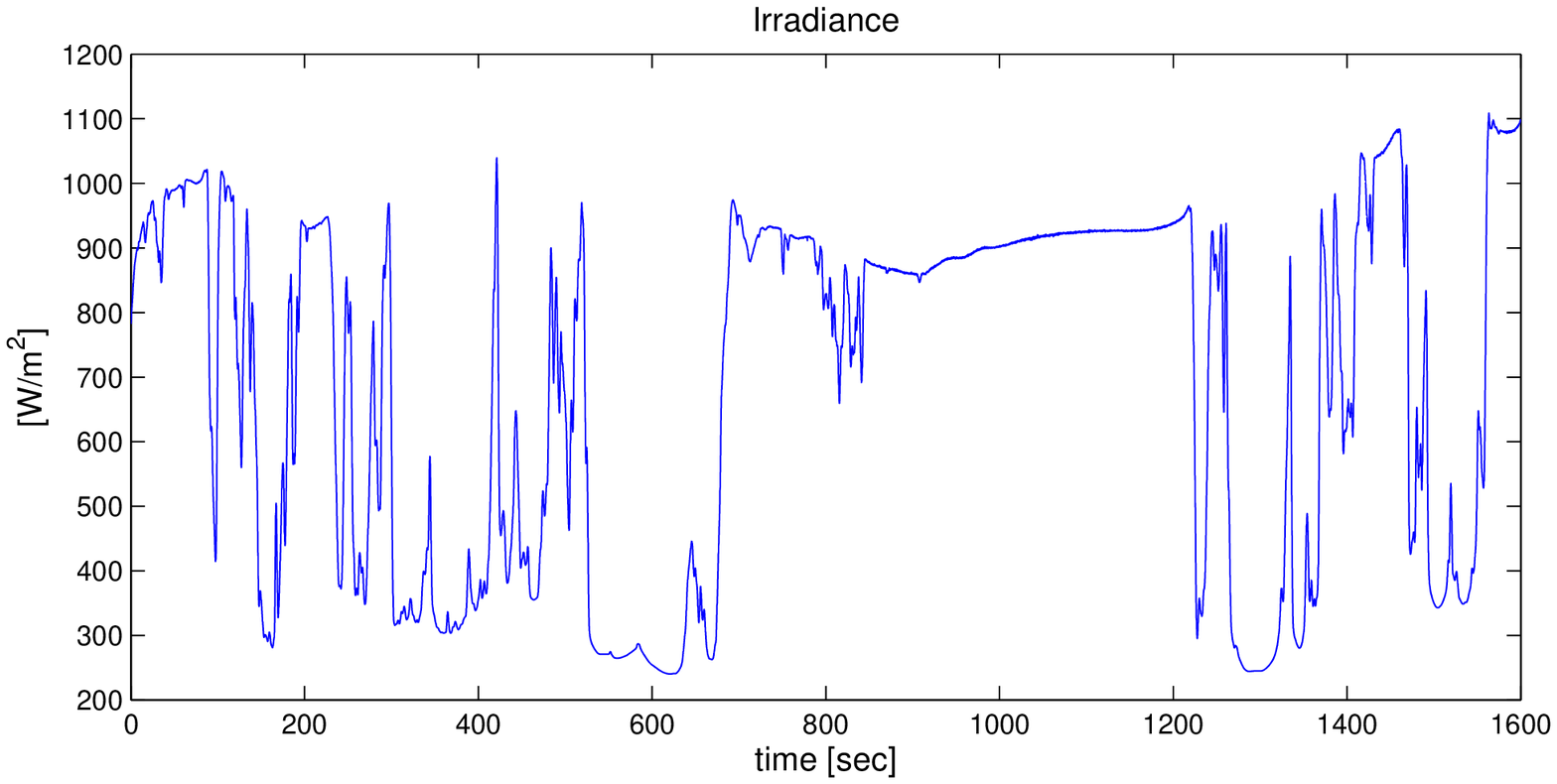}}
\caption{Sources of uncertainty in the case study: UL load profile and solar irradiance.\label{fig:uncertainty}}
\end{figure}

In our case study, there are two sources of uncertainty: the MV-load (UL) power consumption and the solar irradiance (shown in Figure~\ref{fig:uncertainty}(a) and \ref{fig:uncertainty}(b), respectively).
The hypothesis is that all the PV plants are exposed to the same irradiance profile.
The load consumption is characterized by a dynamic behaviour and a low value from minute $11$ onwards, whereas the solar irradiance data is characterized by a highly volatile profile due to the passage of clouds.
The irradiance data is composed of the real measurements performed in the authors' laboratory located in a south western site in Switzerland ($46^{\circ}31'06.20'' N$, $6^{\circ}33'54.56'' E$) on November the $15^{th}$, 2013.
The sampling period used to take the data was $50$ ms.
The used profile is shown in Figure~\ref{fig:uncertainty} (b).
The above quantities represent the forcing functions of the targeted case study.
Lastly, we use the weights shown in Table~\ref{table:sim_param} for Commelec simulations.

\begin{table}[h!]
\caption{Simulation parameters\label{table:sim_param}}
\centering
\subfloat{
\begin{tabular}{l|c}
\hline
Parameter				&	Value\\
\hline
MGA gradient step		&	0.1		\\
MVGA gradient step		&	5E-5		\\
PV$i$ cost weight		&	1		\\
UL$1-2$ cost weight		&	1		\\
WB$1$, WB$3$ cost weight	&	1E-6		\\
WB$2$ cost weight		&	1E-8		\\
\hline
\end{tabular}}
\subfloat{
\begin{tabular}{l|c}
\hline
Parameter				&	Value\\
\hline
ESS$1$ cost weight		&	1E-3		\\
$\mu$H cost weight		&	1		\\
ESS cost weight			&	1E-5		\\
UL cost weight			&	1		\\
SG cost weight			&	0.01		\\
MGA request weight		&	100		\\
\hline
\end{tabular}}
\end{table}

\subsection{Performance Metrics}
In order to assess the performance of the control methods, we use the following metrics:
\begin{inparaenum}[(i)]
\item the distances of node voltages and line currents to their limits, representing the quality of supply and the operational margins of the system;
\item the state of charge of electric and thermal storage devices, representing the reserve of the system;
\item the proportion of curtailed renewables; and
\item the robustness of the method against system collapse.
\end{inparaenum}

\section{Resources Models and Agents} \label{sec:RA}
As anticipated, we consider the following resources:
\begin{inparaenum}[(i)]
	\item \emph{energy storage device} (specifically, a \emph{battery}),
	\item \emph{synchronous generator},
	\item \emph{PV generator} and
	\item \emph{controllable and uncontrollable loads}.
\end{inparaenum}
Depending on their nature and/or internal characteristics, these resources have various degrees of controllability, from \emph{fully controllable} resources (e.g., the battery) to \emph{non-controllable} resources (e.g., uncontrollable load).
The controllability of the resource has a considerable effect on the design of the corresponding resource agent (RA).

RAs are pieces of software usually deployed on a computer, a processor or a microcontroller installed in the vicinity of the resource.
For instance, in a generation or storage unit, the RA can be implemented within its controller, whereas a load agent can be installed in a building computer to monitor and/or control its aggregated power consumption.
RAs might have a simplified or sophisticated view of the internal behaviour of their resources as a function of the RA developer. The better is the resource model, the more accurate advertisement messages will be sent to the GA, and the better would be its decision.

Recall that RA communication messages refer to the power flows at the point of connection with the grid, thus converters are always considered as part of a resource.
As a converter can be used by all kind of resources, we first present a general approach for its model as an interface with the grid.
Further, we present how to implement RAs in detail, specifically, how they manage the requests and produce the advertisements.

\subsection{Power Converter} \label{subsec:ConvModel}

First, we consider that the admissible area of operation of power converters can be modeled with three general constraints:
\begin{trivlist}
	\item[~(i)] \textit{The PQ capability curve of the converter}, which we consider in this paper to be given by
$
\sqrt{P^2+Q^2}\leq S_r,
$
with $S_r$ the rated power of the converter and ($P$,$Q$) the powers on the \emph{AC-side} of the power converter.
Alternatively, it could be considered that a converter is constrained by its nominal current $I_{nom}$.
In such a case the capability curve is given by $\sqrt{P^2+Q^2}\leq VI_{nom}$, where $V$ is the voltage magnitude at the connection point. As $V$ can be affected by other elements of the grid, the resulting power constraint has an uncertainty that has to be reflected in the belief function.
For the sake of simplicity, we work in this paper with power constraints.
\item[~(ii)] \textit{The power factor constraint}, given by
\vspace*{-0.5cm}
\begin{equation} \label{eq:Conv_PowerFactor}
\left| \frac{P}{\sqrt{P^2+Q^2}} \right| \geq \cos_{min}(\phi),
\end{equation}
where $\cos(\phi)$ stands for the power factor of the converter and $\phi$ the phase-shift between voltage and current phasors.
This constraint is relevant, for instance, in the case of PV converters that are required to operate with a minimum power.
\item[~(iii)] \textit{Unidirectional or bidirectional converter} (depending on the nature of the resource). For instance, the grid-tie PV generators usually cannot absorb active power, thus
$
P \geq 0.
$
In the case of an energy storage system (ESS), we consider a unique bidirectional device for charge and discharge.
\end{trivlist}

For simplicity, we assume a constant efficiency ($\eta$) to account for the effect on the DC power ($P_{dc}$),
depending on the power flow direction:
\vspace*{-0.5cm}
\begin{equation}\label{eq:efficiency}
P =
\begin{cases}
\eta P_{dc},  & \text{ if } P_{dc} \geq 0,\\
P_{dc}/\eta, & \text{ if } P_{dc} < 0.\\
\end{cases}
\end{equation}
\vspace*{-0.5cm}

\subsection{Energy Storage System} \label{subsec:ESSA}

For concreteness, we consider the case in which the ESS is represented by a battery.
(However, the concepts and methods can be easily extended to any kind of ESS.)

\textbf{Implementation of Setpoints}. In order to implement a requested power setpoint, the ESS agent (ESSA) needs a model to compute the internal limits this resource must respect for the next time step.
In this paper, we consider that such an agent uses a simple model that can sufficiently represent the dynamic behaviour of the storage system in the considered time frame\footnote{However, more complex models (e.g., \cite{reddy2010linden}) can be easily made compatible with the proposed framework.}.
In particular, assuming that the state of charge (\emph{SoC}) is fixed between two setpoint implementations, we can express the model of the battery as a simple time-varying resistance $R_t$ that is a function of the $dc$ current and voltage measurements of the battery array.
This approximation is reasonable if frequent battery setpoint variations are deployed, enabling a pseudo-continuous computation of $R_t$.

On the contrary, in our simulation environment we use a Two Time Constants (TTC) cell model (e.g., \cite{Batt_Model}) to simulate the internal behaviour of the battery.

Upon receiving a new setpoint request at time $t$, the ESSA computes
\vspace*{-0.5cm}
\begin{equation}\label{eq:R}
R_t= \frac{\Delta V^{dc}}{\Delta I^{dc}} = \frac{V^{dc}_t - V^{dc}_{t - \Delta t}}{I^{dc}_t - I^{dc}_{t - \Delta t}}
\end{equation}
where $\Delta V^{dc}$ and $\Delta I^{dc}$ are the step changes in $dc$ voltage and current measured in the resource at two consecutive requests instants.
(Note that if $\Delta I^{dc}=0$, $R_t$ will not change).
Consequently the ESSA can compute the internal electromotive force of the bank as
$
E_t = R_tI^{dc}_t + V^{dc}_t.
$
Then, by means of this extremely simple model, and considering the limitations on $V^{dc}$ and $I^{dc}$ given by the storage specifications ($V_{min}$, $V_{max}$, $I_{min}$ and $I_{max}$), ESSA computes the $dc$ power bounds for the resource as
\vspace*{-0.5cm}
\[
P^{dc}_{min} = max\left(\frac{V_{max}(E_t - V_{max})}{R_t}, (E_t - R_tI_{min})I_{min}\right),
P^{dc}_{max} = min\left(P^{V_{dc}}_{max}, P^{I_{dc}}_{max}\right)
\]
\vspace*{-0.5cm}
\[
P^{V_{dc}}_{max} =
\begin{cases}
E_t^2/4R_t,  & \text{ if } \frac{E_t}{2} > V_{min},\\
\frac{V_{min}(E_t - V_{min})}{R_t}, & \text{ otherwise, }\\
\end{cases},
P^{I_{dc}}_{max} =
\begin{cases}
E_t^2/4R_t,  & \text{ if } \frac{E_t}{2R_t} > I_{max},\\
(E_t - R_tI_{max})I_{max}, & \text{ otherwise. }\\
\end{cases}
\]
The above $dc$ power bounds \footnote{Note that the arguments of $P^{dc}_{min}$ are always negative.} are combined with the converter model in \eqref{eq:efficiency} to compute the $ac$ active power bounds.
Finally, it projects the requested setpoint onto the set of constraints defined by these bounds and the converter constraints.

\begin{figure}[h!]
\centering
\subfloat[]{
\includegraphics[width=0.3\textwidth]{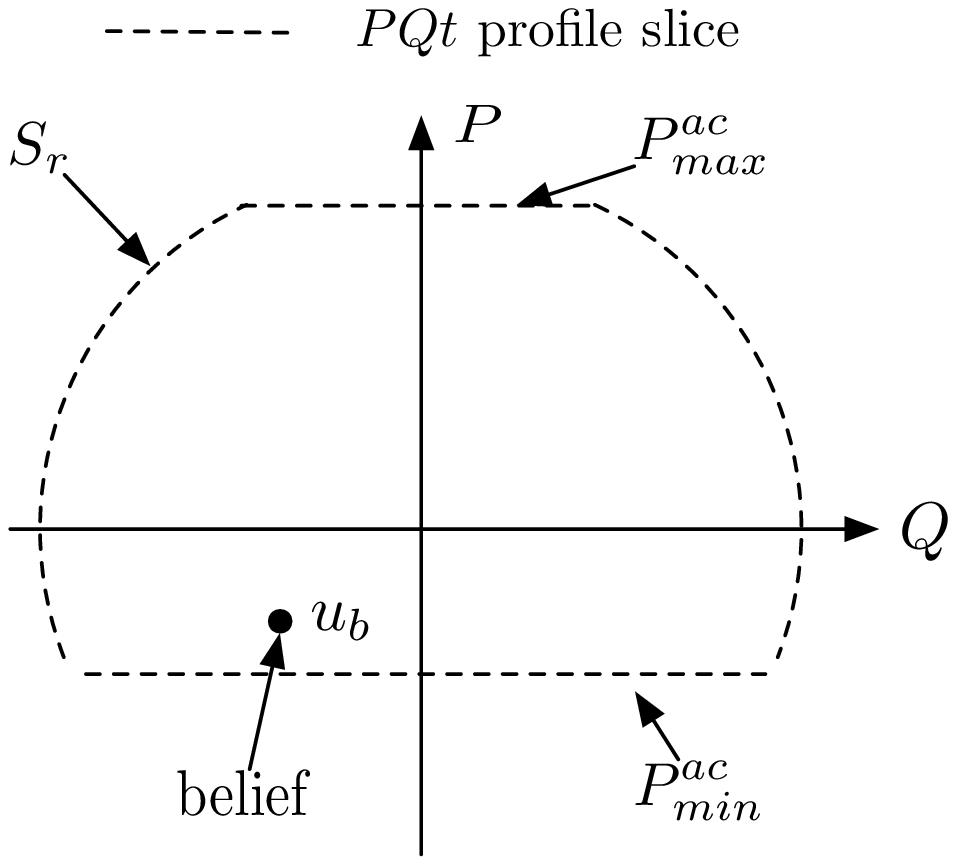}}
\subfloat[]{
\includegraphics[width=0.5\textwidth]{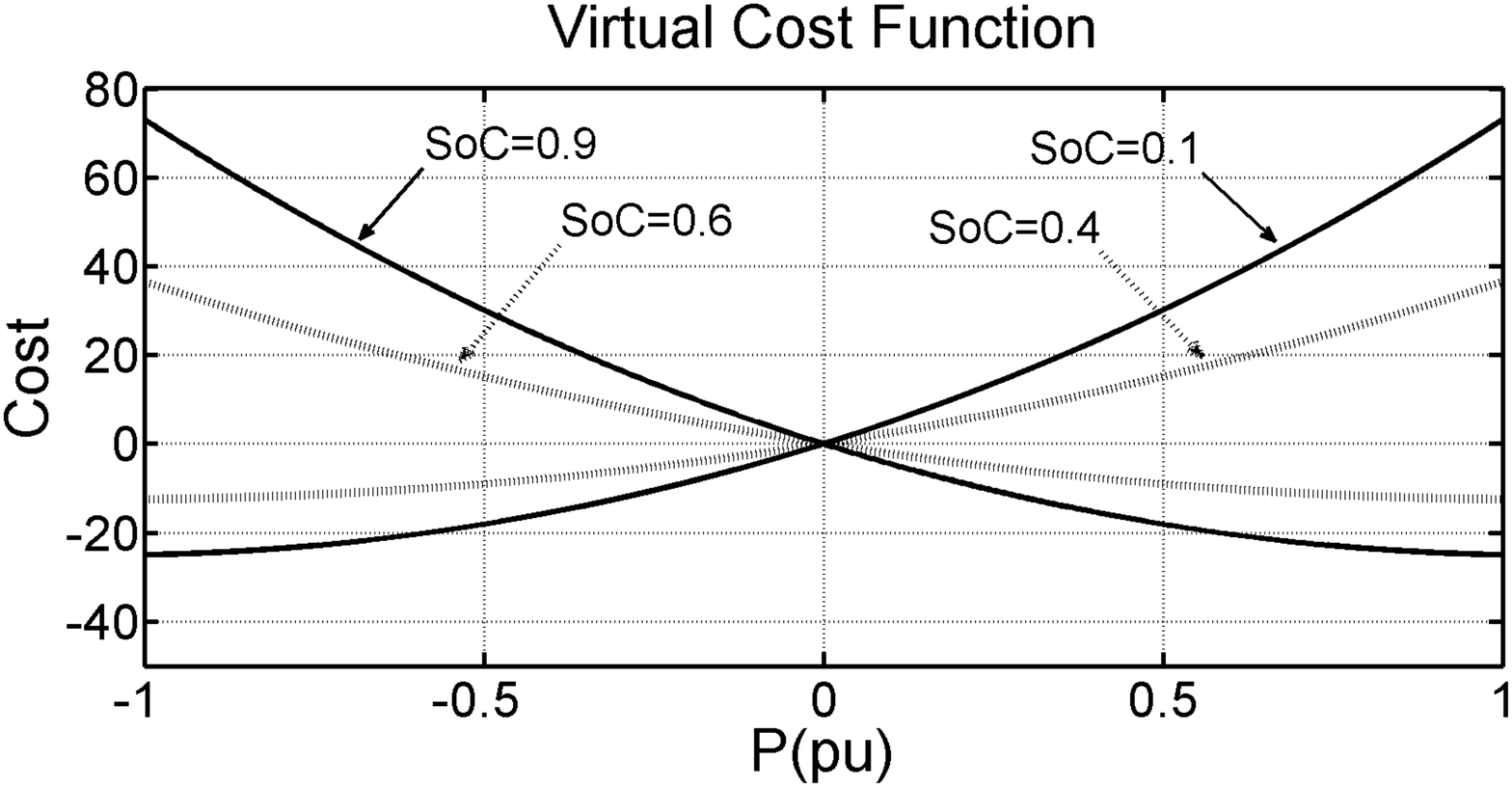}}
\caption{$PQt$ profile, belief function and virtual cost for ESS agent.\label{fig:ADV_ESSA}}
\end{figure}

\textbf{$PQt$ Profile}. As the constant \emph{SoC} assumption is still valid until the next request implementation, all the power bounds for this resource advertised within the $PQt$ profile are fully specified by the aforementioned process.
In this respect, a $PQt$ profile slice for a given time step, $\Al_b$, is shown in Fig.~\ref{fig:ADV_ESSA}(a).

\textbf{Belief Function}. As storage devices are highly controllable, we assume an ideal belief, namely, $BF_b(u_b)=\{u_b\}$ for any $u_b$ belonging to the $PQt$ Profile. Fig.~\ref{fig:ADV_ESSA}(a) shows an example for a given request $u_b$.

\textbf{Virtual Cost}.
The role of the virtual cost function is to measure the tendency of the storage agent to stay within particular zones of the $PQt$ profile.
In this paper, we consider that ESSA tends to steer the \emph{SoC} to a certain target value that represents a suitable (admissible) internal state of the storage obtained from a long-term scheduler\footnote{For electrochemical storage systems, this scheduler may take into account their state-of-health and consequent life.}.
Therefore, if the current \emph{SoC} is larger than a target value, the ESS prefers to be discharged, so the agent advertises a negative cost for discharging (positive $P$) and a positive cost for charging (negative $P$).
This situation is reversed for the case when the \emph{SoC} is lower than the target value.
If the \emph{SoC} is equal to the target, the cost will become zero, as the agent shows no preference of the ESS to be charged or discharged.
We assume that the current \emph{SoC} is measured by the resource using the \emph{SoC} computation presented in \cite{Mario_SoC}.

As an example, the following polynomial function can be used
\vspace*{-0.5cm}
\begin{equation}
\label{eq:final cost}
\\C_b(P,Q)= k\Delta SoC \cdot \left(a_b P^2 + \frac{b_b}{\Delta SoC}  P + c_b\right) \; \cdot P\\,
\end{equation}
where $\Delta SoC = SoC_t-SoC$, $SoC_t$ is the target \emph{SoC}, and  $a_b, b_b, c_b$ and $k$ are positive constants.
This function is chosen so that it presents (i) a positive \emph{cost} when going in the opposite direction of the target \emph{SoC}, and a negative cost (namely, an incentive) when heading towards the target;  and (ii) a higher \emph{price} (that is, the derivative of the cost) for a higher power at constant \emph{SoC}.
An illustration of this function is shown in Fig.~\ref{fig:ADV_ESSA}(b) for different values of \emph{SoC}.
For example, when $\Delta SoC > 0$, the cost for charging is positive with a steep slope and the cost for discharging is negative with gentle slope varying with asked power.
It should be noted that the cost for reactive power for energy storage systems is considered to be zero.

\subsection{Photovoltaic Plant} \label{subsec:PVA}
\textbf{Implementation of Setpoints}. Using measurements in the resource, we assume that the PV agent (PVA) can obtain the current maximum admissible power production $P_{pv}^{max}$.
Then, the PVA controls its resource to set the request $u_{pv}$ projected to the admissible set defined by this bound and the converter limits from section \ref{subsec:ConvModel}.
Afterwards, as described in Part I, the resource tries to deploy such a setpoint. The variation between the requested setpoint ($u_{pv}$) and the actual setpoint ($x_{pv}$) is represented by the belief function of the PVA.

\begin{figure}[h!]
\centering
\subfloat[]{
\includegraphics[width=0.25\textwidth]{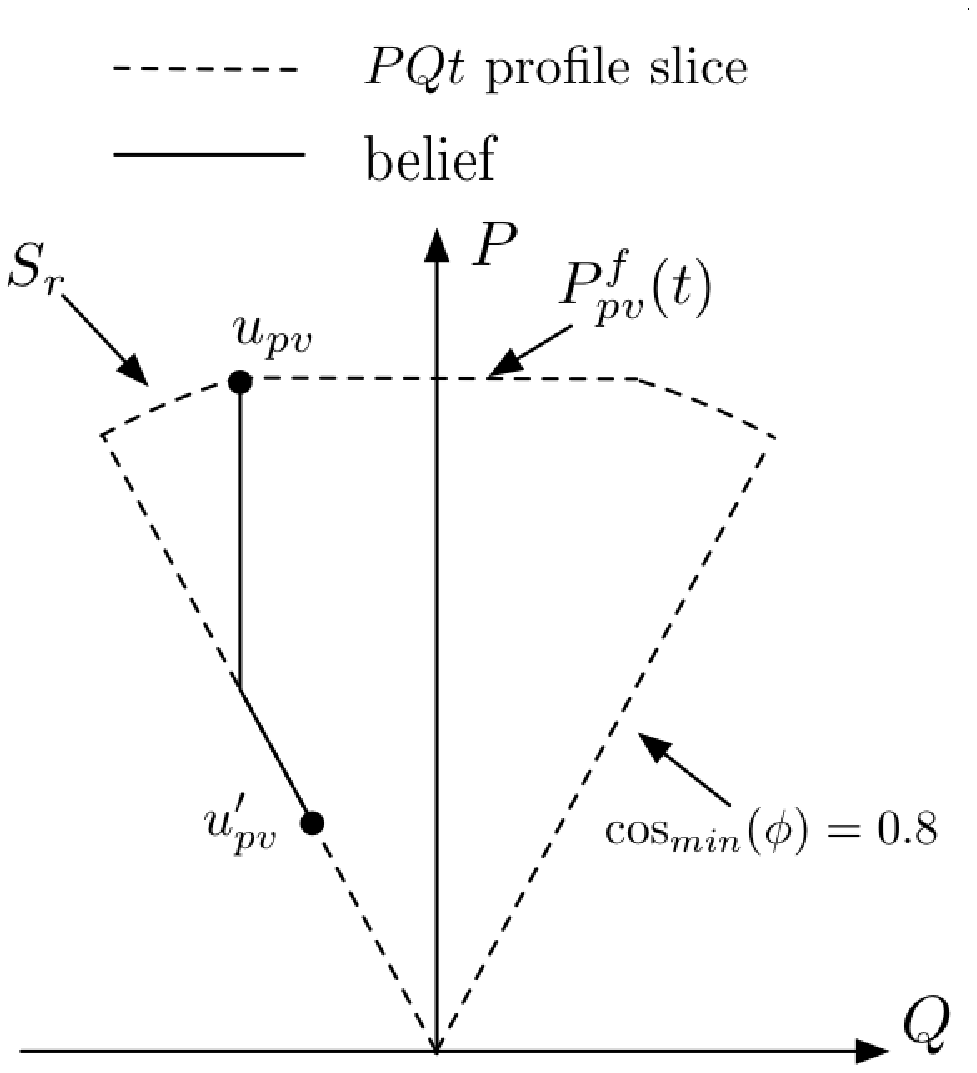}}
\subfloat[]{
\includegraphics[width=0.45\textwidth]{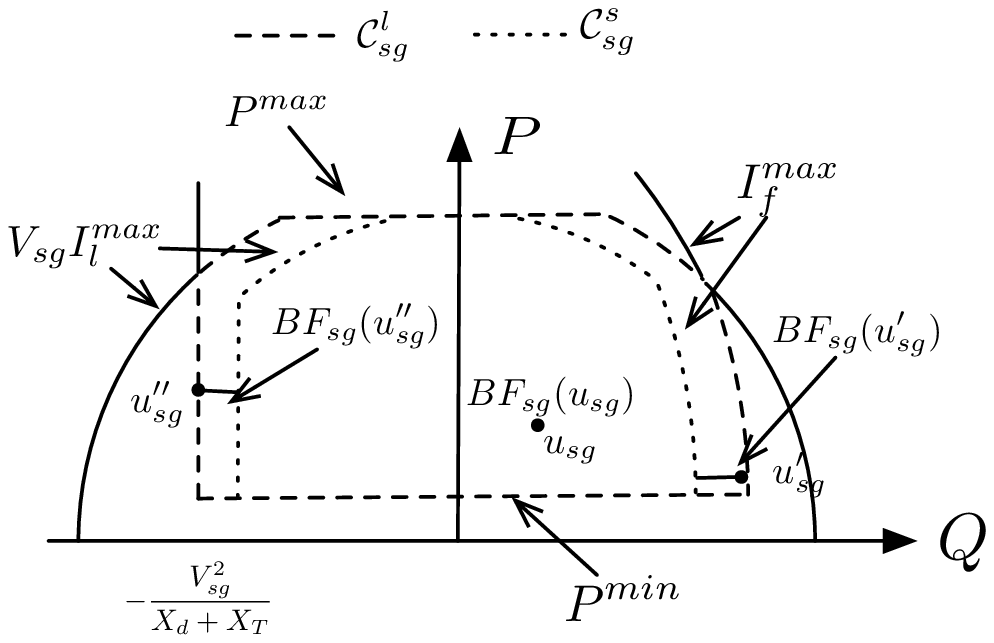}}
\caption{$PQt$ profile and belief function for PV and SG Agents. (a) PVA. (b) SGA.\label{fig:PQtBFPVASGA}}
\end{figure}

\textbf{$PQt$ Profile}. By means of a forecasting tool, and using the converter model (\ref{eq:efficiency}), the PVA computes the maximum power production at time $t \in [t_0, t_0+T]$, $P_{pv}^f(t)$, that can be maintained for any $t' \in [t, t+T]$.
As typically for grid-tie PV converters, we assume a constraint on the reactive power production, given by a minimum power factor (\ref{eq:Conv_PowerFactor}).
For time $t$, a slice of the $PQt$ profile shown in dashed lines in Fig.~\ref{fig:PQtBFPVASGA}(a).

\textbf{Belief Function}. In order to advertise the uncertainty of the solar resource, we assume that the active power production might decrease from the requested setpoint, $u_{pv}$, with a predicted maximum variation $\Delta P_{pv}^{max}$. The value of $\Delta P_{pv}^{max}$ is determined from the worst-case error of the employed forecasting tool.
As the reactive power is controlled by the converter, the belief of $Q$ production is restricted only by its relation with $P$ and the constraints that define the $PQt$ profile.
Hence, $BF_{pv}(u_{pv})$ can be represented as a line that starts in $u_{pv} = (P, Q)$ and finishes in $u'_{pv} = (P', Q')$, with $P' = P - \Delta P_{pv}^{max}$ and
\[
Q' = \begin{cases}
\max \l\{- P' \frac{\sqrt{1 - \cos_{min}^2(\phi)}}{\cos_{min}(\phi)}, \, Q \r\}, & \text{ if } Q < 0, \\
\min \l\{P' \frac{\sqrt{1 - \cos_{min}^2(\phi)}}{\cos_{min}(\phi)}, \, Q \r\}, & \text{ otherwise.}
\end{cases}
\]
An example of $BF_{pv}(u_{pv})$ is shown in Fig. \ref{fig:PQtBFPVASGA}(a).

\textbf{Virtual Cost}. We assume that the PVA seeks to maximize the active power production and minimize the reactive power. Therefore, an example of the advertised virtual cost function is given by
$
C_{pv}(P,Q)=-a_{pv}P+b_{pv}Q^2,
$
with $a_{pv}$, $b_{pv} > 0$.

\subsection{Synchronous Generator} \label{subsec: SGA}
For simplicity, we consider cylindrical rotor machines in both synchronous generators (SG and $\mu$H), whereas the agent uses the basic model for generator (both equivalent circuit and relevant capability curves) as in \cite{Grainger_Stevenson}, and we assume that they are interfaced to the network through an appropriate transformer (in the case of SG) or power converter (in the case of $\mu$H).
We present this section by making specific reference to SG, but is also applicable to $\mu$H.
Furthermore, we assume that the inertia of such machines is small enough to express its behaviour by using algebraic equations.
This assumption appears reasonable in view of the typical capacity of synchronous machines connected to LV grids and as we assume that the slack resource ESS is connected through a power converter with which frequency can be kept constant. As for the MV grid, it is a reasonable assumption that cannot be extended in general.
Indeed, as the proposed Commelec control method does not require the use of the frequency signal, the control of the slack resource can be astatic. For this reason, the above simplification only improves droop simulation results.

\textbf{Implementation of Setpoints}. When receiving a request, the synchronous generator agent (SGA) computes the current internal limits of the resource by using the measurement of the voltage in the connection bus with the grid ($V_{sg}$).
These limits correspond to the well-known capability curves of the synchronous machine defined by the maximum and minimum active power $P^{max}$ and $P^{min}$, the maximum SG field-current $I_f^{max}$, the maximum line current $I_l^{max}$, and the stability limit.
Further, the SGA commands the implementation of the projection to the capability curves.

\textbf{$PQt$ Profile}. As the bounds of this resource are dependent on $V_{sg}$, which is in turn dependent on external variables, the prediction of the limits in the next time slot is a complex task.
Instead, the SGA advertizes the \emph{largest possible} set of power setpoints, $\Cl^l_{sg}$, taking into account all feasible values of $V_{sg}$.

\textbf{Belief Function}. As mentioned above, due to changes in $V_{sg}$, the boundaries of the capability curves may vary at a given time slot. Thus, some setpoints, in this case the nearest to the bounds, might be shifted to the \emph{smallest possible} set of setpoints, $\Cl^s_{sg}$, taking into account all feasible values of $V_{sg}$.
Thus, the belief of a given request $u_{sg}$ is
\[
BF_{sg}(u_{sg}) =
\begin{cases}
u_{sg}, & \text{ if } u_{sg} \in \Cl^s_{sg}, \\
\bigcup\limits_{C_{sg}(V_{sg})} \Pl_{\Cl_{sg}(V_{sg})}\{u_{sg}\},  & \text{ if } u_{sg} \in \Cl^l_{sg} \setminus \Cl^s_{sg},
\end{cases}
\]
where $\Pl_{\Cl_{sg}(V_{sg})}\{ \cdot \} \text{ with } V_{sg} \in \left[V_{min}, V_{max}\right]$ is the Euclidean projection to $\Cl_{sg}(V_{sg})$.
An illustration of both $PQt$ profile and belief is presented in Fig.~\ref{fig:PQtBFPVASGA}(b).

\begin{figure}[h!]
\centering
\includegraphics[width=0.55\textwidth]{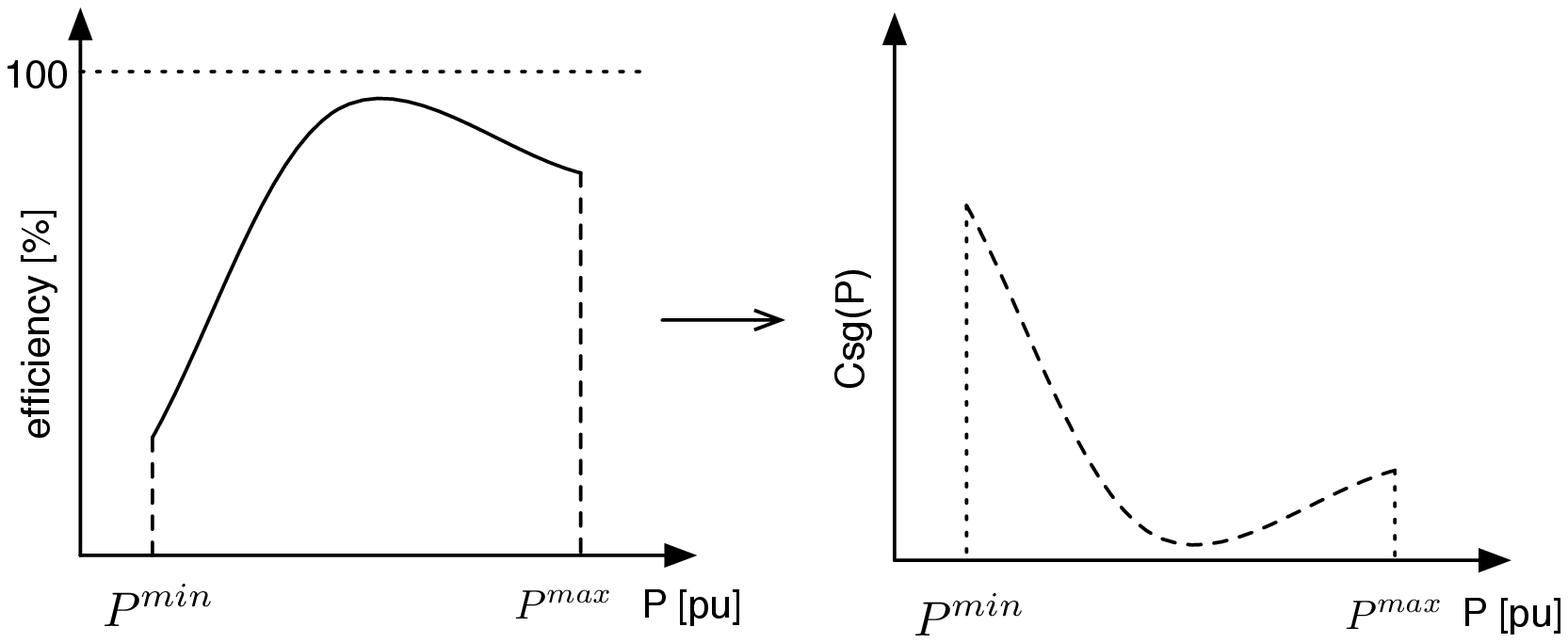}
\caption{Efficiency curve of a turbine (left) and cost function of SG agent (right).\label{fig:CostSG}}
\end{figure}

\textbf{Virtual Cost}. To express the virtual cost, we consider that the SGA operates the resource in order to maximize its overall efficiency.
As the efficiency of the turbine given an electrical produced power, $\eta (P)$, plays the most important role in the overall efficiency, we define the virtual cost as $C_{sg}(P, Q) = a_{sg}(1-\eta(P))$. As an example, the cost function for $\mu$HA is given in Fig. \ref{fig:CostSG}, with $a_{sg}$ a positive constant.

\subsection{Water Boiler} \label{subsec:LA DR}
We consider thermal controllable loads such as \emph{water boilers} (WB). Each controllable load is considered as a \emph{single} boiler capable of estimating its own thermal state. The approach can be extended to \emph{distributed} controllable loads (e.g., \cite{GECN}), but this is out of the scope of this paper. Also, a more sophisticated definition of a controllable load agent representing space heating has been presented in \cite{flexhouse}.

In this section, we consider only active power, with $Q \equiv 0$ throughout.

\textbf{Implementation of Setpoints}. We assume that the internal controller of the WB is capable of any active power in $[0, P_{wb}^{\max}]$.
From our sign convention it follows that $P = 0$ represents the case when the heating device is off, and $P = -P_{wb}^{\max}$ represents the case when the heating device is working at full power.
Next, we assume that the thermal state of the boiler is represented by the total energy stored in it at time $t$, given by
\vspace*{-0.5cm}
\[
\El(t) = \int_{\tau = 0}^t (P_{in}(\tau) - P_{out}(\tau)) d \tau,
\]
where $P_{in}(t)$ is the absolute value of the power injected into the system, and $P_{out}(t)$ is the absolute value of the power drawn from the system. The process $\{P_{out}(t)\}$ is the source of uncertainty in this resource, as it is affected by nature and demand patterns of the users of the boiler. The process $\{P_{in}(t)\}$, on the contrary, is controlled by the WB agent (WBA).

\begin{figure}[h!]
\centering
\subfloat[]{
\includegraphics[width=0.45\textwidth]{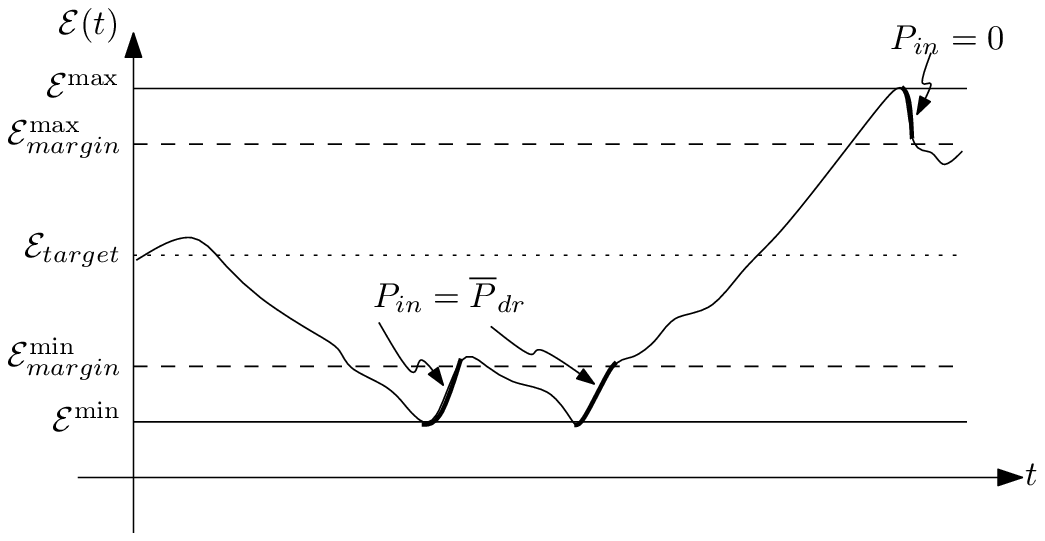}}
\subfloat[]{
\includegraphics[width=0.25\textwidth]{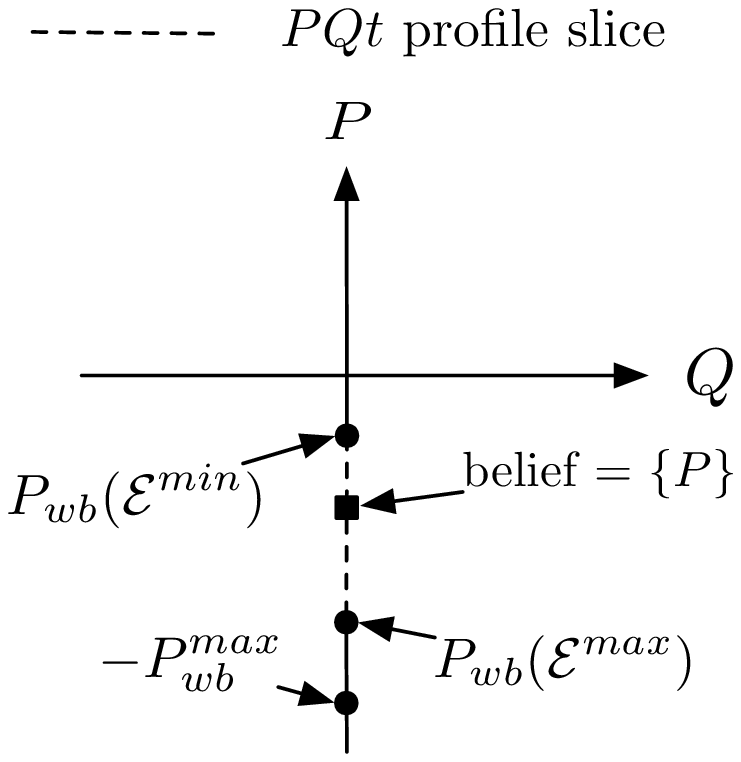}}
\caption{(a) A possible trajectory of boiler energy as a function of time $t$. (b) $PQt$ profile and belief function for a WB agent.\label{fig:energyPQtBFLDR}}
\end{figure}

The constraints on the energy are given by four parameters $\El^{\min} < \El_{margin}^{\min} < \El_{margin}^{\max} < \El^{\max}$ \footnote{It is assumed that there are two levels of stored energy margins: $\El_{margin}^{\min}$ and $\El_{margin}^{\max}$ have been intended to assure an acceptable margin of operation}. Given a requested setpoint $P \in [-P_{wb}^{\max}, 0]$, WBA commands the internal controller to maintain $P_{in}(t) = -P$ as close as possible. Whenever $\El(t) < \El^{\min}$, it switches the setpoint to the maximal heating power (namely, to $P_{in}(t) =  P_{wb}^{\max}$), until $\El(t) \geq \El_{margin}^{\min}$. Then, it switches back to the original request, until the energy constraint is violated again. A similar process is assumed when $\El(t) > \El^{\max}$.
Fig.~\ref{fig:energyPQtBFLDR}(a) shows this concept.

\textbf{$PQt$ Profile}. We assume that the
stored energy is constant until the next request implementation (such an assumption is assumed to be always satisfied in view of the large difference between the period of computation of the setpoints $T$ and the resource's time constants).
Hence, the $PQt$ profile is specified by the interval $[0, -P_{wb}^{\max}]$.
An example of $PQt$ profile is shown in Fig. \ref{fig:energyPQtBFLDR}(b).

\textbf{Belief Function}.
In contrast to a regular storage device, the WB load can be highly uncertain.
To account for this uncertainty, we assume that the WBA has a forecasting tool to predict the load profile. Let $[\underline{P}_{out}^f(t), \overline{P}_{out}^f(t)]$ denote
the confidence interval of the forecast at time $t$. To compute the belief set at time $t$ for a given setpoint $P$, the WBA first computes the worst cases ($\hat{\El}^{\min}(t')$, $\hat{\El}^{\max}(t')$) of the estimated energy at times $t' \in [t, t + T]$ by using the confidence interval. The belief $BF_{wb}(P)$ is then given by either $\{P, 0\}$, $\{-P_{wb}^{\max}, P\}$, or $\{-P_{wb}^{\max}, P, 0\}$ depending on whether, for some $t'$, $\hat{\El}^{\max}(t') > \El^{\max}$ or $\hat{\El}^{\min}(t') < \El^{\min}$, or both. Otherwise, if no violation occurs, the belief is $BF_{wb}(P) = \{P\}$.
An example of belief function is given in Fig. \ref{fig:energyPQtBFLDR}(b).

\textbf{Virtual Cost}. Similarly to the ESS agents, we assume that the basic goal of WBA is to keep the stored energy at a certain target level $\El_{target}$. Therefore, the virtual cost function advertised by WBA is similar to that advertised by  ESSA as shown in Fig.~\ref{fig:ADV_ESSA}(b), but centered around the forecasted value of the demand given by $P_{center} = -P_{out}^f(t)$.

\subsection{Uncontrollable Load} \label{subsec:LA}

\textbf{Implementation of Setpoints}. The UL agent (ULA) does not take into account the requested setpoint as it does not have any way to set it.

\textbf{$PQt$ Profile}.
We implement the simplest case, where the $PQt$ profile is given by $\{x_l^f(t)=(P^f, Q^f)\}_{t=t_0}^{t_0+T}$.
Specifically, for each time step, the $PQt$ profile is defined by a \emph{single point} $x_l^f(t)$ given by \emph{a demand forecasting tool}.

\begin{figure}[h!]
\centering
\includegraphics[width=0.3\textwidth]{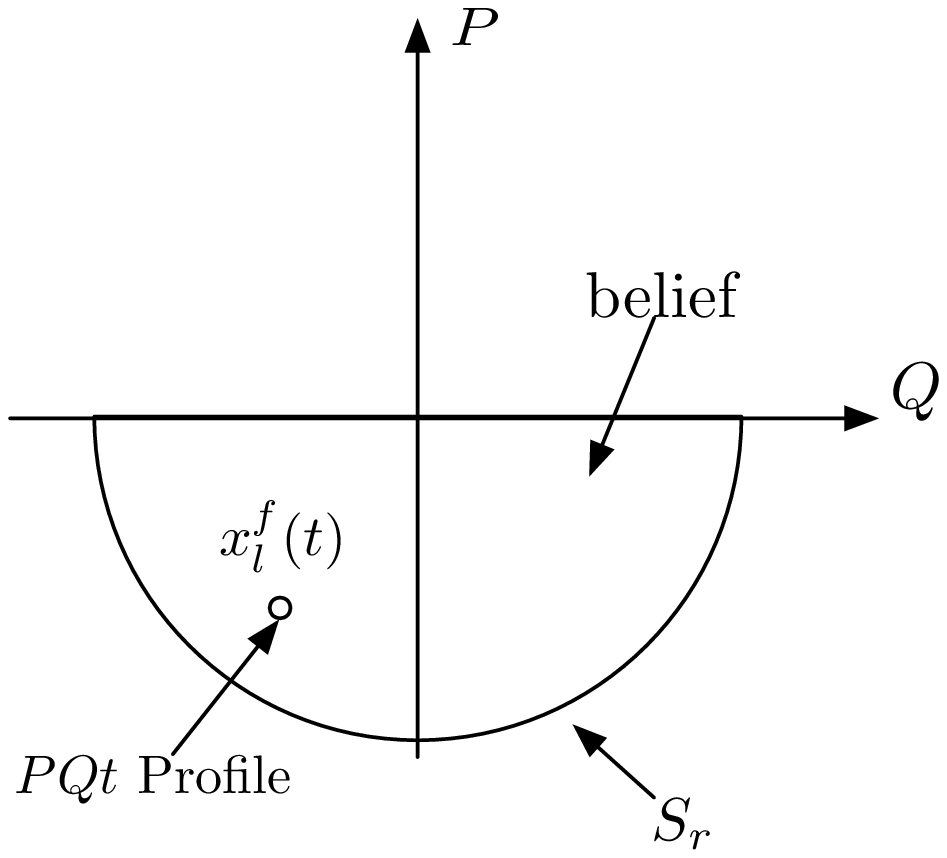}
\caption{$PQt$ profile and belief function for UL agent.\label{fig:PQtBFL}}
\end{figure}

\textbf{Belief Function}.
In this paper, we assume that the UL can change to \emph{any} admissible value at any moment.
Hence, the belief is considered as the complete area of operation of the UL.
We assume that the consumption of the UL is always inside the semi-circle defined by its maximum apparent power $S_r$ (or $\rho_{max}$ in polar coordinates); that is, it can consume active power and to inject or absorb reactive power.
With this representation, the belief is defined by $BF_l= \{(\rho, \theta): \rho \in [0, \rho_{\max}], \theta \in \l[180^{\circ}, 360^{\circ}\r]\}$, as can be seen in Fig. \ref{fig:PQtBFL}.

\textbf{Virtual Cost}. As the ULA cannot control its resource, we set the advertized virtual cost to
$
C_{ul}(P,Q)=0.
$
We note that, in our implementation of the grid agent, only the \emph{gradient of the cost} is used by the employed gradient descent algorithm (see Subsection~5.2.1 in Part~I for details). Hence, the uncontrollable load agents can have any \emph{constant} cost without influencing the setpoints computation procedure.

For the different resources here presented, we use the parameters shown in Table \ref{table:resources_param}.

\begin{table}[h!]
\caption{Resources Parameters\label{table:resources_param}}
\subfloat{
\begin{tabular}{l|c|c}
\hline
Parameter	&	Value	 	&	Unit		\\
\hline
\multicolumn{3}{c}{Solar Plants}	\\
\hline
Rated power PV$1-3$ 	&	10	&	$[kW]$	\\
Rated power PV$4$					&	3	&	$[kW]$	\\
\hline
\multicolumn{3}{c}{Uncontrollable Loads}	\\
\hline
Rated power UL			&	250	&	$[kVA]$	\\
Rated power UL$1-2$		&	15	&	$[kVA]$	\\
\hline
\multicolumn{3}{c}{Water Boilers}	\\
\hline
Max power WB$1$	&	50	&	$[kW]$	\\
Max power WB$2$	&	47	&	$[kW]$	\\
Max power WB$3$	&	72	&	$[kW]$	\\
Max energy		&	20	&	$[kWh]$	\\
Min energy		&	1	&	$[kWh]$	\\
Upper margin			&	19	&	$[kWh]$	\\
Lower margin			&	2	&	$[kWh]$	\\
\hline
\end{tabular}}
\subfloat{
\begin{tabular}{l|c|c}
\hline
Parameter	&	Value	 	&	Unit		\\
\hline
\multicolumn{3}{c}{Energy Storage Systems}	\\
\hline
Rated power	ESS					&	250	&	$[kVA]$	\\
Rated energy	ESS					&	500	&	$[kWh]$	\\
Converter efficiency ESS			&	98	&	$[\%]$	\\
Rated power	ESS$1$				&	30	&	$[kVA]$	\\
Rated energy	ESS$1$				&	30	&	$[kWh]$	\\
Converter efficiency ESS$1$		&	97	&	$[\%]$	\\
\hline
\multicolumn{3}{c}{Synchronous Generators}	\\
\hline
Rated power SG			&	250	&	$[kVA]$	\\
Rated power $\mu$H		&	30	&	$[kVA]$	\\
Minimum active power		&	1	&	$[pu]$	\\
Maximum active power		&	0.2	&	$[pu]$	\\
Synchronous reactance	&	3.07&	$[pu]$	\\
Transformer reactance	&	0.1	&	$[pu]$	\\
Exc. current no-load		&	1	&	$[A]$	\\
Exc. current load		&	3.6	&	$[A]$	\\
\hline
\end{tabular}}
\end{table}

\section{Simulation Results} \label{sec:results}
Below we present the comparison between the behaviour of Commelec and the two above mentioned droop-based control strategies, which is followed by the validation of the employed aggregation methods.

\subsection{Short-Term Behaviour}
In this section, we compare the results obtained in the scenario described in Section \ref{sec:scen}, with three different control methods: Commelec, Droop with only primary frequency control control (DP) and primary voltage control only in the slack resource (in our case the ESS), and Droop with additional secondary frequency control (DPS) at the slack resource. The focus here is on the \emph{dynamic short-term behaviour}. In particular, the results are presented over the time horizon of $1600$ seconds.

\subsubsection{Control of the Reserve of the Storage Systems}
\begin{figure}[h!]
\centering
\includegraphics[width=1\textwidth]{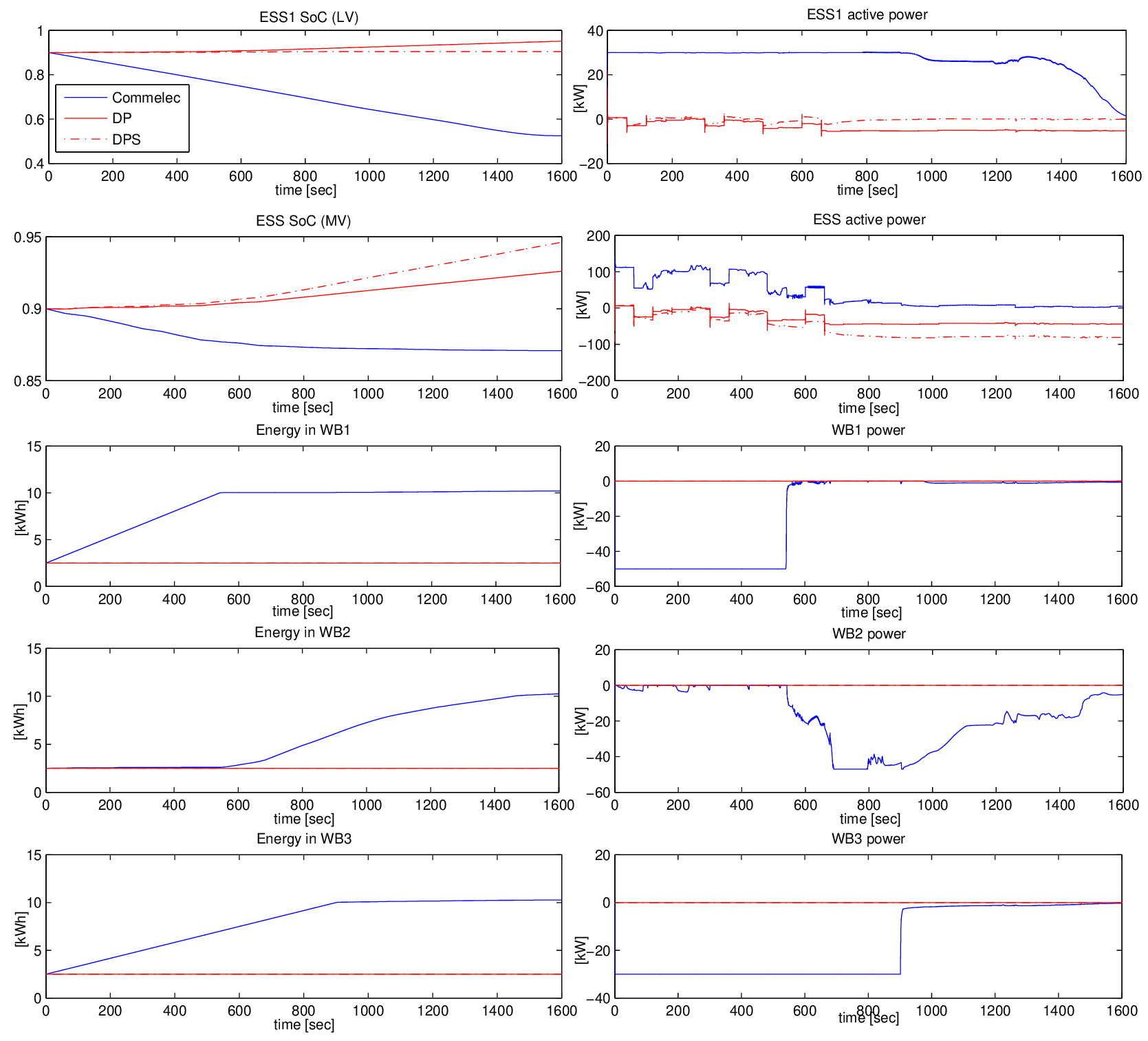}
\caption{Comparison between Commelec and both droop strategies. The left column presents the state-of-charge of the battery systems and the stored energy in the water boilers, while the power profile of the same elements is presented in the right column.\label{fig:SoC_CvsD}}
\end{figure}

The evolution of the state of charge (\emph{SoC}) of both battery systems is shown in Figure \ref{fig:SoC_CvsD}.
Note that in the case of Commelec, the \emph{SoC} decreases towards the target value ($SoC = 0.5$) as opposed to DP/DPS, in both LV and MV networks.
In the case of the LV battery, when using Commelec, the \emph{SoC} decreases much faster because this resource is being requested to discharge mostly at full power; whereas in the case of the MV battery, it is discharging but subject to the fact that this resource is the slack bus of the system (therefore its power production/absorption is the result of all other resources).

The evolution of the \emph{SoC} of the water boilers is also presented in Figure \ref{fig:SoC_CvsD}.
It can be seen that the boilers are locally controlled to react to power variations in the network while following their willing to be charged.
WB$1$ and WB$3$ are being charged from the beginning at full power, whereas WB$2$ is charged when possible.
On the contrary, in DP/DPS, the boilers are not charged at all.

\subsubsection{Reduced Curtailment of Renewables}
\begin{figure}[h!]
\centering
\includegraphics[width=1\textwidth]{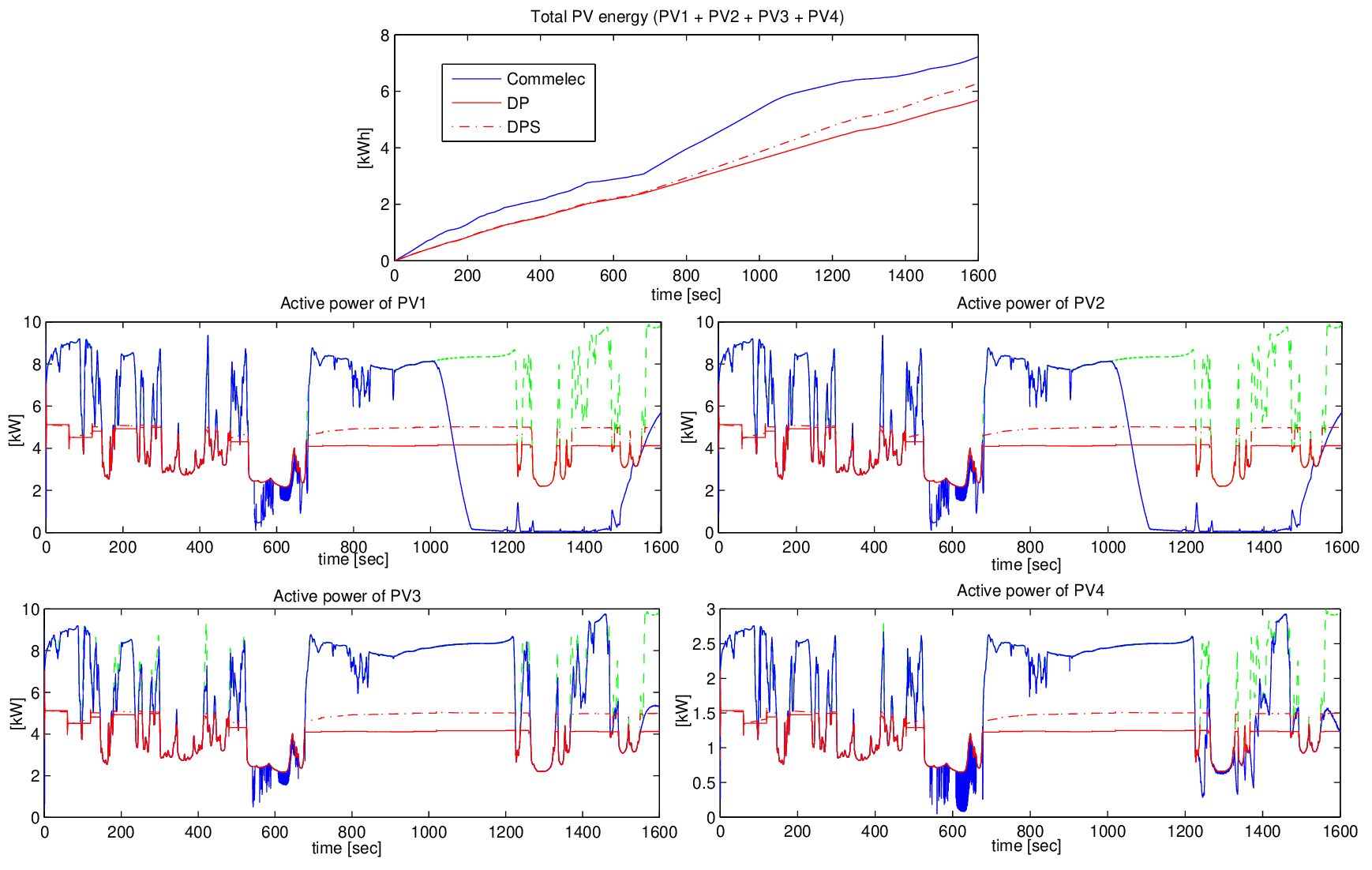}
\caption{Comparison between Commelec and both droop strategies. Total produced energy for the four PV plants connected to the LV microgrid and the power production for each. The dashed green line represents the maximum power production following directly the irradiance profile.\label{fig:PV_CvsD}}
\end{figure}

Figure \ref{fig:PV_CvsD} shows the production of the PVs, by means of the PV active power and the total produced PV energy.
It can be seen that in Commelec, the PVs produce at maximum available power most of the time; whereas in DP/DPS, the PV power is curtailed given the excess of power in the network assessed by the frequency signal.
In this respect, with the proposed method the renewables production is maximized even with high variability profiles and it is curtailed only when it affects the power quality or there is not enough storing capacity in the system.

\subsubsection{Local Power Compensations and Exploitation of Degrees of Freedom}
\begin{figure}[h!]
\centering
\includegraphics[width=0.9\textwidth]{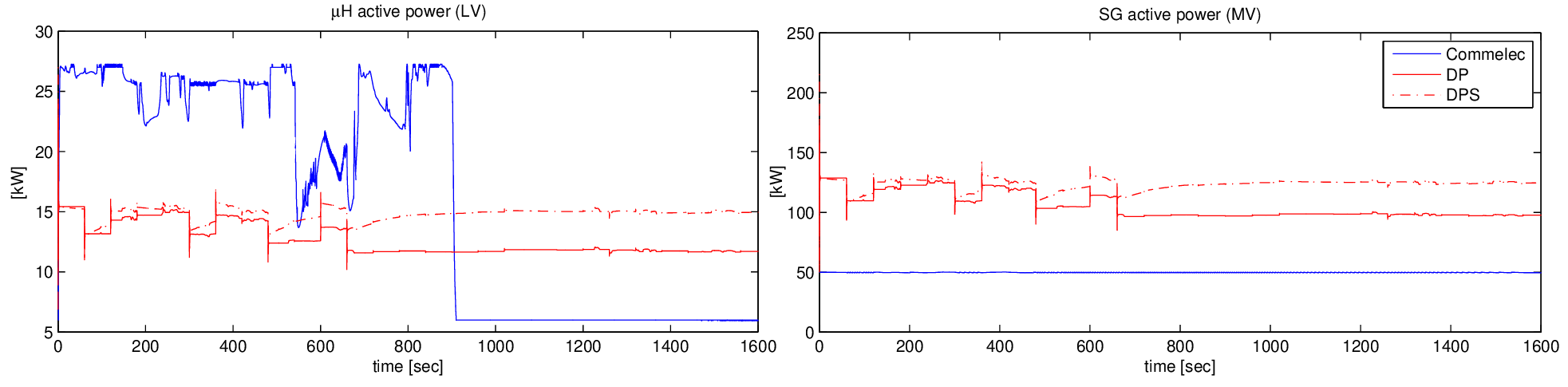}
\caption{Comparison between Commelec and both droop strategies. Active power production of SG and $\mu$H.\label{fig:SG_CvsD}}
\end{figure}

Figure \ref{fig:SG_CvsD} shows the production of the synchronous generators (SG and $\mu$H).
It is worth noting that in the case of Commelec, the power variations in the LV grid are compensated locally by the means of $\mu$H,  while maintaining the MV SG at minimum power.
In the droop simulation, on the contrary, both machines react in the same way.
The main reason for this difference is that Commelec exports and use the internal state of the resources, whereas in DP/DPS the control is performed via the global frequency signal.

\begin{figure}[h!]
\centering
\includegraphics[width=0.9\textwidth]{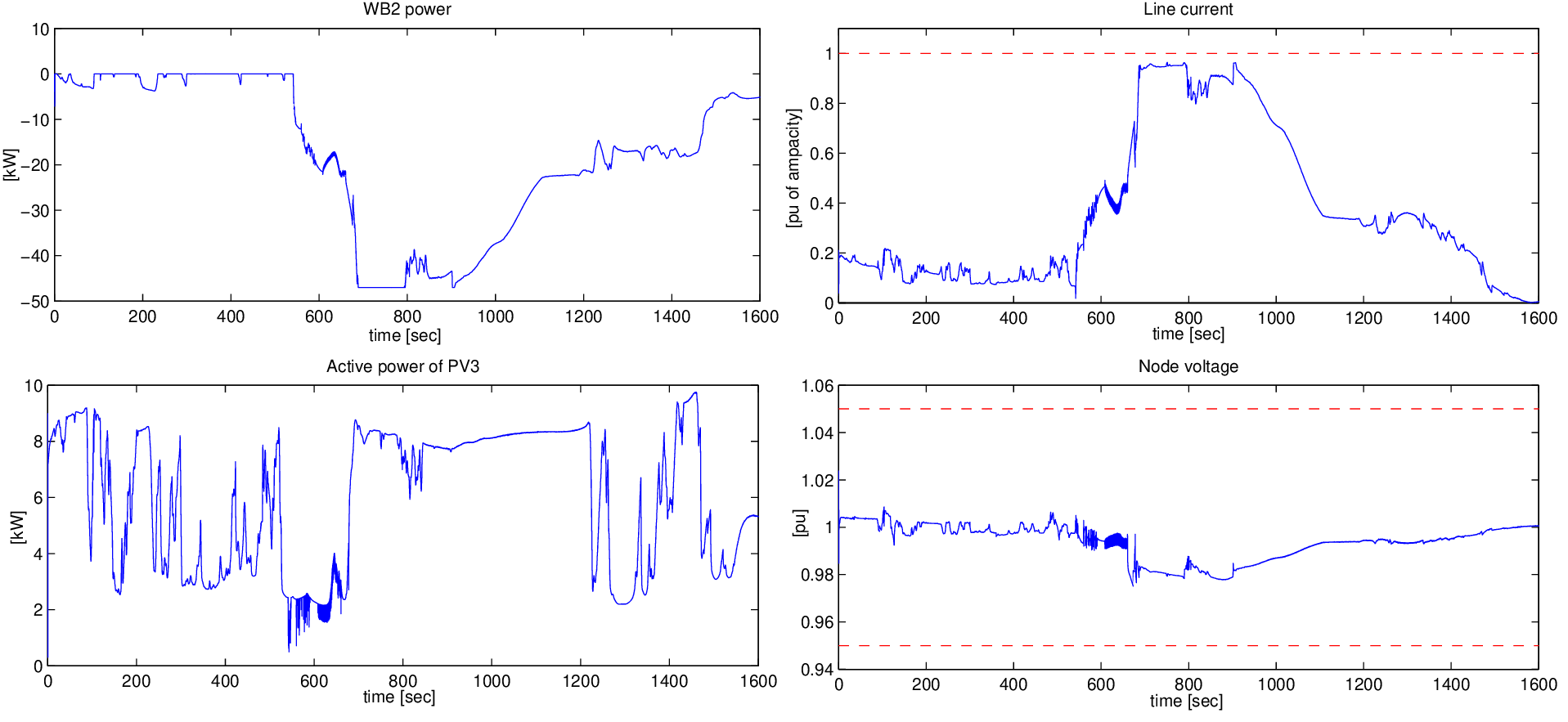}
\caption{Local power management between WB$2$ and PB$3$. The left column shows the power profiles while the right column shows the current of the line connected to both resources and the voltage of the common node. The dashed red lines represent the bounds.\label{fig:WB2_PV3}}
\end{figure}

It is interesting to observe specifically the case of WB2, which is connected to the same bus with PV3 (see Fig.~\ref{fig:Grid&AgentsCaseStudy}).
This node is then connected to the main feeder of the LV network by a line with an ampacity close to the current being absorbed by WB2 at its rated power.
We show the dynamic behaviour of these two devices in Figure \ref{fig:WB2_PV3}.
It can be seen that WB2 starts charging around $t = 550[sec]$.
This becomes possible due to the overall state of the system, and in particular due to the fact that WB1 stops charging at this time (see Figure \ref{fig:SoC_CvsD}).
However, due to low production from PV3 at this time and the weakness of the line that connects both devices to the network, the charging is not at the maximum possible power.
When the production of PV3 increases at around $t = 650[sec]$, WB2 starts charging at maximum power.
We note that the line current remains below the ampacitity during the whole process.
This case illustrates again the ability of our method to compensate for power imbalances locally and to exploit the various degrees of freedom of the system by using the advertised information about the internal state of the devices.

\subsubsection{Quality of Service and Stable Frequency}
\begin{figure}[h!]
\centering
\includegraphics[width=0.7\textwidth]{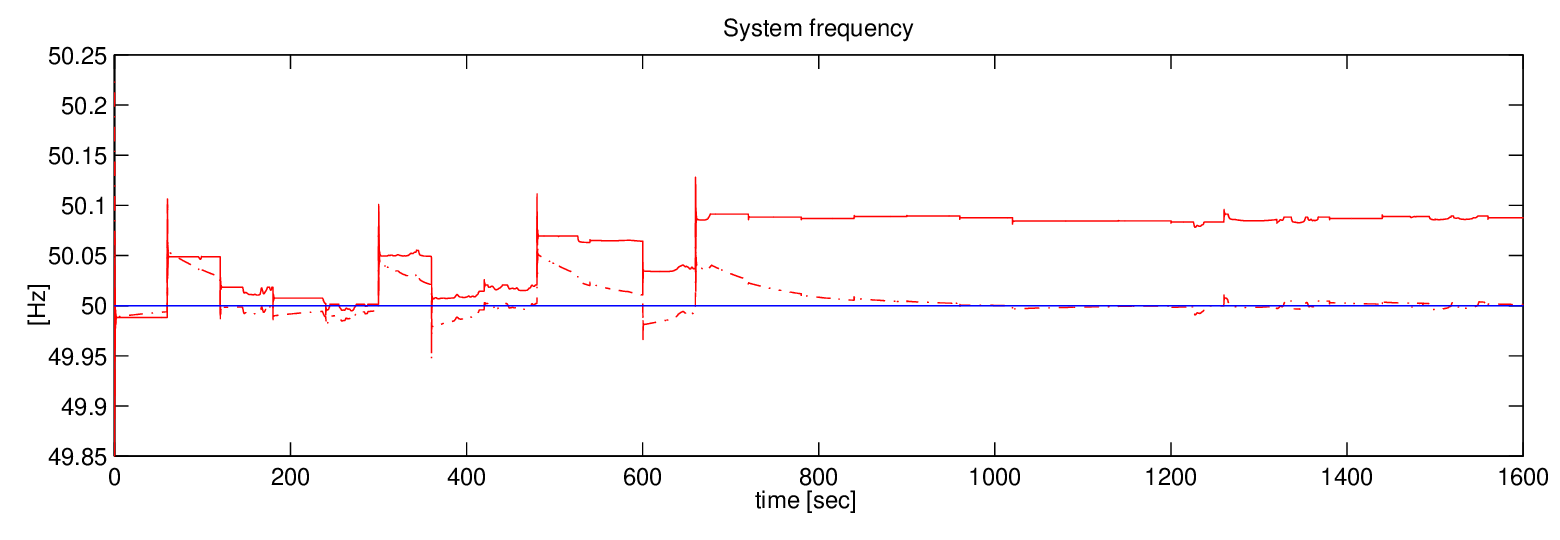}
\caption{Comparison between Commelec and both droop strategies. System frequency.\label{fig:freq}}
\end{figure}

In Figure \ref{fig:freq}, the system frequency is presented.
Recall that the slack bus is the MV storage system (ESS).
As the Commelec method is explicit, the slack works at a constant frequency (i.e. $50[Hz]$).

In the case of DP, on the contrary, the frequency reacts to the changes in UL; whereas in the case of DPS, it tries to return to the reference value.
It is important to note that the frequency variations are highly dependent on the droop parameters of different devices and can be very high when there is a sudden change in the network.
Therefore, by keeping the frequency constant, Commelec allows for a more accurate control of the speed of electrical machines. This is true especially in a microgrid that, when controlled using standard droop-based strategies, is expected to face high variability of the frequency signal due to the uncertainty of the renewables.

\begin{figure}[h!]
\centering
\includegraphics[width=0.95\textwidth]{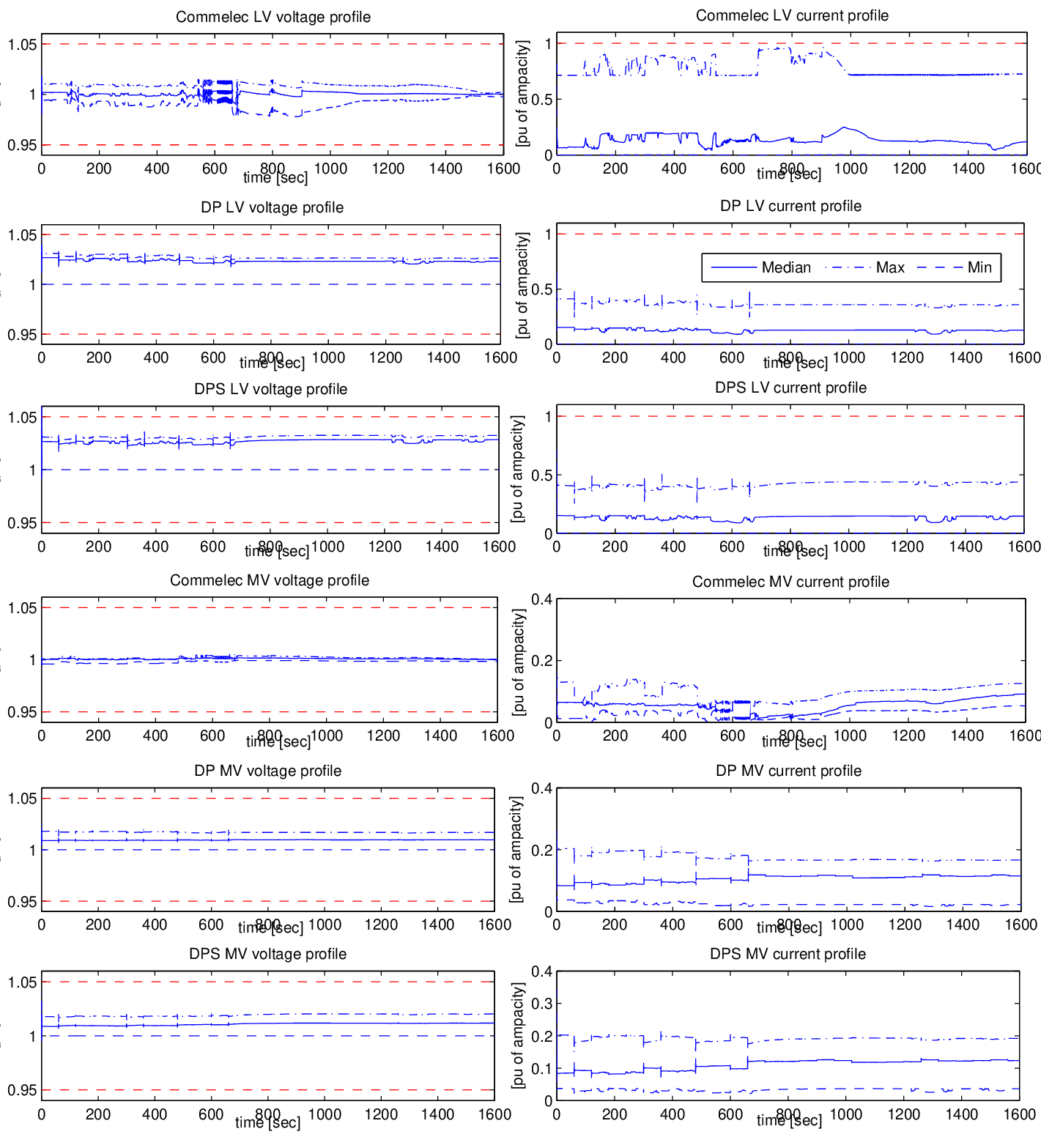}
\caption{Comparison between Commelec and both droop strategies. The red dashed lines represent the predefined bounds for voltage and line congestions.\label{fig:QoS}}
\end{figure}

In Figure \ref{fig:QoS}, we present the aggregated voltage and current profiles for both networks (i.e., median, minimum and maximum values of all node voltages and line currents).
It can be seen that the improvement in the overall operation obtained by using our method does not affect the quality of service.
The voltage and current magnitudes are always maintained within the acceptable regions.
Note that in Figure~\ref{fig:QoS}, the maximum LV current profile for Commelec is always close to the ampacity.
This specific case is related to WB$2$ and PV$3$ as explained before.
Observe, however, that the median value is much lower during the entire simulation run.

\subsection{Medium-Term Behaviour and System Collapse}
\begin{figure}[h!]
\includegraphics[width=0.95\textwidth]{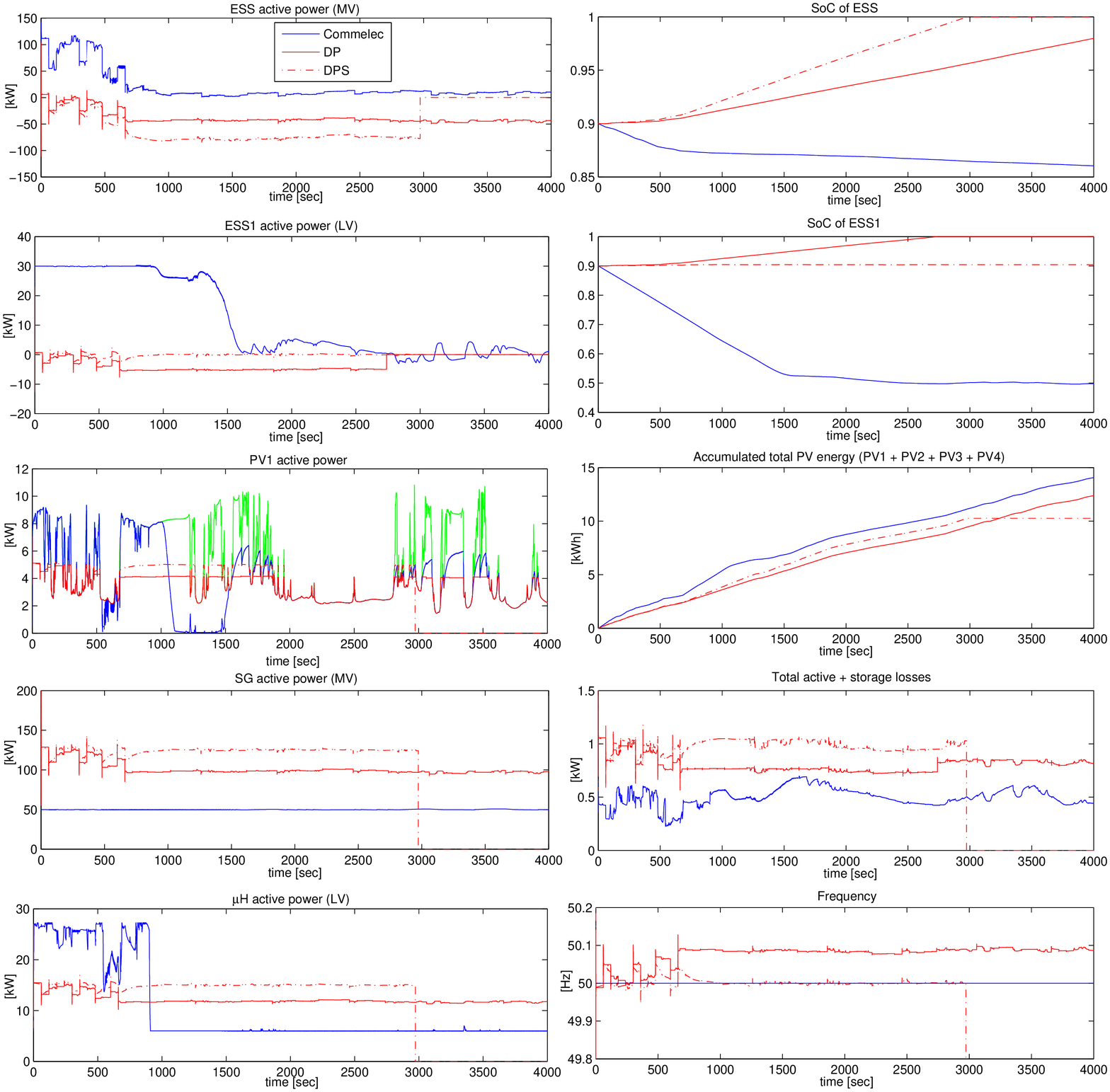}
\caption{Comparison between Commelec and both droop strategies. Medium-term comparison where the batteries are overcharged using DP/DPS strategies due to the production of renewables, even when curtailing their production. \label{fig:collapse}}
\end{figure}

In this subsection, we illustrate the medium-term system behaviour in the critical case corresponding to the overproduction from renewables with initial high value of the \emph{SoC} of the batteries and minimum load.
Specifically, we present the \emph{SoC}, the production of a PV, and the injection of SG and $\mu$H in Figure \ref{fig:collapse}, over the time horizon of $4000$ seconds (namely, around $1$ hour).
It can be seen that both DP and DPS control strategies lead to the overcharge of the MV battery, essentially causing the \emph{collapse of the system}.
In particular, when the power is injected into the ESS with $SoC=1$, the local controller of the resource trips its breaker, with the consequent loss of the slack resource provoking the collapse.
The main reason for this behaviour is that the droop strategies force the generators to overproduce power regardless of the \emph{SoC} of the slack resource.
It is worth noting that in DP, as there is a permanent positive frequency error, the LV battery (ESS$1$) is always being charged. Hence, it trips even before the MV battery (ESS).
The early loss of ESS$1$ can be also interpreted as a lack of autonomy of the microgrid if islanded.
In the case of DPS, the secondary frequency control allows for a larger production of the generation units, and therefore the \emph{SoC} of ESS$1$ is essentially constant.
As a result, the MV battery is charged without restriction.
On the contrary, Commelec keeps the \emph{SoC} of both ESS and ESS$1$ away from the margins by using internal information from each resource and controlling explicitly their power setpoints.

\subsection{Unexpected Disconnection of a Device}
In this subsection, we demonstrate how the Commelec method is able to cope with an unexpected disconnection of a device. In particular, at $t = 1000$ s the resource PV1 and its agent are disconnected. Immediately, the slack resource (ESS) reacts to cope with the imbalance. We remind that, as mentioned above, the control of the slack resource can be astatic. As a consequence, the compensation performed by the slack has no impact on the system frequency. Afterwards, Commelec takes over. Note that PV1 is directly connected to WB1 and PV2. As WB1 is already close to its minimum power, and PV2 aims at producing at maximum, the algorithm also reduces the consumption of WB2 and WB3 (connected to different nodes) to assist the maneuver.
The simulation shows how Commelec handles unexpected disconnection by assisting the slack bus in redistributing the power imbalance between the resources and by
keeping the overall state of the grid feasible.

\begin{figure}[h!]
\includegraphics[width=0.95\textwidth]{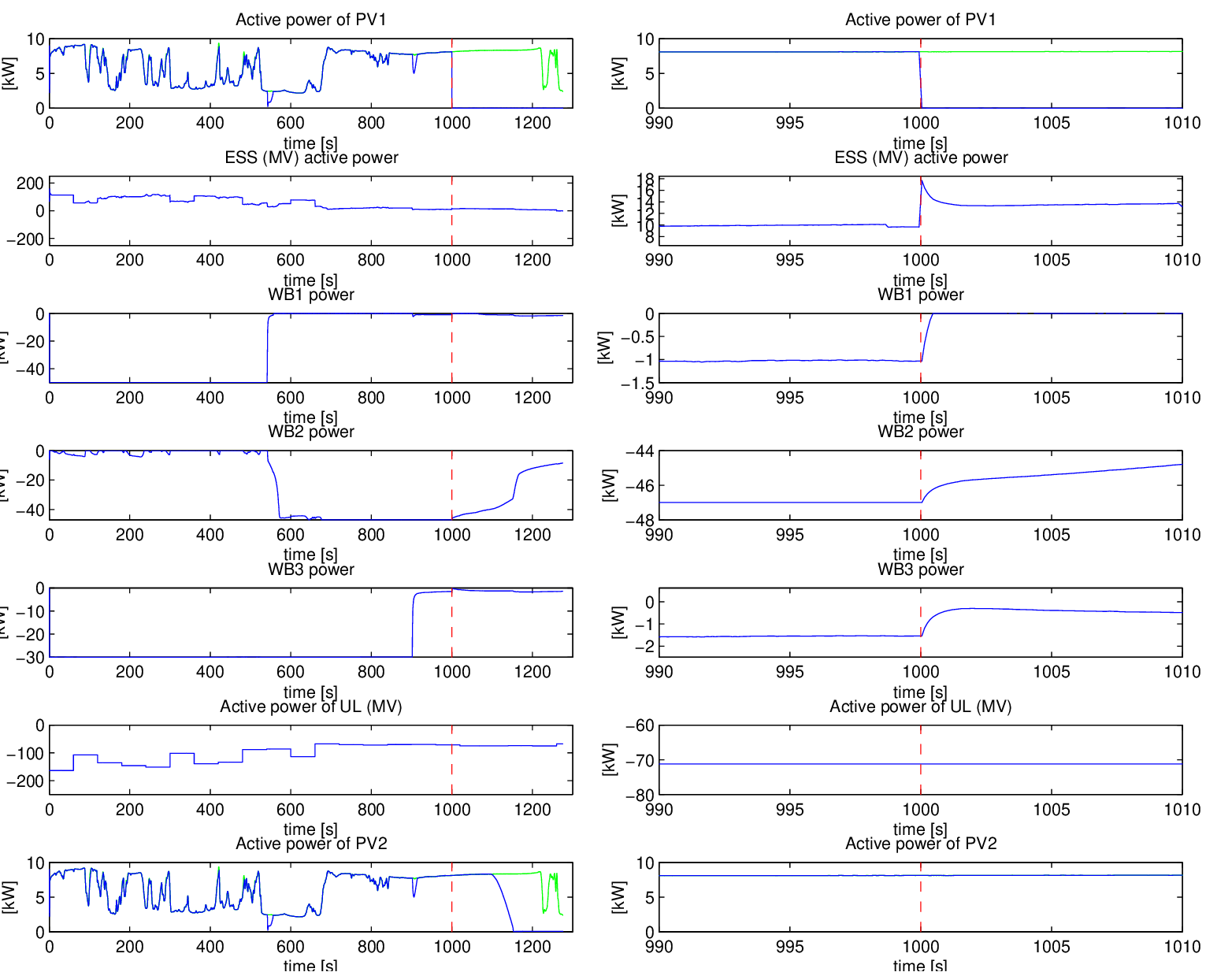}
\caption{Unexpected disconnection of PV1 at $t = 1000$ s. The right column shows a zoom on the left column in the time window of $[990,1010]$ s. \label{fig:disc}}
\end{figure}

\subsection{Validation of the Aggregation Methods}
\label{subsec:val_aggr}

In this section, we numerically validate the aggregation methods described in Section 6 of Part I of this paper.
To this end, we performed a simulation of the ``flat'' setting of agents shown in Figure \ref{fig:AgentsCaseStudy}(b).
In order to make a fair comparison between the results obtained in the standard (hierarchical) setting of agents (Figure \ref{fig:AgentsCaseStudy}(a)) and those obtained in the flat setting, we adjusted the weights of the objective function and the step-size parameters accordingly. In particular, the step-size parameter in the flat case was set to that of the MVGA, and the weights of the microgrid resources were multiplied by the ratio between the step-size of the LVGA and MVGA.

\begin{figure}[h!]
\includegraphics[width=0.95\textwidth]{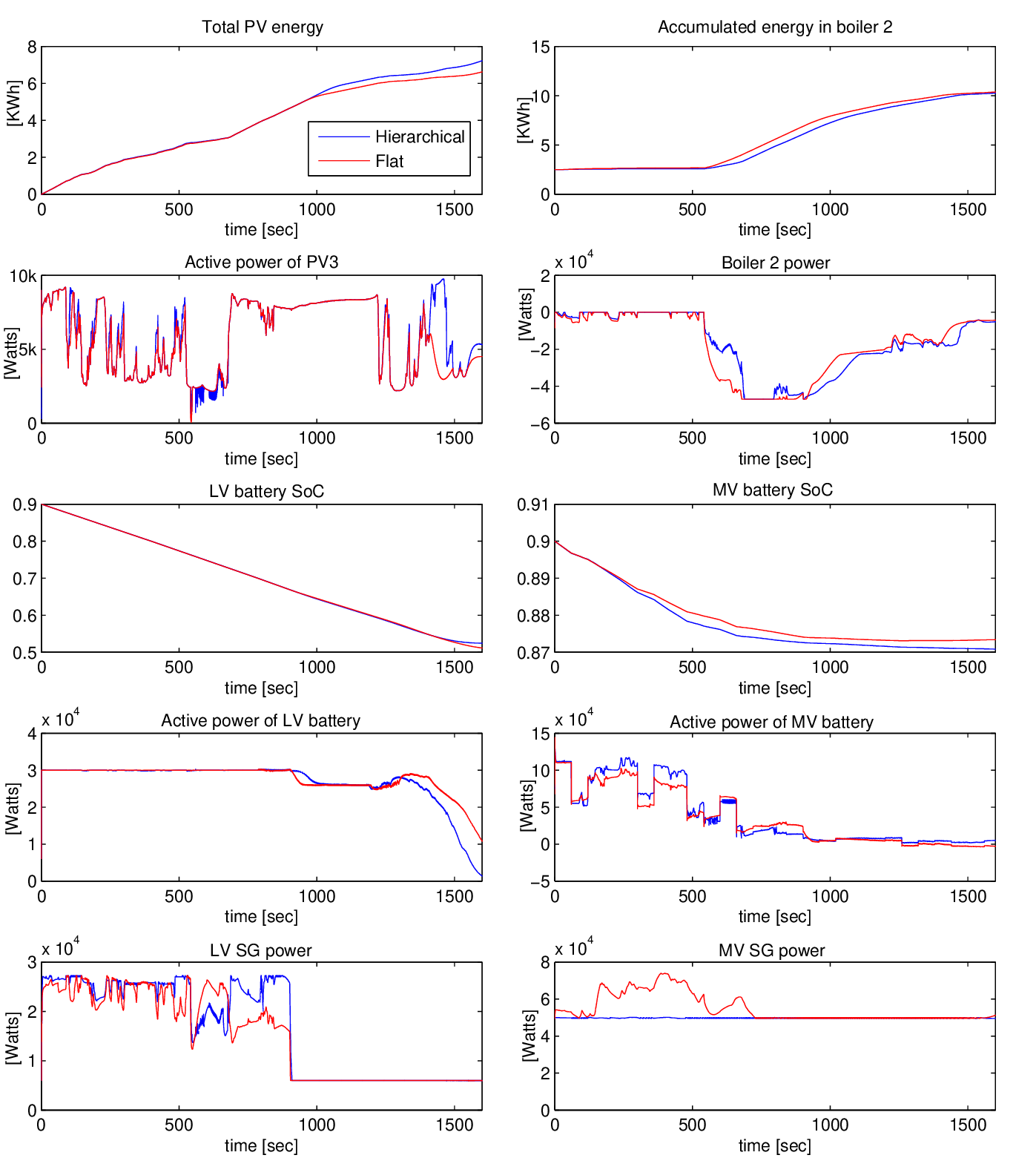}
\caption{Comparison between the hierarchical and flat agents architecture using the Commelec protocol. \label{fig:HvsF}}
\end{figure}

Figure \ref{fig:HvsF} presents a comparison between the results obtained in the two settings.
As it can be observed, the overall behaviour is similar. The main difference is in the profiles of the synchronous generators, where a difference of up to 20 kW can be observed in the injection of SG.
However, the contribution of this difference to the overall behaviour is negligible, as can be inferred from the presented energy metrics (\emph{SoC}, and PV and boilers energy).

As shown in Proposition 5.1 in Part~I, the two settings are equivalent under the ``ideal'' conditions stated there. In our implementation, however, there are three main reasons for the observed difference. First, there is a natural difference due to the approximate methods used for aggregation. Second, recall that we implement a gradient-based algorithm rather than solve an exact optimization. Moreover, in the hierarchical setting, the LVGA is requested to provide a certain fixed power at the connection point, whereas in the flat setting this power can vary without any prescribed restrictions.
Third, the projection algorithms used to compute the control are \emph{randomized}. In particular, in Algorithm 2 presented in Appendix~A in Part~I, in order to efficiently find the direction of minimum violation, we draw setpoints uniformly and randomly.

\section{Discussion} \label{sec:disc}
This section presents a general discussion of the proposed control framework in this two-part paper, with a focus on extensions and future directions.

\subsection{Extension to Higher-Level Grids}
As shown in this part of the paper, the proposed framework achieves several desirable performance goals in an islanded distribution network, in the presence of highly volatile resources.
These goals are achieved through a simple and generic method, with a key property of composability. It can be seen that this property allows us to easily extend our method to higher levels of power grids, up to the transmission level.

\subsection{Partially Controlled Grids} \label{sec:part_contr}
In this paper, we assume that a grid agent faces resources that are fully controlled by Commelec agents. In a practical deployment, however, it is envisioned that certain parts of the grid cannot be controlled explicitly.
It is possible to extend our framework to this ``partially controlled'' case. In particular, a ``shadow agent'' can be attached to each ``non-Commelec'' part of the grid. This shadow agent will take place of the regular Commelec agent. It will monitor the behaviour of that part, and estimate its model (by using, e.g, network equivalence methods \cite{net_equiv} or heuristic forecasting tools). By using this model, the agent can forecast the power production/consumption of that part of the grid and capture the uncertainty of this forecast in a belief function (similarly to the uncontrolled load agent), which in turn can be used in the optimization performed by the GA.

\subsection{Robustness to Faults and Their Treatment}
We note that the proposed method naturally relies on the communication infrastructure for transmitting messages. In this paper, in the simulated case study, we assumed a perfect communication channel, with no message losses.
Moreover, we did not take into account a possible failure of agents to produce advertisement messages, which might lead to incomplete information at the leader side.
In this subsection, we outline how our method can be extended in order to make the communication between the agents more robust, and how to treat the failures when they occur.

First, we envision a \emph{triplicated implementation} of the grid agents (and possibly some important resource agents, e.g., those responsible for storage devices).
In a normal mode of operation, each copy will perform the same computation and issue the same requests and advertisements. The triplicated data will be then sent to the receiver agent that will validate it using standard methods for validation of triplicated data (e.g., using voting mechanism based on some distance metric between the messages). Provided that the GA copies are synchronized to a certain extent, such a method would ensure correctness in the face of a failure of a single copy and/or loss of a single message. The communication protocol can be easily adapted to account for triplicated messages.

Second, a special communication networking infrastructure is envisioned to prevent packet losses; traffic engineering will be used to reduce congestion losses; source coding for long messages will mitigate the effect of packet losses; and a parallel redundancy protocol \cite{iprp} will be used to provide instant packet loss repair.

Next we outline how failures can be treated when they occur. From the point of view of a follower agent, if there is no valid request setpoint from the leader (either due to the loss of messages or due to validation failure), the agent can move to its \emph{backup mode}.
In this mode, the agent will produce setpoints according to some internal decision process, within the feasibility of the system, with the little information that is available.
For instance, a resource agent can use a droop-based control method, whereas a grid agent can operate in a similar way described in this paper but without the term penalizing the deviation from the request at the connection point.

From the point of view of a leader agent, a ``shadow agent'' will be attached to each of its followers. If a failure is detected (e.g., if the advertisement message from a follower is not received for a long period of time, or if the message validation procedure fails),
the shadow agent will take place of the real agent. These shadow agents will have a functionality and goals similar to those responsible for the uncontrolled part of the grid discussed in Section \ref{sec:part_contr}.

\subsection{Islanding Maneuver and Choice of Local Slack Resources}
In this paper, we focus on an islanded system in order to show that our method is able to operate a microgrid in an autonomous way, locally compensating for power imbalances. However, the proposed framework can be extended in order to allow for the \emph{islanding maneuver} of a connected active distribution network. In particular, given a command from the leader to perform this maneuver, the grid agent will steer the system towards the state with $0$ power at the connection point. At the same time, it will perform a ``negotiation'' with its followers in order to choose \emph{a set of slack resources}. We note that the grid agent can take its decision based solely on the advertised information from the followers (e.g., using a metric as in \cite{Mario_Islanding_Slack}). In particular, it will prefer to choose a resource with (i) a ``good'' belief function (e.g., battery or SG), (ii) a large range of available power as represented by the current $PQt$ profile, and (iii) an internal state far away from the margins as represented by the advertised cost function.

Similarly, a \emph{reconnection maneuver} can be implemented by steering the system towards a common frequency and voltage phasor at the connection point. In particular, we can add a further term in the GA objective function $J(y)$ accounting for the difference of the voltage phasors between the microgrid and the upper (larger) network, and we can instruct the slack resource in the microgrid to follow a common frequency.

\subsection{Slack Voltage Control}
In the current implementation, we assumed that the voltage at the slack bus of the system is fixed. Hence, it is not considered as a control variable.
Moreover, the grid agent responsible for the slack does not have a way to decide which slack voltage is good or bad for the system, because it does not receive any related information from its followers. For instance, in our case study, the LVGA may prefer to increase the voltage due to high consumption in the microgrid, but the MVGA does not have a way to obtain this information. Still, we can easily adapt our framework to treat this case. Specifically, a follower agent can export an additional cost function to its leader, which gives a cost to specific value of the voltage magnitude at the connection point. Then, the leader that is responsible for the slack bus can incorporate these functions in the overall optimization problem in order to choose an optimal voltage at the slack bus.

\subsection{Incorporation of Long-Term Objectives}
We note that when considering resources equipped with storage systems (such as batteries, hot water boilers, heating systems, etc.), the related long-term objectives can be incorporated easily in our framework by using the advertised cost functions.
This can be achieved using a stand-alone ``trip planner'' (that is not necessarily part of the specific resource agent) that works on a much slower time scale.
For example, consider a trip planner responsible for controlling a residential building. Typically, it will have access to long-term forecasts of consumption and production patterns. It can compute an optimal control strategy by solving a multi-time step optimization problem using methods such as Model Predictive Control (see, e.g., \cite{MPC1, MPC2}). This computation is done usually on a time scale of tens of minutes. The trip planner can then ``feed'' the Commelec resource agent with a cost function that represents this long-term control strategy. Hence, the agent can advertise this information to its leader in order to be able to ``steer'' towards the trajectory prescribed by this strategy.

\subsection{Probabilistic Profiles}
Recall that the $PQt$ profiles sent by the followers are assumed to be deterministic sets in the $PQ$ plane. As a result, the grid agent performs deterministic optimization under constraints imposed by the advertised belief functions.
However, as the agents of volatile resources, such as PVs and loads, usually base their profiles on \emph{forecasts}, probabilistic profiles can be considered. In particular, a probabilistic $PQt$ profile can be considered as a collection of conditional probability distributions $p_t(x|u)$, with the interpretation that the actual setpoint $x$ at time $t$ is distributed according to $p_t(\cdot|u)$ whenever the control is $u$. Then, the grid agent will perform \emph{stochastic optimization} using the expected value of the objective function with respect to the advertised distributions. In this case, the constraints using the belief function can be replaced by \emph{chance constraints} using the distributions $p_t(x|u)$.

In this paper, however, we choose not to pursue this direction because the stochastic optimization framework poses several fundamental problems. First, it usually assumes independence of the underlying random variables, thus not taking into account the correlation between different resources (e.g., two PV farms located at the same geographical area). Moreover, normal distributions and noise independence are usually assumed to make the computation feasible. However, these assumptions do not necessarily hold in practice.
On the contrary, developing a stochastic controller that accounts for the correlation between different resources as well as non-normal distribution of the volatilities requires the use of numerical approaches, and thus will be computationally prohibitive for our real-time application.

\section{Conclusion} \label{sec:conc}
In this sequence of papers, we have introduced a method that uses explicit power setpoints in order to perform real-time control of electrical grids in a scalable and reliable way.
The two main features of the proposed method are \emph{correctness by construction} and \emph{composability}.
The applicability of the method was verified via simulations performed on a case study composed of a low voltage microgrid benchmark (proposed by Cigr\'{e} Task Force C6.04.02) connected to a generic medium voltage feeder.
The selected case study is characterized by (i) the typical level of complexity of distribution networks, (ii) a pervasive penetration of renewable energy resources, (iii) the presence of distributed storage systems, and (iv) the fact that most of the inertia comes from storage and thermal loads rather than rotating machines.

The results of the performed simulations suggest that the proposed real-time control framework is able to efficiently steer such a system in the presence of extremely volatile energy resources.
In particular, our findings show that (a.) the method is able to indirectly control the reserve of the storage systems, thus maximizing the autonomy of the islanding operation, (b.) it dramatically reduces the curtailment of renewables and is able to implicitly identify local power compensation, (c.) it keeps the system in feasible operation conditions while better exploring the various degrees of freedom that characterize the system, and (d.) that it maintains the system's power equilibrium without using he frequency as a global variable being able to do so in inertia-less systems.
Most importantly, it prevents the system collapse in the case of overproduction from renewables.
We have also proven, by simulations, that the composability property of the proposed method holds.
This specific peculiarity will potentially enable its application to generic and more complex power systems and further research efforts are expected in this respect. It can be concluded that the proposed control scheme represents an effective \emph{actuation method} for the sub-second control of active distribution networks capable of accounting for the main requirements associated with the evolution of these grids.

\appendix
{

\singlespacing
\bibliographystyle{elsarticle-num}
\bibliography{Bib_Commelec_Paper}

} 